\documentclass[aos,preprint]{imsart}
  
\RequirePackage[OT1]{fontenc}
\RequirePackage{amsthm,amsmath}
\RequirePackage[colorlinks,citecolor=blue,urlcolor=blue]{hyperref}

\usepackage{amsgen,amsfonts,amsmath,amsthm,amstext,amsbsy,amsopn,amssymb}
\usepackage[pdftex]{graphicx}
\usepackage{verbatim}
\usepackage{comment}
\usepackage{blkarray}
\usepackage{booktabs}
\usepackage[dvipsnames,table,dvipsnames*, svgnames*]{xcolor}
\newcommand{\red }{\color{red}}

\usepackage{epstopdf}
  
\usepackage{natbib} 
\usepackage{algorithm}
\usepackage{algpseudocode}

\definecolor{blue}{RGB}{000,000,200}
\definecolor{green}{RGB}{000,150,100}
\definecolor{purple}{RGB}{220,040,250}

\def\red{\color{red}}

\newtheorem{Theorem}{Theorem}

{
	\theoremstyle{definition}

	\newtheorem{Remark}{Remark}
}

\newtheorem{Lemma}{Lemma}
\newtheorem{Corollary}{Corollary}

\newtheorem{Proposition}{Proposition}

\newcommand{\be}{\begin{equation}}
\newcommand{\ee}{\end{equation}}

\newcommand{\Var}{{\rm Var}}

\newcommand{\Cov}{{\rm Cov}}

\newcommand{\X}{{\mathbf{X}}}
\newcommand{\Y}{{\mathbf{Y}}}
\newcommand{\Z}{{\mathbf{Z}}}

\newcommand{\Bdelta}{\boldsymbol{\delta}}

\newcommand{\SSigma}{\boldsymbol{\Sigma}}
\newcommand{\vvarXi}{\boldsymbol{\varXi}}
\newcommand{\tr}{{\rm tr}}

\newcommand{\poly}{{\rm poly}}
\newcommand{\Trun}{{\rm Trun}}

\newcommand{\argmin}{\mathop{\rm arg\min}}

\newcommand{\E}{\mathbb{E}}

\def\mybox#1{\vskip1mm \begin{center} \bf \red
		\hspace{.0\textwidth}\vbox{\hrule\hbox{\vrule\kern6pt
				\parbox{.95\textwidth}{\kern6pt#1\vskip6pt}\kern6pt\vrule}\hrule}
	\end{center} \vskip-5mm}

\arxiv{arXiv:0000.0000}

\startlocaldefs
\numberwithin{equation}{section}
\theoremstyle{plain}

\endlocaldefs

\begin{document}

\begin{frontmatter}
\title{Semi-supervised Inference: General Theory and Estimation of Means\thanksref{T0}}
\runtitle{Semi-supervised Inference}
\thankstext{T0}{In Memory of Lawrence D. Brown.}
\thankstext{T1}{The research of Anru Zhang was supported in part by NSF Grant DMS-1811868.}
\thankstext{T2}{The research of Lawrence Brown was supported in part by NSF Grant DMS-10-07657.}
\thankstext{T3}{The research of Tony Cai was supported in part by NSF Grants DMS-1208982 and DMS-1403708, and NIH Grant R01 CA127334.}

\begin{aug}
\author{\fnms{Anru} \snm{Zhang}\thanksref{T1}\ead[label=e1]{anruzhang@stat.wisc.edu},
}
\author{\fnms{Lawrence D.} \snm{Brown}\thanksref{T2}\ead[label=e2]{lbrown@wharton.upenn.edu}\ead[label=u2,url]{http://www-stat.wharton.upenn.edu/{\raise.17ex\hbox{$\scriptstyle\sim$}}lbrown/}}
\and
\author{\fnms{T. Tony} \snm{Cai}\thanksref{T3}\ead[label=e3]{tcai@wharton.upenn.edu}\ead[label=u3,url]{http://www-stat.wharton.upenn.edu/{\raise.17ex\hbox{$\scriptstyle\sim$}}tcai/}}

\runauthor{A. Zhang, L. D. Brown and T. T. Cai}

\affiliation{University of Wisconsin-Madison and University of Pennsylvania}

\address{Department of Statistics\\
	University of Wisconsin-Madison\\
	Madison, WI, 53706.\\
	\printead{e1}\\
}

\address{Department of Statistics\\
The Wharton School\\
University of Pennsylvania\\
Philadelphia, PA, 19104.\\
\printead{e2}\\
\printead{u2}\\
\printead{e3}\\
\printead{u3}\\
}

\end{aug}

\begin{abstract}
	We propose a general semi-supervised inference framework focused on the estimation of the population mean. As usual in semi-supervised settings, there exists an unlabeled sample of covariate vectors and a labeled sample consisting of covariate vectors along with real-valued responses (``labels"). Otherwise the formulation is ``assumption-lean" in that no major conditions are imposed on the statistical or functional form of the data. We consider both the ideal semi-supervised setting where infinitely many unlabeled samples are available, as well as the ordinary semi-supervised setting in which only a finite number of unlabeled samples is available. 
	
	Estimators are proposed along with corresponding confidence intervals for the population mean. Theoretical analysis on both the asymptotic distribution and $\ell_2$-risk for the proposed procedures are given. Surprisingly, the proposed estimators, based on a simple form of the least squares method, outperform the ordinary sample mean. The simple, transparent form of the estimator lends confidence to the perception that its asymptotic improvement over the ordinary sample mean also nearly holds even for moderate size samples. The method is further extended to a nonparametric setting, in which the oracle rate can be achieved asymptotically. The proposed estimators are further illustrated by simulation studies and a real data example involving estimation of the homeless population.
\end{abstract}

\begin{keyword}[class=MSC]
\kwd[Primary ]{62F10}
\kwd{62J05}
\kwd[; secondary ]{62F12}
\kwd{62G08}
\end{keyword}

\begin{keyword}
\kwd{confidence interval}
\kwd{efficiency}  
\kwd{estimation of mean} 
\kwd{limiting distribution}
\kwd{semi-supervised Inference} 
\end{keyword}

\end{frontmatter}
 
\section{Introduction}
\label{sec.intro}
 
Semi-supervised learning arises naturally in statistics and machine learning when the labels are more difficult or more expensive to acquire than the unlabeled data. While numerous algorithms have been proposed for semi-supervised learning, they are mostly focused on classification, where the labels are discrete values representing the classes to which the samples belong (see, e.g., \cite{blum1998combining,zhu2005semi,ando2005framework,ando2007two,wang2007large,wang2008probability, wang2009efficient,zhu2009introduction,wang2009efficient,vapnik2013nature}). 
The setting with continuous valued $y$ has also been discussed in the literature, see, e.g., \cite{johnson2008graph}, \cite{wasserman2007statistical} and \cite{chakrabortty2017efficient}. For a survey of recent development in semi-supervised learning, readers are referred to \cite{zhu2009introduction} and the references therein. 

The general semi-supervised model can be formulated as follows. Let $(Y, X_1, X_2, \cdots, X_p)$ be a $(p+1)$-dimensional random vector following an unknown joint distribution $P = P(dy, dx_1, \ldots, dx_p)$. Denote by $P_X$ the marginal distribution of $X=(X_1, X_2, \cdots, X_p)$.
Suppose one observes $n$ ``labeled" samples from $P$,
\begin{equation}\label{eq:labeled_sample}
[\Y, \X] = \left\{Y_k, X_{k1}, X_{k2}, \cdots, X_{kp}\right\}_{k=1}^n,
\end{equation}
and, in addition, $m$ ``unlabeled" samples from the marginal distribution $P_X$
\begin{equation}\label{eq:unlabeled_sample}
\X_{\rm add} = \left\{X_{k1}, X_{k2}, \cdots, X_{kp}\right\}_{k=n+1}^{n+m}.
\end{equation}
In this paper, we focus on estimation and statistical inference for one of the simplest features, namely the population mean $\theta = \E Y$. No specific distributional or marginal assumptions relating $X$ and $Y$ are made.

This inference of population mean under a general semi-supervised learning framework has a variety of applications. We discuss the estimation of treatment effect (ATE) in Section \ref{sec.ATE} and a prototypical example involving survey data in Section \ref{sec.real_data}. It is noteworthy that for some other problems that do not at first look like mean estimation, one can recast them as mean estimation, possibly after an appropriate transformation. Examples include estimation of the variance of $Y$ or covariance between $Y$ and a given $X_i$. In work that builds on a portion of the present paper, \cite{azriel2016semi} considers construction of linear predictors in semi-supervised learning settings. 

To estimate $\theta = \E Y$, the most straight-forward estimator is the sample average $\bar{\Y} := \frac{1}{n}\sum_{k=1}^n Y_k$. Surprisingly, as we show later, in the semi-supervised setting, a simple adjusted-least-squares estimator, which exploits the unknown association of $Y$ and $X$, outperforms $\bar{\Y}$. We first consider an ideal setting where there are infinitely many unlabeled samples, i.e., $m = \infty$. This is equivalent to the case of known marginal distribution $P_X$. We refer to this case as {\bf ideal semi-supervised inference}. In this case,  our proposed estimator is
\begin{equation}\label{eq:proposed1}
\hat \theta = \bar{\Y} - \hat{\beta}_{(2)}^\top (\bar{\X} - \mu),
\end{equation}
where $\bar{\X} \in \mathbb{R}^{p}$ such that $\bar{\X}_i = \frac{1}{n}\sum_{k=1}^n X_{ki}$, $\hat{\beta}_{(2)}$ is the $p$-dimensional least squares estimator for the regression slopes, and $\mu=\E X$ is the population mean of $X$. We emphasize again that although the estimator \eqref{eq:proposed1} has a linear structure, we are not assuming that  $\E(Y|X)$ is linearly related to $X$. 
This estimator is analyzed in detail in Section \ref{sec.ideal}. 

We then consider the more realistic setting where there are a finite number of unlabeled samples, i.e., $m < \infty$. Here one has only partial information about $P_X$. 
We call this case  {\bf ordinary semi-supervised inference}. In this setting, we propose to estimate $\theta$ by
\begin{equation}\label{eq:proposed2}
\hat \theta = \bar{\Y} - \hat{\beta}_{(2)}^\top (\bar{\X} - \hat\mu),
\end{equation}
where $\hat{\mu}$ denotes the sample average of both the labeled and unlabeled $X$'s. The detailed analysis of this estimator is given in Section \ref{sec.ordinary}. 

We will investigate the properties of  these estimators and in particular establish their asymptotic distributions and the $\ell_2$ risk bounds. The limiting distribution results allow us to construct an asymptotically valid confidence interval based on the proposed estimators that is shorter than the traditional sample-mean-based confidence interval. Both the case of a fixed number of covariates and the case of a growing number of covariates are considered. The basic asymptotic theory in Section \ref{sec.procedure} begins with a setting in which the dimension, $p$, of $X$, is fixed and $n\to \infty$ (see Theorem \ref{th:theta_LS_asymptotic}). For ordinary semi-supervised learning, the asymptotic results are of non-trivial interest whenever $\liminf_{n\to\infty}(m_n/n) > 0$ (see Theorem \ref{th:asymp_imperfect}(i)). We then formulate and prove asymptotic results in the setting where $p$ also grows with $n$. In general, these results require the assumption that $p = o(\sqrt{n})$ (see Theorems \ref{th:theta_LS_asymptotic_p_grow} and \ref{th:asymp_imperfect}(ii)). 

In Section \ref{sec.nonparametric} we propose a methodology for improving the results of Section \ref{sec.procedure} by introducing additional covariates as functions of those given in the original problem. We show the proposed estimator achieves an oracle rate asymptotically.  This can be viewed as a nonparametric regression estimation procedure.

There are results in the sample-survey literature that are qualitatively related to what we propose. The earliest citation we are aware of is \citet[Chapter 7]{cochran1953sampling} for sample survey. See also \cite{deng1987estimation} and more recently \citet[Chapter 3.2]{lohr2009sampling}. In these references one collects a finite sample, without replacement, from a (large) finite population. There is a response $Y$ and a single, real covariate, $X$. The distribution of $X$ within the finite population is known. The sample-survey target of estimation is the mean of $Y$ within the full population.  In the case in which the size of this population is infinitely large, sampling without replacement and sampling with replacement are indistinguishable. In that case the results from this sampling theory literature coincide with our results for the ideal semi-supervised scenario with $p=1$, both in terms of the proposed estimator and its asymptotic variance. Our work also relates to the control variates in Monte Carlo simulation \citep{bratley1987guide,fishman1996monte,hickernell2005control}. Suppose one is interested in evaluating the integral $\int_{\Omega} f(x)dx$, where  $f$ is a integrable function and $\Omega$ is a subset in the Euclidean space. The regular Monte Carlo estimator is $\frac{1}{n}\sum_{k=1}^n f(X_i)$, if $X_1,\ldots, X_n$ are i.i.d. uniform samples from $\Omega$. One can further sharpen the estimator if one or more control variates $\{h_1(x),\ldots, h_p(x)\}$ and their integrals $\{\int_\Omega h_1(x)dx,\ldots, \int_\Omega h_p(x)dx\}$ are available a priori. From this perspective, the results from control variates Monte Carlo can be viewed as a special case in the ideal semi-supervised and noiseless response setting, i.e., $\mathbb{E}X$ is known and $\mathbb{\Var}(Y|X)=0$. Otherwise the sample-survey and Monte Carlo theory results differ from those within our formulation, although there is a conceptual relationship. In particular the theoretical population mean that is our target is different from the finite population mean that is the target of the sample-survey methods. In addition we allow both the noisy response and $p>1$, and as noted above, we also have asymptotic results for $p$ growing with $n$. Most notably, our formulation includes the possibility of semi-supervised learning. We believe it should be possible, and sometimes of practical interest, to include semi-supervised sampling within a sampling survey and Monte Carlo simulation framework, but we do not do so in the present treatment.

Remarks at the end of Section \ref{sec.nonparametric} discuss in some detail the relation of our proposal to results in the semiparametric efficiency literature. In brief, it is known that $\bar{\Y}$ is not asymptotically semiparametric efficient. See \cite{hasminskii1983asymptotic} and \cite{bickel1991efficient} for an asymptotically efficient estimator in the case of ideal semi-supervision. \cite{chakrabortty2017efficient} deal with ideal semi-supervision and situations that are asymptotically equivalent to the ideal situation. They propose an estimators that is asymptotically efficient in this setting under mild regularity conditions. For situations in which there are many covariates their estimator may not perform well in practice, and they propose a number of alternative estimators.

Our current primary objective is rather different. We describe simply expressed, easily implemented, effective improvements on $\bar{\Y}$. Our basic estimator asymptotically improves on $\bar{\Y}$, but is not asymptotically efficient. Virtually no regularity conditions are imposed for the asymptotic improvement in distribution. (Asymptotic improvement in quadratic risk requires a little more care.) Because of their simple form as well as the nature of our proofs it is heuristically clear that with finite samples our estimators usually improve on $\bar{\Y}$ even for quite moderate sample sizes. This is seen in the simulations reported in Tables \ref{tb:simulation_1} and \ref{tb:simulation_CI}. The series estimator we propose in section \ref{sec.nonparametric} is semiparametric efficient under regularity conditions. (See Remarks \ref{rm:literature-efficiency} and \ref{rm:efficiency}) But this is not a primary focus of our paper, so we do not concentrate on stating that asymptotic efficiency under the weakest possible conditions.

The rest of the paper is organized as follows. We introduce the fixed covariate procedures in Section \ref{sec.procedure}. Specifically, ideal semi-supervised learning and ordinary semi-supervised learning are considered respectively in Sections \ref{sec.ideal} and \ref{sec.ordinary}, where we analyze the asymptotic properties for both estimators. We further give the $\ell_2$-risk upper bounds for the two proposed estimators in Section \ref{sec.l_2 risk}. We extend the analysis in Section \ref{sec.nonparametric}  to nonparametric regression model, where we show the proposed procedure achieves an oracle rate asymptotically. Simulation results are reported in Section \ref{sec.simulaton}. Applications to the estimation of Average Treatment Effect is discussed in Section \ref{sec.ATE}, and Section \ref{sec.real_data} describes a real data illustration involving estimation of the homeless population in a geographical region. 
The proofs of the main theorems are given in Section \ref{proofs.sec} and additional technical results are proved in the Supplement.

\section{Procedures}
\label{sec.procedure}

We propose in this section a least squares estimator for the population mean in the semi-supervised inference framework. To better characterize the problem, we begin with a brief introduction of the random design regression model. More details of the model can be found in, e.g., \cite{buja2014models,buja2016models}.

\subsection{A Random Design Regression Model}

Let $(Y, X) \sim P$ represent the population response and predictors. Assume all second moments are finite. Denote $\vec{X} = (1, X^\top)^\top \in \mathbb{R}^{p+1}$ as the predictor with intercept. The following is a linear analysis, even though no corresponding linearity assumption is made about the true distribution $P$ of $(X, Y)$. 

Some notation and definitions are needed. Let 
$$\beta = \argmin_{\gamma\in \mathbb{R}^{p+1}} \E  \left(Y - \vec{X}^\top\gamma\right)^2$$ 
be the {\it population slopes}, and $\delta = Y - \beta^\top\vec{X}$ is called the {\it total deviation}. We also denote 
\begin{equation}\label{eq:variance_deviation}
\begin{split}
& \tau^2 :=  \E \delta^2, \quad \mu :=  \E X\in \mathbb{R}^p,\quad \vec{\mu} :=  \E \vec{X}=(1, \mu^\top)^\top,\\ 
& \vec{\varXi} := \E\vec{X}\vec{X}^\top, \quad \Sigma := \mathbb{E}(X-\mu)(X-\mu)^\top.
\end{split}
\end{equation}
It should be noted that under our general model, there is no independence assumption between $X$ and $\delta$, and $\mathbb{E}(\delta|X)$ is not necessarily zero. This is different from classical regression literature.

For sample of observations $(Y_k, X_{k1}, X_{k2}, \cdots, X_{kp}) \overset{iid}{\sim} P $, $k= 1,\cdots, n$, let $\vec{X}_i = (1, \vec{X}_i^\top)^\top$ and denote the design matrix $\vec{\X}\in\mathbb{R}^{n\times (p+1)}$ as follows
$$\vec{\X} := \begin{bmatrix}
\vec{X}_1^\top\\
\cdots\\
\cdots\\
\vec{X}_n^\top
\end{bmatrix} := \begin{bmatrix}
1 & X_{11} & X_{12} & \cdots & X_{1p}\\
\vdots & \vdots & \vdots &  & \vdots\\
1 & X_{n1} & X_{n2} & \cdots & X_{np}
\end{bmatrix}. $$
In our notation, $\vec{\cdot}$ means that the vector/matrix contains the intercept term; boldface indicates that the symbol is related to a multiple sample of observations. Meanwhile, denote the sample response and deviation as $\Y = (Y_1, \cdots, Y_n)^\top$ and $\Bdelta = (\delta_1, \cdots, \delta_n)^\top$. Now $\Y$ and $\X$ are connected by a regression model:
\begin{equation} 
\Y = \vec{\X}\beta + \Bdelta,\quad \text{and} \quad Y_k = \vec{X}_k^\top\beta + \delta_k,\quad  k = 1,\cdots, n.
\end{equation}
Let $\hat \beta = (\hat{\beta}_1, \cdots, \hat\beta_{p+1})^\top$ be the usual least squares estimator, i.e. $\hat\beta = (\vec{\X}^\top\vec{\X})^{-1}\vec{\X}^\top\Y$.
$\beta$ and $\hat{\beta}$ can be further split into two parts, 
\begin{equation}\label{eq:beta_split}
\beta = \begin{blockarray}{c}
\begin{block}{[c]}
\beta_1  \\
\beta_{(2)} \\
\end{block}
\end{blockarray},
\quad 
\hat{\beta} = \begin{blockarray}{c}
\begin{block}{[c]}
\hat{\beta}_1 \\
\hat{\beta}_{(2)} \\
\end{block}
\end{blockarray}, \quad \beta_1, \hat{\beta}_1\in \mathbb{R},\quad \beta_{(2)}, \hat{\beta}_{(2)}\in \mathbb{R}^p.
\end{equation}
$\beta_1, \hat{\beta}_1$ and $\beta_{(2)}, \hat{\beta}_{(2)}$ play different roles in the analysis as we will see later. The $\ell_2$ risk of the sample average $\bar{\Y}$ about the population mean $\theta = \E Y$ has the following decomposition.
\begin{Proposition}\label{th:nvar(bar_Y)}
	$\bar{\Y}$ is an unbiased estimator of $\theta$ and
	\begin{equation}\label{eq:var_bar_Y} 
	n\E (\bar{\Y} - \theta)^2 = n \Var(\bar \Y) = \tau^2 + \beta_{(2)}^\top \Sigma \beta_{(2)}.
	\end{equation}
\end{Proposition}
From \eqref{eq:var_bar_Y}, we can see that as long as $\beta_{(2)} \neq 0$, i.e., there is a significant linear relationship between $Y$ and $X$, then the risk of $\bar \Y$ will be significantly greater than $\tau^2$. 

In the next two subsections, we discuss separately under the ideal semi-supervised setting and the ordinary semi-supervised setting.

\subsection{Improved Estimator under the Ideal Semi-supervised Setting}
\label{sec.ideal}

We first consider the ideal setting where there are infinitely many unlabeled samples, or equivalently $P_X$ is known. To improve $\bar{\Y}$, we propose the \emph{least squares estimator},
\begin{equation}\label{eq:delta_LS}
\hat \theta_{\rm LS} :=  \vec{\mu}^\top \hat\beta = \hat\beta_1 + \mu^\top \hat\beta_{(2)} = \bar \Y - \hat\beta_{(2)}^\top (\bar \X - \mu),
\end{equation}
where $\hat \beta = (\hat\beta_1, \hat\beta_{(2)}^\top )^\top $ is the usual least square estimator. 

When $(Y_i, X_i) \overset{iid}{\sim} P$ with no specific assumptions imposed on the relationship between $Y_i$ and $X_i$, the following theorem provides the asymptotic distribution of the least squares estimator under the minimal conditions that $[Y, X]$ have finite second moments, $\vec{\varXi}=\E \vec{X}\vec{X}^\top$ be non-singular and $\tau^2=\E\delta^2>0$. In addition, a Berry-Esseen  bound is given under the finite fourth moment condition.
\begin{Theorem}[Asymptotic Distribution of $\hat{\theta}_{\rm LS}$, fixed $p$]\label{th:theta_LS_asymptotic}
	Let $(Y_1, X_1)$, $\cdots$,  $(Y_n, X_n)$ be i.i.d. copies from $P$, and assume that $[Y, X]$ has finite second moments, $\vec{\varXi}$ is non-singular and $\tau^2 > 0$. 
	Then, under the setting that $P$ is fixed and $n\to\infty$,
	\begin{equation}\label{eq:asym_th1}
	\frac{\hat{\theta}_{\rm LS} - \theta}{\tau/\sqrt{n}} \overset{d}{\to} N(0, 1),
	\end{equation}
	and
	\begin{equation}\label{eq:asym_th2}
	MSE/\tau^2 \overset{d}{\to} 1,\quad \text{where}\quad  MSE := \frac{\sum_{i=1}^n (Y_i - \vec X_i^\top \hat \beta)^2}{n-p-1}.
	\end{equation}
	Denote the cumulative distribution functions of $\frac{\hat{\theta}_{\rm LS}-\theta}{\tau/\sqrt{n}}$ and the standard normal variable by $F_n$ and $\Phi$, respectively. If $P$ has finite fourth moment, then we further have
		$$\left|F_n(x) - \Phi(x)\right| \leq Cn^{-1/4},$$
	where $C$ is a constant not depending on $n$.
\end{Theorem}

In the more general setting where $P = P_{n, p}$ varies and $p=p_n$ grows, we need stronger conditions to analyze the asymptotic behavior of $\hat \theta_{\rm LS}$. Recall $\E X = \mu$, $\E (X-\mu)(X-\mu)^\top  = \Sigma$, we consider the standardization of $X$ as 
\begin{equation}\label{eq:Z_standard}
Z\in \mathbb{R}^p, \quad Z = \Sigma^{-1/2}(X - \mu).
\end{equation}
Clearly, $\E Z = 0, \E ZZ^\top  = I_p$. For this setting we assume that $Z, \delta$ satisfy the following moment conditions for constants $M_1, M_2, M_3$:
\begin{equation}\label{ineq:th3_assump0}
\text{for some $\kappa>0$}, \quad \frac{\E \delta^{2+2\kappa}}{(\E \delta^2)^{1+\kappa}} \leq M_1;
\end{equation}
\begin{equation}\label{ineq:th3_assump1}
\forall v\in \mathbb{R}^p,\quad \E |\langle v, Z \rangle|^{2+\kappa} \leq M_2;
\end{equation}
\begin{equation}\label{ineq:th3_assump2}
\frac{\E \left(\|Z\|_2^2\delta^2\right)}{\left(\E \|Z\|_2^2\right) \cdot \left(\E \delta^2\right)}\leq M_3.
\end{equation} 
\begin{Theorem}[Asymptotic result, growing $p$]\label{th:theta_LS_asymptotic_p_grow}
	Let $(Y_1, X_1), \cdots, (Y_n, X_n)$ be i.i.d. copies from $P = P_{n,p}$, $p = p_n = o(\sqrt{n})$. Assume that the matrix of the second moments of $X$ exists and is non-singular and the standardized random variable $Z$ given in \eqref{eq:Z_standard} satisfies \eqref{ineq:th3_assump0}, \eqref{ineq:th3_assump1} and \eqref{ineq:th3_assump2}, then the asymptotic behavior results \eqref{eq:asym_th1} and \eqref{eq:asym_th2} still hold.
\end{Theorem}

Based on Theorems \ref{th:theta_LS_asymptotic} and \ref{th:theta_LS_asymptotic_p_grow}, we can construct the asymptotic $(1-\alpha)$-level confidence interval for $\theta$ as
\begin{equation}\label{eq:CI_theta_LS}
\left[\hat{\theta}_{\rm LS} - z_{1-\alpha/2} \sqrt{\frac{MSE}{n}}\quad , \quad  \hat{\theta}_{\rm LS} + z_{1-\alpha/2} \sqrt{\frac{MSE}{n}}\right].
\end{equation}
\begin{Remark}\rm
	It is not difficult to see that, under the setting in Theorem \ref{th:theta_LS_asymptotic_p_grow},
	$$MSE \overset{d}{\to} \tau^2, \quad\hat\sigma^2_Y \overset{d}{\to} \Var(Y) = \tau^2 + \beta_{(2)}^\top \Sigma \beta_{(2)}.$$ 
	Then the traditional $z$-interval for the mean of $Y$, 
	\begin{equation}\label{eq:CI_tranditional}
	\left[\bar \Y - z_{1-\alpha/2}\sqrt{\frac{\hat\sigma^2_Y}{n}}\quad, \quad\bar \Y + z_{1-\alpha/2}\sqrt{\frac{\hat\sigma^2_Y}{n}}\right],
	\end{equation}
	is asymptotically longer than \eqref{eq:CI_theta_LS}, which implies that the proposed least squares estimator is asymptotically more accurate than the sample mean.
\end{Remark}

\subsection{Improved Estimator under the Ordinary Semi-supervised Inference Setting}
\label{sec.ordinary}

In the last section, we discussed the estimation of $\theta$ based on $n$ full observations $Y_k, X_k, k=1,\cdots, n$ with infinitely many unlabeled samples $\{X_k, k = n+1, \cdots\}$ (or equivalently with known marginal distribution $P_X$). However, having $P_X$ known is rare in practice. A more realistic practical setting would assume that distribution $P_X$ is unknown and we only have finitely many i.i.d. samples $(X_{i+1}, X_{i+2}, \cdots, X_{i+m})$ without corresponding $Y$. This problem relates to the one in previous section since we are able to obtain partial information of $P_X$ from the additional unlabeled samples. 

When $\mu$ or $\vec \mu$ is unknown, we estimate by 
\begin{equation}
\hat\mu = \frac{1}{n+m} \sum_{k=1}^{n+m} X_k, \quad \hat{\vec{\mu}} = (1, \hat\mu^\top )^\top .
\end{equation}
Recall that $\hat \beta = (\hat \beta_1, \beta_{(2)}^\top )^\top $ is the ordinary least squares estimator. Now, we propose the \emph{\underline{s}emi-\underline{s}upervised \underline{l}east \underline{s}quares estimator} $\hat{\theta}_{\rm SSLS}$,
\begin{equation}\label{eq:hat_theta_SPLS}
\hat\theta_{\rm SSLS} =\hat{\vec \mu}^\top \hat{\beta} = \bar \Y - \hat \beta_{(2)}^\top  \left(\frac{\sum_{i=1}^{n}X_i}{n} - \frac{\sum_{i=1}^{n+m}X_i}{n+m}\right).
\end{equation}
$\hat\theta_{\rm SSLS}$ has the following properties:
\begin{itemize}
	\item when $m = \infty$, $\hat{\vec\mu} = \vec{\mu}$. Then $\hat{\theta}_{\rm SSLS}$ exactly equals $\hat\theta_{\rm LS}$ in \eqref{eq:delta_LS};
	\item when $m=0$, $\hat{\theta}_{\rm SSLS}$ exactly equals $\bar{\Y}$. As there are no additional samples of $X$ so that no extra information for $P_X$ is available, it is natural to use $\bar \Y$ to estimate $\theta$.
	\item In the last term of \eqref{eq:hat_theta_SPLS}, it is important to use $\frac{1}{n+m}\sum_{i=1}^{n+m}X_i$ rather than $\frac{1}{m}\sum_{i=1}^m X_i $, in spite of the fact that the latter might seem more natural because it is independent of the term $\frac{\sum_{i=1}^n X_i}{n}$ that precedes it.
\end{itemize}
Under the same conditions as Theorems \ref{th:theta_LS_asymptotic}, \ref{th:theta_LS_asymptotic_p_grow}, we can show the following asymptotic results for $\hat{\theta}_{\rm SSLS}$, which relates to the ordinary semi-supervised setting described in the introduction. 
The labeled sample size $n \to \infty$, the unlabeled sample size is $m = m_n \geq 0$ and the distribution $P$ is fixed (but unknown) which, in particular, implies that $p$ is a fixed dimension, not dependent on $n$. Let
$$\nu^2 = \sqrt{\tau^2 + \frac{n}{n+m}\beta_{(2)}^\top \Sigma \beta_{(2)}}.$$

\begin{Theorem}[Asymptotic distribution of $\hat\theta_{\rm SSLS}$, fixed $p$]\label{th:asymp_imperfect} 
Let $(Y_1, X_1)$, $\cdots$,  $(Y_n, X_n)$  be i.i.d. labeled samples from $P$, and let $X_{n+1}, \cdots, X_{n+m}$ be $m$ additional unlabeled independent samples from $P_X$. Suppose $\vec{\varXi}$ is non-singular and $\tau^2>0$. If $P$ is fixed and $n\to\infty$, then
	\begin{equation}\label{eq:asymp1_imperfect}
	\frac{\sqrt{n}(\hat{\theta}_{\rm SSLS} - \theta)}{\nu} \overset{d}{\to} N(0, 1),
	\end{equation}
	and
	\begin{equation}\label{eq:asymp2_imperfect}
	\frac{\hat{\nu}^2}{\nu^2} \overset{d}{\to} 1
	\end{equation}		
	where $\hat{\nu}^2 = \frac{m}{m+n}MSE + \frac{n}{m+n} \hat\sigma_Y^2$ with $MSE = {1\over n-p-1}\sum_{k=1}^n(Y_i - \vec X_k^\top  \hat{\beta})^2$ and  $\hat \sigma_Y^2 ={1\over n-1} \sum_{k=1}^n (Y_i -\bar{\Y})^2$.
\end{Theorem}

The following statement refers to a setting in which $P=P_n$ and $p=p_n$ may depend on $n$ as $n\to\infty$. Consequently, $\vec{\varXi} = \vec{\varXi}_n$, $\Sigma = \Sigma_n$ and $Z = Z_n$ (defined at \eqref{eq:Z_standard}) may also depend on $n$.

\begin{Theorem}[Asymptotic distribution of $\hat\theta_{\rm SSLS}$, growing $p$]\label{th:asymp_imperfect_growing_p}
	Let $n \to \infty$, $P = P_{n}$, and $p= p_n=o(\sqrt{n})$. Suppose $\vec{\varXi}_n$ is non-singular, $\tau_n^2>0$ and the standardized random variable $Z$ satisfies \eqref{ineq:th3_assump0}, \eqref{ineq:th3_assump1} and \eqref{ineq:th3_assump2}. 
	Then \eqref{eq:asymp1_imperfect} and \eqref{eq:asymp2_imperfect} hold.
\end{Theorem}

We can obtain asymptotic confidence interval for $\theta$ based on Theorems \ref{th:asymp_imperfect} or \ref{th:asymp_imperfect_growing_p}.
\begin{Corollary}
	The $(1-\alpha)$-level asymptotic confidence interval for $\theta$ can be written as
	\begin{equation}\label{eq:CI_semi}
	\left[\hat \theta_{\rm SSLS} - z_{1-\alpha/2}\frac{\hat{\nu}}{\sqrt{n}}\quad ,\quad \hat \theta_{\rm SSLS} + z_{1-\alpha/2}\frac{\hat{\nu}}{\sqrt{n}} \right].
	\end{equation}
	Since $MSE \leq \hat{\sigma}_Y^2$ asymptotically (with equality only when $\beta_{(2)} = 0$), it follows that when $\beta_{(2)}\neq 0$ the asymptotic CI in \eqref{eq:CI_semi} is shorter than the traditional sample-mean-based CI \eqref{eq:CI_tranditional}.
\end{Corollary}

\subsection{$\ell_2$ Risk for the Proposed Estimators}
\label{sec.l_2 risk}

In this subsection, we analyze the $\ell_2$ risk for both $\hat{\theta}_{\rm LS}$ and $\hat{\theta}_{\rm SSLS}$. Since the calculation of the proposed estimators involves the unstable process of inverting the Gram matrix $\vec{\X}^\top \vec{\X}$, for the merely theoretical purpose of obtaining the $\ell_2$ risks we again consider the refinement 
\begin{equation}\label{eq:hat_theta_LS^1}
\hat{\theta}_{\rm LS}^1 :=  \Trun_{\Y}(\hat\theta_{\rm LS}),\quad \text{and}\quad \hat{\theta}_{\rm SSLS}^1 :=  \Trun_{\Y}(\hat\theta_{\rm SSLS}),
\end{equation}
where
\begin{equation}\label{eq:truncation}
\Trun_{\Y}(x) = \left\{
\begin{array}{ll} 
(n+1)y_{\max} -ny_{\min}, & \text{if } x >  (n+1) y_{\max} - n y_{\min}, \\
x, & \text{if } |x - \frac{y_{\max}+y_{\min}}{2}| \leq (n + \frac{1}{2})(y_{\max} - y_{\min}),\\
(n+1)y_{\min} - ny_{\max}, & \text{if } x < (n+1)y_{\min} - ny_{\max},
\end{array}\right.
\end{equation}
$y_{\max} = \max_{1\leq k\leq n}Y_k, y_{\min} = \min_{1\leq k\leq n}Y_k$. We emphasize that this refinement is mainly for theoretical reasons and is often not necessary in practice.

The regularization assumptions we need for analyzing the $\ell_2$ risk are formally stated as below. 
\begin{enumerate}
	\item {\it(Moment conditions on $\delta$)}  There exists constant $M_4>0$ such that
	\begin{equation}\label{ineq:assump1_l2_risk}
	\E \delta^4 = \E \delta_n^4 \leq M_4;\\
	\end{equation}
	\item {\it (sub-Gaussian condition)} 
	Let $Z=Z_n$ be the standardization of $X=X_n$,
	$$Z_n\in \mathbb{R}^p, \quad Z_n = \Sigma_n^{-1/2}(X_n-\mu_n),\quad \Sigma_n = \E(X_n - \mu_n)(X_n-\mu_n)^\top.$$
	Assume $Z_n$ satisfies
	\begin{equation}\label{ineq:assump2_l2_risk}
	\forall u\in \{u\in \mathbb{R}^{p+1}: \|u\|_2=1\}, \quad  \left\|u^\top  Z_n\right\|_{\psi_2} \leq M_5
	\end{equation}
	for constant $M_t>0$. Here $\|\cdot\|_{\psi_2}$ is defined as $ \|x\|_{\psi_2} = \sup_{q\geq 1} q^{-1/2} (\E |x|^q)^{1/q}$ for any random variable $x$.
	\item[2'] {\it (Bounded condition)} 
	The standardization $Z_n$ satisfies
	\begin{equation}\label{ineq:assumption3_l2_risk}
	\|Z_n\|_{\infty} \leq M_5, \quad \text{almost surely.}
	\end{equation}
	(If the dimension $p$ remains bounded as $n\to \infty$ then \eqref{ineq:assumption3_l2_risk} implies \eqref{ineq:assump2_l2_risk}. However if $p$ increases without bound, as in section 3, then there are rather unusual examples in which \eqref{ineq:assumption3_l2_risk} holds but \eqref{ineq:assump2_l2_risk} does not.)
\end{enumerate}

We also note $ \Sigma_{\delta 1}=\E (X-\mu)\delta(X-\mu)^\top$, $\Sigma_{\delta 2}=\E (X-\mu)\delta^2(X-\mu)^\top  $. Under the regularization assumptions above, we provide the $\ell_2$ risks for $\hat{\theta}_{\rm LS}^1$ and $\hat{\theta}_{\rm SSLS}^1$ respectively in the next two theorems.
\begin{Theorem}[$\ell_2$ Risk of $\hat{\theta}_{\rm LS}^1$]\label{th:MSE_delta} 
	Let $(Y_1, X_1), \cdots, (Y_n, X_n)$ be i.i.d. copies from $P_n$. Assume Assumptions 1 holds. In addition, either Assumptions 2 or 2' hold, 
	$p = p_n = o(\sqrt{n})$. Recall $\tau^2 = \tau_n^2 = \E (Y - \vec{X}\beta)^2$ depends on $n$. Then we have the following estimate for the risk of $\hat{\theta}_{\rm LS}^1$,
	\begin{equation}\label{eq:risk_delta_LS}
	n\E \left(\hat{\theta}_{\rm LS}^1 - \theta\right)^2 = \tau^2_n + s_n,
	\end{equation}
	where
	\begin{equation}\label{eq:risk_delta_LS_s_n}
	s_n = \frac{p^2}{n}A_{n, p} + \frac{p^2}{n^{5/4}}B_{n, p}, \quad \max(|A_{n, p}|, |B_{n, p}|) \leq C
	\end{equation}
	for a constant $C$ that depends on $M_0, M_1$ and $M_2$. The formula for $A_{n, p}$ is
	\begin{equation}\label{eq:s_n}
	\begin{split}
	A_{n, p} = & \frac{1}{p^2} \Bigg( [\tr(\Sigma^{-1}\Sigma_{\delta 1})]^2 + 3\|\Sigma^{-1}\Sigma_{\delta 1} \|_F^2 - \tr (\Sigma^{-1}\Sigma_{\delta 2})\\
	& + 2 \E \left(\delta^2 (X-\mu)^\top\right) \cdot \E \left(\Sigma^{-1}(X-\mu)(X-\mu)^\top \Sigma^{-1}(X-\mu)\right) + 2p\tau^2 \Bigg).
	\end{split}
	\end{equation}
\end{Theorem}
\begin{Theorem}[$\ell_2$ risk of $\hat\theta_{\rm SSLS}^1$]\label{th:MSE_SPLS}  
Let $(Y_1, X_1), \cdots, (Y_n, X_n)$ be i.i.d. labeled samples from $P$, and let $X_{n+1}, \cdots, X_{n+m}$ be additional $m$ unlabeled independent samples from $P_X$. 
 Assume Assumptions 1 holds. In addition, either Assumptions 2 or 2' hold, $p=o(\sqrt{n})$. We have the following estimate of the risk for $\hat \theta_{\rm SSLS}^1$,
	\begin{equation}
	n\E \left(\hat\theta_{\rm SSLS}^1 - \theta\right)^2 = \tau^2_n + \frac{n}{n+m}\beta_{(2), n}^\top \Sigma_n \beta_{(2), n} + s_{n, m}
	\end{equation}
	where 
	\begin{equation}
	|s_{n, m}| \leq \frac{Cp^2}{n}.
	\end{equation}
	for constant $C$ only depends on $M_0, M_1$ and $M_2$ in Assumptions \eqref{ineq:assump1_l2_risk}-\eqref{ineq:assumption3_l2_risk}.
\end{Theorem}

\begin{Remark}\rm
Comparing Proposition \ref{th:nvar(bar_Y)}, Theorems \ref{th:MSE_delta} and \ref{th:MSE_SPLS}, we can see as long as 
$$\beta_{(2), n}^\top \Sigma_n \beta_{(2), n} >0,$$ 
i.e., $\E (Y|X)$ has non-zero correlation with $X$, $\hat\theta_{\rm LS}^1$  and $\hat{\theta}_{\rm SSLS}^1$ outperform $\bar{\Y}$ asymptotically in $\ell_2$-risk.
	We can also see the risk of $\hat\theta_{\rm SSLS}$ is approximately a linear combination of $\bar \Y$ and $\hat\theta_{\rm LS}$ with weight based on $m$ and $n$,
	$$\E \left(\hat\theta_{\rm SSLS}^1 - \theta\right)^2 \approx \frac{n}{n+m}\E \left(\bar \Y - \theta\right)^2 + \frac{m}{m+n}\E \left( \hat\theta_{\rm LS}^1 - \theta\right)^2 $$
\end{Remark}

\begin{Remark}
	The proposed $\hat{\theta}_{\rm SSLS}$ is a direct and simple estimator that achieves good finite sample performance for both estimation and confidence interval. An improved semi-supervised least square estimator that achieves semiparametric efficiency will be further introduced and discussed later in Section \ref{sec:improved procedure}. 
\end{Remark}

\begin{Remark}[Gaussian Design]\rm\label{rm:ideal_gaussian}
	Theorems \ref{th:MSE_delta} and \ref{th:MSE_SPLS} only provide upper bound of the $\ell_2$ risks since only moment conditions on the distribution of $Y, X$ are assumed. In fact, under Gaussian design of $Y, X$, we can obtain an exact expression for the $\ell_2$-risk of both $\hat\theta_{\rm LS}$ and $\hat{\theta}_{\rm SSLS}$. It is noteworthy that the truncation refinement is not necessary for both estimators under Gaussian design. All results are non-asymptotic.
	\begin{Proposition}\label{pr:Gaussian}
		Assume $X\sim N_{p}(\mu, \Sigma)$ and $Y|X \sim N_{p}(X\beta, \tau^2I)$, where $\Sigma$ is non-singular. If $\{Y_k, X_k\}_{k=1}^n$ are $n$ i.i.d. copies, then
		\begin{equation}\label{eq:expectation_loss_Gaussian}
		n \E \left(\hat\theta_{\rm LS} - \theta\right)^2 = \tau^2 + \frac{p\tau^2}{(n-p-2)}.
		\end{equation}
		If we further have $m$ additional unlabeled samples $\{X_k\}_{k=n+1}^{n+m}$, we also have
		\begin{equation}\label{eq:expectation_loss_Gaussian_semi}
		\begin{split}
		n\E \left(\hat\theta_{\rm SSLS} - \theta\right)^2 = & \tau^2 + \frac{m}{n+m} \frac{p\tau^2}{n-p-2} + \frac{n}{n+m}\beta_{(2)}^\top \Sigma \beta_{(2)}.
		\end{split}
		\end{equation}
	\end{Proposition}
	
	\ \par
	
	The result in Proposition \ref{pr:Gaussian} matches with the general expression of \eqref{eq:risk_delta_LS} and \eqref{eq:s_n} as $ \frac{p\tau^2}{(n-p-2)} =  \frac{p\tau^2}{n} + O\left(\frac{p^2}{n^2}\right)$ if $p = o(\sqrt{n})$.
	By comparing \eqref{eq:expectation_loss_Gaussian}, \eqref{eq:expectation_loss_Gaussian_semi}, we can also see
	$$n\E \left(\hat{\theta}_{\rm SSLS} - \theta\right)^2 =  \frac{n}{n+m} n\E (\bar \Y - \theta)^2 + \frac{m}{n+m} n \E (\hat\theta_{\rm LS} - \theta)^2.$$
\end{Remark}

\section{Further Improvements -- Oracle Optimality}
\label{sec.nonparametric}

In the previous sections, we proposed and analyzed $\hat\theta_{\rm LS}$ and $\hat{\theta}_{\rm SSLS}$ under the semi-supervised learning settings. These estimators are based on linear regression and best linear approximation of $Y$ by $X$. 
We consider further improvement in this section. Before we illustrate how the improved estimator works, it is helpful to take a look at the oracle risk for estimating the mean $\theta = \E Y$, which can serve as a benchmark for the performance of the improved estimator.

\subsection{Oracle Estimator and Risk}\label{sec:oracle}

Define $\xi(X) = \E_P(Y|X)$ as the response surface and suppose
$$\xi(x) = \xi_0(x) + c + o(1/\sqrt{n}) $$
for some unknown constant $c$. Here, the $o(1/\sqrt{n})$ term is uniform in $X$ and $\xi_0(x)$ represents any approximately ``location-free shape" of $\xi_0(x)$ in the sense that $\xi(x) - \xi_0(x)$ is nearly a constant: $|\xi(x)-\xi_0(x)-c|\leq o(1/\sqrt{n})$.
Given samples $\{(Y_k, X_k)\}_{k=1}^n$, our goal is to estimate $\E Y = \theta$. Now assume an oracle has knowledge of $\xi_0(x)$, but not of $\theta = \E(Y)$, $c$, nor the distribution of $Y - \xi_0(X)$. 
In this case, the model can be written as
\begin{equation}
\begin{split}
Y_k - \xi_0(X_k) = & c + \varepsilon_k, \quad k=1,\cdots, n,\quad \text{where} \quad \E\varepsilon_k = o(1/\sqrt{n});\\
\theta = & \E \xi_0(X) + c + o(1/\sqrt{n}).
\end{split}
\end{equation}
Under the ideal semi-supervised setting,  $P_X$, $\xi_0$ and $\E \xi_0(X)$ are known. To estimate $\theta$, the natural idea is to use the following estimator
\begin{equation}\label{eq:oracle_estimator}
\hat\theta^\ast =   \bar{\Y} - \bar{\xi}_0 + \E\xi_0(X) = \frac{1}{n}\sum_{k=1}^n\left(Y_k - \xi_0(X_k)\right) + \E\xi_0(X).
\end{equation}
Consider a sample $\{Y_i: i = 1,\ldots, n\}$ with no covariates. It is known that $\bar{\Y}$ is an asymptotically efficient estimator of $\mathbb{E}(Y)$, locally on a neighborhood of the true distribution of $Y$. In much the same way, $\bar{Y} - \overline{\xi_0(X)}$ is an asymptotically efficient estimator of $\mathbb{E}(Y-\xi_0(X))$, even when the ancillary statistics, $\{X_i\}$, are also observed. For details see the proof of Proposition 3 in the supplement. Thus, $\hat{\theta}^\ast$ is an asymptotically efficient estimator of $\mathbb{E}(Y)=\mathbb{E}(Y-\xi_0(X)) + E(\xi_0(X))$. And,

\begin{equation}\label{eq:oracle_risk}
\begin{split}
n\E \left(\hat\theta^\ast - \theta\right)^2 & = n\Var\left(\frac{1}{n}\sum_{i=1}^n (Y_i - \xi_0(X_i)) \right) = \Var\left(Y_i  - \xi(X_i)\right)\\
& = \E _X\left(\E _Y\left(Y - \xi(X)|X\right)^2\right) := \sigma^2.
\end{split}
\end{equation}
This defines the oracle risk for population mean estimation under the ideal semi-supervised setting as $\sigma^2 = \E_X\left(\E_Y(Y - \E(Y|X))^2\right)$.

For the ordinary semi-supervised setting, where $P_X$ is unknown but $m$ additional unlabeled samples $\{X_k\}_{k=n+1}^{n+m}$ are available, we propose the semi-supervised oracle estimator as
$$\hat{\theta}^\ast_{\rm ss} = \bar{\Y} - \frac{1}{n}\sum_{k=1}^n\xi_0(X_k) + \frac{1}{n+m}\sum_{k=1}^{n+m}\xi_0(X_k). $$
Then one can calculate that 
\begin{equation}\label{eq:oracle_semi_supervised_risk}
\begin{split}
& n\E \left(\hat{\theta}^\ast_{ss} - \theta\right)^2 = \sigma^2 + \frac{n}{n+m}\Var_{P_X}(\xi(X)).
\end{split}
\end{equation}
The detailed calculation of \eqref{eq:oracle_semi_supervised_risk} is provided in the Supplement.

The preceding motivation for $\sigma^2$ and $\sigma^2 + \frac{n}{n+m}\Var_{P_X}(\xi(X))$ as the oracle risks are partly heuristic, but it corresponds to formal minimax statements, as in the following Propositions \ref{pr:oracle_lower_bound} and \ref{pr:oracle_lower_bound-efficiency}. Particularly, Proposition \ref{pr:oracle_lower_bound} proposes the general lower bounds for both ideal and semiparametric settings. Proposition \ref{pr:oracle_lower_bound-efficiency} develops the asymptotic lower bound on a more restrictive set, i.e. the least favorable one-dimensional family of conditional means of any specific distribution $P$, under ideal semi-supervision. 
Proposition \ref{pr:oracle_lower_bound-efficiency} further yields an asymptotic semiparametric efficiency result as we will illustrate later in Remark \ref{rm:efficiency}.

\begin{Proposition}[Oracle Lower Bound]\label{pr:oracle_lower_bound}
	Let $\sigma^2>0$, $\xi_0:\mathbb{R}^p\to \mathbb{R}$ be a measurable function, and $P_X$ be a $p$-dimensional distribution of $X$. Suppose
	\begin{equation*}
	\begin{split}
	\mathcal{P}_{\xi_0(\cdot), P_X, \sigma^2} = \Big\{P: & P_X \text{is the marginal distribution of $P$},\\ 
	& \E_P(Y|X=x) = \xi_0(x) + c, \sigma^2 = \E_X\left(\E_Y(Y - \E(Y|X))^2\right)\Big\}. 
	\end{split}
	\end{equation*}
	Then based on observations $\{Y_i, X_i\}_{i=1}^n$ and known marginal distribution $P_X$,
	\begin{equation}\label{eq:variance_oracle}
	\inf_{\tilde{\theta}} \sup_{P\in \mathcal{P}_{P_X, \xi_0, \sigma^2}} \left[\E_P\left(n\left(\tilde{\theta} - \theta\right)^2\right)\right] \geq \sigma^2.
	\end{equation}

	Let $\sigma^2, \sigma_\xi^2>0$, $\xi_0(X):\mathbb{R}^p\to \mathbb{R}$ be a linear function,
	\begin{equation*}
	\begin{split}
	\mathcal{P}^{\rm ss}_{\xi_0, \sigma_\xi^2, \sigma^2} = \Big\{P: & \xi_0(x) = \E(Y|X=x)-c, \sigma_\xi^2 = \Var(\xi(X)),\\ 
	& \sigma^2 = \E_X\left(\E_Y(Y - \E(Y|X))^2\right)\Big\}, 
	\end{split}
	\end{equation*}
	based on observations $\{Y_i, X_i\}_{i=1}^n$ and $\{X_i\}_{i=n+1}^{n+m}$,
	\begin{equation}\label{eq:variance_oracle_semi}
	\inf_{\tilde{\theta}} \sup_{P\in \mathcal{P}^{\rm ss}_{\xi_0, \sigma_\xi^2, \sigma^2}} \left[\E_P\left(n\left(\tilde{\theta} - \theta\right)^2\right)\right] \geq \sigma^2 + \frac{n}{n+m} \sigma_\xi^2.
	\end{equation}
\end{Proposition}

\begin{Proposition}[Asymptotic Oracle Lower Bound for ideal semi-supervised setting]\label{pr:oracle_lower_bound-efficiency}
Let $\sigma^2 = \mathbb{E}_P\left[(Y - \xi(X))^2\right]>0$ and 
\begin{equation*}
\mathcal{P}_K = \left\{P: \begin{array}{l}
P_X = P_X^0, |\mathbb{E}_P(Y|X) - \xi^0(X)| \leq K\sigma^2(X)/\sigma^2 + 1/(K\sqrt{n}),\\
\left|\mathbb{E}_P\left[(Y - \xi(X))^2\right]-\sigma^2\right| < 1/K, |c|<K
\end{array}\right\}.
\end{equation*}
Then
\begin{equation}\label{eq:efficiency-oracle-lower-bound}
\lim_{K\to\infty} \liminf_{n\to \infty}\inf_{\hat{\theta}}\sup_{P\in \mathcal{P}_k}n\mathbb{E}\left(\hat{\theta} - \mathbb{E}_P(Y)\right)^2 \geq \sigma^2.
\end{equation}

\end{Proposition}

\subsection{Improved Procedure}\label{sec:improved procedure}

In order to approach oracle optimality we propose to augment the set of covariates $X_1, \ldots, X_p$ with additional covariates $g_1(X),\ldots, g_q(X)$. (Of course these additional covariates need to be chosen without knowledge of $\xi_0$. We will discuss their choice later in this section.) In all there are now $p^\bullet = p + q$ covariates, say
$$X^\bullet = (X_1^\bullet, \ldots, X_p^\bullet, X_{p+1}^\bullet, \ldots, X_{p+q}^\bullet) = (X_1, \ldots, X_p, g_1(X), \ldots, g_q(X)).$$
For both ideal and ordinary semi-supervision we propose to let $q = q_n$ as $n \to \infty$, and to use the estimator $\hat{\theta}_{\rm LS}^\bullet$ and $\hat{\theta}_{\rm SSLS}^{\bullet}$. For merely theoretical purpose of $\ell_2$ risks we consider the refinement again
$$\hat{\theta}_{\rm LS}^{\bullet 1} = \Trun_{\Y}(\hat{\theta}_{\rm LS}^{\bullet}) \quad\text{   and   }\quad \hat{\theta}_{\rm SSLS}^{\bullet 1} = \Trun_{\Y}(\hat{\theta}_{\rm SSLS}^{\bullet}),$$ 
where $\Trun_{\Y}(\cdot)$ is defined as \eqref{eq:truncation}. Apply previous theorems for asymptotic distributions and moments. 
For convenience of statement and proof we assume that the support of $X$ is compact, $\xi(X)$ is bounded and $Y$ is sub-Gaussian. These assumptions can each be somewhat relaxed at the cost of additional technical assumptions and complications. Here is a formal statement of the result. 

\begin{Theorem}\label{th:oracle_upper_bound}
	Assume the support of $X$ is compact,  $\xi(X) = \E(Y|X)$ is bounded, and $Y$ is sub-Gaussian. Consider asymptotics as $n\to \infty$ for the case of both ideal and ordinary semi-supervision. Assume also that either (i) $\xi(X)$ is continuous or (ii) that $P_X$ is absolutely continuous with respect to Lebesgue measure on $\{X\}$. Let $\{g_k(x): k=1,\ldots\}$ be a bounded basis for the continuous functions on $\{X\}$ in case (i) and be a bounded basis for the ordinary $\ell_2$ Hilbert space on $\{X\}$ in case (ii). 
	Suppose $q_n \to \infty$ satisfying $q_n = o(\sqrt{n})$, Assumptions 1 holds, and either Assumptions 2 or 2' are satisfied, then
	\begin{itemize}
		\item the estimator $\hat{\theta}_{LS}^{\bullet 1}$ for the problem with observations $\{Y_i, X_{p+q_n}^\bullet: i=1,\ldots, n\}$ asymptotically achieves the ideal oracle risk, i.e.
		\begin{equation}\label{eq:oracle_upper_bound}
		\lim_{n\to \infty} n \E\left(\hat{\theta}_{\rm LS}^{\bullet 1} - \theta\right)^2 = \sigma^2. 
		\end{equation}
		\item Now we suppose $\lim_{n\to \infty} \frac{n}{n+m_n} = \rho$ for some fixed value $0\leq \rho \leq 1$. Applying the estimator $\hat{\theta}_{\rm SSLS}^\bullet$ for the problem with observations $\{Y_i, X_{p+q_n}^\bullet: i=1,\ldots, n\}$ and $\{X_{i}^\bullet\}_{i=n+1}^{n+m_n}$. Then
		\begin{equation}\label{eq:oracle_upper_bound_semi}
		\lim_{n\to \infty} n\E \left(\hat{\theta}_{\rm SSLS}^{\bullet 1} - \theta\right)^2 = \sigma^2 + \rho\Var_{P_X}(\xi(X)). 
		\end{equation}
	\end{itemize}
	Finally, $\hat{\theta}_{\rm LS}^\bullet$ and $\hat{\theta}_{\rm SSLS}^\bullet$ are asymptotically unbiased and normal with the corresponding variances.
\end{Theorem}

\ \par

\eqref{eq:oracle_upper_bound} and \eqref{eq:oracle_upper_bound_semi} show that the proposed estimators asymptotically achieve the oracle values in \eqref{eq:variance_oracle} and \eqref{eq:variance_oracle_semi}.
On the other hand, one could use the simpler ordinary estimators $\hat{\theta}_{\rm LS}^\bullet$ and $\hat{\theta}_{\rm SSLS}^\bullet$ in place of $\hat{\theta}_{\rm LS}^{\bullet 1}$ and $\hat{\theta}_{\rm SSLS}^{\bullet 1}$ in practice, since $\hat{\theta}_{\rm LS}^{\bullet}$ and $\hat{\theta}_{\rm SSLS}^{\bullet}$ converge in distribution with asymptotic variance as in \eqref{eq:variance_oracle} and \eqref{eq:variance_oracle_semi}.

	\begin{Remark}\label{rm:literature-efficiency}
	There are several results in the semiparametric regression literature \citep{hasminskii1983asymptotic,bickel1991efficient,bickel1998efficient,peng2002efficient,Vaart2002semiparametric,hansen2017econometrics,chakrabortty2017efficient} that show similar aspects to our results. 
	For example \cite{bickel1991efficient} discusses semiparametric inference for the joint	distribution of bivariate $(Y, X)\in \mathbb{R}^2$ given known marginal distributions $P_X$ and/or
	$P_Y$. With $P_X$ known and $P_Y$ unknown this corresponds to our ideal semi-supervised setting. Their estimator is built from a suitable, binned nonparametric regression estimator of $\xi(x)$. It can be shown using comments in Section 4 of their article that their procedure will yield a semiparametric efficient estimator of $\E Y$ for our ideal semi-supervised problem when $X$ is real. (Generalization to multivariate $X$ is relatively straightforward.) \cite{chakrabortty2017efficient} develop several different semiparametric efficient estimators for the population regression slopes in ideal semi-supervised semiparametric regression, or when $m/n\to\infty$. It can be shown with a little extra work that these also yield semiparametric efficient estimators of the mean of $Y$ for such a setting. Though it shares some common features with each of these approaches our series estimator also shows some fundamental differences	to any of these proposals. We remark below that our series estimator is also semiparametric efficient under suitable regularity conditions. But our emphasis remains on its directness, simplicity, and the implications of this for good finite	sample performance (including confidence intervals) relative to $\bar{\Y}$.

\end{Remark}

	\begin{Remark}\label{rm:efficiency}	

	The oracle optimality in Proposition \ref{pr:oracle_lower_bound-efficiency} involves an asymptotically least favorable one-dimensional family of conditional means under ideal semi-supervised setting. It also places no special restriction on the conditional distribution of $Y-\xi_0(X)$. Consequently, the conditional sample mean (if a large conditional sample were available) would be the asymptotically efficient estimator of $\xi_0(X)$. It follows that the oracle optimal rates in \eqref{eq:oracle_risk} is equal to the asymptotic semiparametric efficiency bound. Hence the series estimator $\hat{\theta}^{\bullet}_{\rm LS}$ of section \ref{sec:improved procedure} is asymptotically efficient under the regularity conditions of Theorem \ref{th:oracle_upper_bound}. We believe Proposition \ref{pr:oracle_lower_bound-efficiency} can be further extended to a version applying to ordinary semi-supervision and yields the corresponding semiparametric efficiency bound. 
			
		Although the preceding argument is informal, it can be made precise. \cite{bickel1991efficient} and \cite{chakrabortty2017efficient} contain more detailed, conventional arguments for estimating regression slopes in the ideal semi-supervised case, and the result for estimating $\E Y$ can be drawn from there via standard reasoning.	Some remarks in the latter paper about MAR data can be used to extend the	treatment to ordinary semi-supervision, as can a specialization of the MAR results in \cite{graham2011efficiency}. A detailed argument for all cases can be found in \cite{kuchibhotla2017research}.
		
\end{Remark}

\begin{Remark}
Theorem \ref{th:oracle_upper_bound} suggests that the number of terms in the series should be $q_n = o(\sqrt{n})$. As a crude rule of thumb we could suggest choosing $q_n \approx n^{1/3}$ . Hence, if $n =100$ one could choose $q_n =5$. Our estimator in a problem having such $n$ and $q$ is not optimal in any sense, but one can be fairly confident that it will at least be noticeably better than $\bar{\Y}$.
\end{Remark}

\section{Simulation Results}
\label{sec.simulaton}

In this section, we investigate the numerical performance of the proposed estimators in various settings in terms of estimation errors and coverage probability as well as length of confidence intervals. All the simulations are repeated for 1000 times.

We analyze the linear least squares estimators $\hat{\theta}_{\rm LS}$ and $\hat{\theta}_{\rm SSLS}$ proposed in Section \ref{sec.procedure} in the following three settings.
\begin{enumerate}
	\item (Gaussian $X$ and quadratic $\xi$) We generate the design and parameters as follows, $\mu \sim N(0, I_p)$, $\Sigma \in \mathbb{R}^{p\times p}$, $\Sigma_{ij} = I\{i = j\} + \frac{1}{2p} I\{i\neq j\}$, $\beta \sim N(0, I_{p+1})$. Then we draw i.i.d. samples $\Y, \X$ as 
	$$X_k \sim N(\mu, \Sigma),\quad Y_k = \xi(X_k) + \varepsilon_k, $$
	where 
	$$\xi(X_k) = (\|X_k\|_2^2 - p) + \vec{\X}^\top  \beta, \quad  \varepsilon_k\sim N\left(0, 2 \|X_k\|_2^2/p\right). $$
	It is easy to calculate that $\theta = \E Y= \beta_1$ in this setting.
	
	\item (Heavy tailed $X$ and $Y$) We randomly generate
	\begin{equation*}
	\{X_{ki}\}_{1\leq k\leq n, 1\leq i\leq p} \overset{iid}{\sim} P_3, \quad  Y_k = \sum_{i=1}^p \left(\sin(X_{ki}) + X_{ki}\right) +  .5\cdot \varepsilon_k, \quad \varepsilon_k \overset{iid}{\sim} P_3.
	\end{equation*}
	where $P_3$ has density $f_{P_3}(x) = \frac{1}{1 + |x|^3}$, $-\infty <x <\infty$. Here, the distribution $P_3$ has no third or higher moments. In this case, $\mu = \E X = 0$, $\theta = \E  Y = 0$. 
	\item (Poisson $X$ and $Y$) Then we also consider a setting where 
	$$\{X_{ki}\}_{1\leq k\leq n, 1\leq i\leq p} \overset{iid}{\sim} {\rm Poisson}(10),\quad Y_k|X_k \overset{iid}{\sim} {\rm Poisson}(10X_{k1}).$$ 
	In this case, $\mu = \E X = (10, \ldots, 10)^\top \in \mathbb{R}^p$, $\theta = \E\Y = 100$.
\end{enumerate}

We compare the average $\ell_2$-loss of $\bar\Y$, $\hat{\theta}_{\rm LS}$ and $\hat{\theta}_{\rm SSLS}$ for various choices of $n, p$ and $m$. The results are summarized in Table \ref{tb:simulation_1}. The primary message to notice is that in every case, our estimator is preferable to Y-bar. An interesting aspect is even when $p$ grows faster than $n^{1/2}$, $\hat{\theta}_{\rm LS}$ and $\hat{\theta}_{\rm SSLS}$ are still preferable estimators to $\bar{\Y}$. It is also noteworthy that although our theoretical analysis for the $\ell_2$-risk focused on the refined estimators $\hat{\theta}_{\rm LS}^1$ and $\hat{\theta}_{\rm SSLS}^1$ with bounded or sub-Gaussian designs, the refinement and assumptions are for technical asymptotic needs, which might not be necessary in practice as we can see from these examples. 

We also compute the 95\%-confidence interval for each setting above and list the average length and coverage probability in Table \ref{tb:simulation_CI}. It can be seen that under the condition $p = o(n^{1/2})$, the proposed confidence intervals based on $\hat{\theta}_{\rm LS}$ and $\hat{\theta}_{\rm SSLS}$ are close to valid and shorter on average than the traditional $z$-confidence interval centered at $\bar{\Y}$.
\begin{table}[htbp]
	\begin{center}
		\begin{tabular}{cccccc}
			\toprule
			$(p,n)$ & $(\bar{\Y} - \theta)^2$ & \multicolumn{3}{c}{$(\hat{\theta}_{\rm SSLS}-\theta)^2$} & $(\hat{\theta}_{\rm LS}-\theta)^2$\\
			\cmidrule{3-5}
			&  & $m= 100$ & $m=1000$ & $m =10000$ & \\  \hline
			\multicolumn{6}{c}{Setting 1: Gaussian $X$ and Quadratic $\xi$}\\
			(1, 100) & 0.304& 0.184& 0.075 & 0.063 & 0.056\\
			(10, 100) & 2.73& 1.529& 0.518 & 0.313 & 0.296\\
			(50, 100) & 13.397  &  7.961 & 3.967 & 2.988 &  2.868\\
			(10, 500) & 0.526  & 0.464 & 0.211 & 0.067 & 0.045\\
			(50, 500) & 2.668  & 2.278 & 1.089 & 0.373 & 0.273\\
			(200, 500) & 10.743  &  9.135 &  4.615 &  2.345 &  1.949 \\ \hline
			\multicolumn{6}{c}{Setting 2: Heavy tailed $X$ and $Y$}\\
			(1, 100) & 0.732  & 0.410 & 0.244 & 0.196 & 0.188\\
			(10, 100) & 7.791 & 5.428 & 2.505 & 1.959 & 1.831\\
			(50, 100) & 107.363 & 47.036 & 17.754 & 14.201 & 13.435\\
			(10, 500) & 2.575 & 2.097 & 0.988 & 0.354 & 0.261\\
			(50, 500) & 12.569  & 10.481 & 5.619 & 2.342 & 1.780\\
			(200, 500) & 43.997 & 36.123 & 30.856 & 13.175 & 9.642\\\hline
			\multicolumn{6}{c}{Setting 3: Poisson $X$ and $Y$}\\
			(1, 100) & 97.912 & 50.510 & 10.168 & 2.036 & 1.015 \\
			(10, 100) & 98.337 & 50.772 & 10.535 & 2.085 & 1.061 \\
			(50, 100) & 94.475 & 52.166 & 10.951 & 3.146 & 2.100 \\
			(10, 500) & 20.062 & 16.765 & 6.890 & 1.104 & 0.186 \\
			(50, 500) & 19.915 & 15.793 & 6.541 & 1.165 & 0.225 \\
			(200, 500) & 20.933 & 17.639 & 7.159 & 1.300 & 0.333 \\\hline
		\end{tabular}
	\end{center}
	\caption{Average squared loss of sample mean estimator $\bar{\Y}$, the least squares estimator $\hat{\theta}_{\rm LS}$ and the semi-supervised least squares estimators $\hat{\theta}_{\rm SSLS}$ under different values of $(p, n)$ and various settings.}
	\label{tb:simulation_1}
\end{table}
 
\begin{table}[htbp]
	\begin{center} 
		\begin{tabular}{cccccc}
			\toprule
			$(p,n)$ & $\bar{\Y}$ &  \multicolumn{3}{c}{$\hat{\theta}_{\rm SSLS}$} & $\hat{\theta}_{\rm LS}$\\
			\cmidrule{3-5}
			&  & $m= 100$ & $m=1000$ & $m =10000$ & \\ \hline
			\multicolumn{6}{c}{Setting 1: Gaussian $X$ and Quadratic $\xi$}\\
			(1, 100) & 1.902(0.945)  & 1.521(0.954) & 1.074(0.951) & 0.940(0.939) & 0.921(0.936)\\
			(5, 100) & 4.430(0.942)  & 3.301(0.930) & 1.911(0.945) & 1.467(0.941) & 1.400(0.931) \\
			(10, 100) & 6.318(0.952)  & 4.678(0.942) & 2.655 (0.937) & 2.010(0.924) & 1.913(0.916) \\
			(1, 500) & 0.845(0.959)  & 0.793(0.958) & 0.608(0.959) & 0.451(0.958) & 0.413(0.954) \\
			(10, 500) & 2.818(0.955)  & 2.596(0.959) & 1.768(0.952) & 1.023(0.949) & 0.832(0.936)\\
			(25, 500) & 4.558(0.949)  & 4.194(0.961) & 2.837(0.946) & 1.606(0.942) & 1.288(0.922) \\  \hline
			\multicolumn{6}{c}{Setting 2: Heavy tailed $X$ and $Y$}\\
			(1, 100) & 3.349(0.961)  & 2.069(0.941) & 1.596(0.939) & 1.446(0.956) & 1.420(0.962)\\
			(5, 100) & 7.332(0.950)  & 4.885(0.918) & 3.384(0.933) & 2.920(0.937) & 2.847(0.952)\\
			(10, 100) & 11.292(0.956)  & 7.436(0.921) & 5.073(0.922) & 4.343(0.943) & 4.225(0.956)\\
			(1, 500) & 1.573(0.954)  & 1.205(0.945) & 0.970(0.923) & 0.773(0.937) & 0.723(0.942)\\
			(10, 500) & 5.947(0.957)  & 4.427(0.939) & 3.217(0.916) & 2.180(0.931) & 1.904(0.953)\\
			(25, 500) & 8.582(0.960)  & 7.079(0.945) & 5.197(0.928) & 3.617(0.931) & 3.229(0.953) \\\hline
			\multicolumn{6}{c}{Setting 3: Poisson $X$ and $Y$}\\
			(1, 100) & 39.164(0.937) & 27.831(0.939) & 12.386(0.944) & 5.506(0.953) & 3.895(0.925) \\
			(5, 100) & 39.396(0.947) & 28.003(0.957) & 12.485(0.933) & 5.600(0.938) & 4.004(0.930) \\
			(10, 100)& 39.143(0.935) & 27.832(0.946) & 12.443(0.936) & 5.655(0.942) & 4.105(0.937)\\
			(1, 500) & 17.548(0.946) & 16.035(0.946) & 10.232(0.950) & 4.195(0.957) & 1.753(0.946)\\
			(10, 500) & 17.621(0.947) & 16.102(0.938) & 10.276(0.952) & 4.216(0.950) & 1.768(0.957)\\
			(25, 500) & 17.632(0.947) & 16.113(0.948) & 10.285(0.949) & 4.229(0.955) & 1.795(0.939)\\
			\hline
		\end{tabular}
	\end{center}
	\caption{Average length and coverage probability (in the parenthesis) 95\%-CI based on $\bar{\Y}$, $\hat{\theta}_{\rm LS}$ and $\hat{\theta}_{\rm SSLS}$ under different values of $(p, n)$ and various settings.}
	\label{tb:simulation_CI}
\end{table}

\section{Applications}
\label{application.sec}

In this section, we apply the proposed procedures to the average treatment effect estimation and a real data example on homeless population.

\subsection{Application to Average Treatment Effect Estimation}
\label{sec.ATE}

We first discuss an application of the proposed least squares estimator to Average Treatment Effect (ATE) estimation. Suppose $Y_T$ and $Y_C$ are the responses for the treatment population and control population respectively, then ATE is then defined as
\begin{equation}
d = \E Y_T - \E Y_C.
\end{equation} 
Under Neyman's paradigm \citep{splawa1990application,rubin1990application}, a total number of $(n_t+n_c)$ subjects are randomly assigned to the treatment group and control group. Suppose $Y_{t,1},\cdots, Y_{t, n_t}$ are the responses under treatment, while $Y_{t, 1}, \cdots, Y_{t, n_c}$ are the responses of the control group. The straight forward idea for estimating ATE is the sample average treatment effect (SATE), which simply takes the difference of average effects between the two groups. In addition, the covariates associated with the responses are often available and helpful to improve the estimation of ATE.

In the estimation of ATE, we follow the model setting of \cite{pitkin2013improved}. Suppose $n_t, n_c$ people are from treatment group and control group respectively, where their response and predictor satisfies
$$(Y_t, X_t) \overset{iid}{\sim} P^t, \quad (Y_c, X_c)\overset{iid}{\sim} P^c.$$ 
Here due to the randomization setting, it is reasonable to assume $P^t$ and $P^c$ share the same marginal distribution of $X$: $P_X^t = P_X^c=P_X$. There are also $m$ additional samples possibly coming from drop-outs or any other subjects that also represent the population $P_X$. In summary, the available samples include
\begin{equation}
\{(Y_{t, k}, X_{t, k})\}_{k=1}^{n_t}, \quad \{(Y_{c, k}, X_{c, k})\}_{k=1}^{n_c},\quad  \{(X_{a, k})\}_{k=1}^{m}.
\end{equation}
We again introduce the population slope for both treatment and control group to measure the relationship between $Y_t, X_t$ and $Y_c, X_c$ respectively
\begin{equation}
\beta_t = \argmin_{\gamma\in \mathbb{R}^{p+1}} \E \left(Y_t - \vec{X}_t^\top \gamma \right)^2, \quad \beta_c = \argmin_{\gamma\in \mathbb{R}^{p+1}} \E \left(Y_c - \vec{X}_c^\top \gamma \right)^2.
\end{equation}
Based on Lemma \ref{lm:basic_properties_delta}, $\beta_t, \beta_c$ has the following close form when $P_t, P_c$ have non-degenerate second moment:
\begin{equation}
\beta_t = \left(\E \vec {X}_t \vec{X}_t^\top \right)^{-1} \left(\E \vec{X}_t Y_t\right),\quad \beta_c = \left(\E \vec {X}_c \vec{X}_c^\top \right)^{-1} \left(\E \vec{X}_c Y_c\right).
\end{equation}
Our target, the population ATE, is defined as $d = \E Y_c - \E Y_t$. We propose the corresponding semi-supervised least squares estimator
\begin{equation}
\hat d_{\rm SSLS} = \hat{\mu}^\top  \left(\hat{\beta}_t - \hat{\beta}_c\right).
\end{equation}
Here $\hat{\beta}_t, \hat{\beta}_c\in \mathbb{R}^{p+1}$ are the least squares estimators for treatment and control group respectively; $\hat{\vec{\mu}}$ is the mean of all available predictors,
\begin{equation}
\hat{\beta}_t = \left(\vec{\X}_t^\top \vec{\X}_t\right)^{-1} \vec{\X}_t^\top\Y_t ,\quad \hat{\beta}_c = \left(\vec{\X}_c^\top \vec{\X}_c\right)^{-1} \vec{\X}_c^\top \Y_c, 
\end{equation}
\begin{equation}
\text{where} \quad \hat{\vec{\mu}} = \begin{pmatrix}
1\\
\hat{\mu}
\end{pmatrix},\quad \hat{\mu} = \frac{1}{n_t+n_c + m}\left(\sum_{k=1}^{n_t}X_{t, k} + \sum_{k=1}^{n_c}X_{c, k} + \sum_{k=1}^m X_{a, k}\right).
\end{equation}

Based on the analysis we have in the previous section, the proposed $\hat{d}_{\rm SSLS}$ has the following asymptotic distribution with a fixed $p$, $P^t$ and $P^c$.
\begin{Theorem}[Asymptotic behavior of $\hat{d}_{\rm SSLS}$]\label{th:d_SPLS_asymptotic}
	Suppose $P^t, P^c$ are fixed distribution with finite and non-degenerate second moments, then we have the following asymptotic distribution if the sample size $n_t, t_c$ grow to infinity:
	\begin{equation}
	\frac{\hat{d}_{\rm SSLS} - d}{V}\overset{d}{\to} N(0, 1),\quad
	\frac{\hat V^2}{V^2} \overset{d}{\to} 1.
	\end{equation}
	Here 
	\begin{equation}\label{eq:V^2}
	V^2 = \frac{\tau_t^2}{n_t} + \frac{\tau_c^2}{n_c} + \frac{1}{n_t+n_c+m}(\beta_{t, (2)}-\beta_{c, (2)})^\top \E (X - \mu)(X-\mu)^\top  (\beta_{t, (2)} - \beta_{c, (2)}),
	\end{equation}
	\begin{equation}\label{eq:hat V^2}
	\hat V^2 = \frac{MSE_t}{n_t} + \frac{MSE_c}{n_c} + \frac{1}{n_t+n_c+m} (\hat{\beta}_t - \hat{\beta}_c)^\top \hat{\SSigma}_X(\hat{\beta}_t - \hat{\beta}_c),
	\end{equation}
	$$MSE_t = \frac{1}{n_t-p-1} \sum_{k=1}^{n_t} (Y_{t, k} - \vec{X}_{t, k}^\top \hat\beta_{t})^2, \quad MSE_c = \frac{1}{n_c-p-1} \sum_{k=1}^{n_c} (Y_{c, k} - \vec{X}_{c, k}^\top \hat\beta_{c})^2, $$
	\begin{equation*}
	\begin{split}
	\hat{\SSigma}_X = \frac{1}{n_t+n_c+m}\Big(& \sum_{k=1}^{n_t} (X_{t, k} - \hat{\mu})(X_{t, k} - \hat{\mu})^\top +\sum_{k=1}^{n_c} (X_k - \hat{\mu})(X_k - \hat{\mu})^\top \\
	& +\sum_{k=1}^m (X_k - \hat{\mu})(X_k - \hat{\mu})^\top \Big).
	\end{split}
	\end{equation*}
\end{Theorem}
\begin{Remark}\rm 
	Similarly to the procedure in Proposition \ref{th:nvar(bar_Y)}, we can calculate that for the sample average treatment effect, i.e.,
	$$\hat{d} = \sum_{k=1}^{n_t} \frac{Y_{t, k}}{n_t} - \sum_{k=1}^{n_c}\frac{Y_{t, c}}{n_c}, $$
	$$\Var(\hat{d}) = \frac{\tau_t^2 + \beta_{t, (2)}^\top \E (X - \mu)(X - \mu)^\top\beta_{t, (2)}}{n_t} + \frac{\tau_c^2 + \beta_{c, (2)}^\top \E (X - \mu)(X - \mu)^\top\beta_{t, (2)}}{n_c}. $$
	We can check that asymptotically $V^2 \leq \Var(\hat{d})$, which also shows the merit of the proposed semi-supervised least squares estimator.
\end{Remark}

\begin{Remark}\rm
	The asymptotic behavior of $\hat{d}_{\rm SSLS}$ and the $\ell_2$ risk for a refined $\hat{d}_{\rm SSLS}$ for growing $p$ can be elaborated similarly to the previous sections.
\end{Remark}

\subsection{Real Data Example: Estimating Homeless in Los Angeles County}
\label{sec.real_data}

We now consider an application to estimate the number of homeless people in Los Angeles County. Homelessness has been a significant public issue for the United States since nearly a century ago \citep{rossi1991strategies}. A natural question for the demographers is to estimate the number of homeless in a certain region. Estimating the number of homeless in metropolitan area is an important but difficult task due to the following reasons. In a typical design of U.S. Census, demographers visit people through their place of residence. In this case, most of the homeless will not be contacted \citep{rossi1991strategies} through this process. Visiting homeless shelters or homeless service centers may collect some information of the homeless, but a large number of homeless still cannot be found since they may use the service anonymously or simply not use the service.

Los Angeles County includes land of 2000 square miles, total population of 10 million and 2,054 census tracts. In 2004-2005, the Los Angeles Homeless Services Authority (LAHSA) conducted a study of the homeless population. Due to the cost of performing street visits for all census tracts, LAHSA used a stratified sampling plan.
First, 244 tracts that were believed to have large amount of homeless were pre-selected and visited. Next for the rest of the tracts, 265 of them were randomly selected and visited. This design leaves 1,545 tracts unvisited. Besides the number of homeless, some predictors were available for all 2,054 tracts. In our analysis, 7 of them were included, \verb|Perc.Industrial|, \verb|Perc.Residential|, \verb|Perc.Vacant|, \verb|Perc.Commercial|, \verb|Perc.OwnerOcc|, \verb|Perc.Minority|, \verb|MedianHouseholdIncome|. These predictors have been used and were known to have a useful correlation with the response \cite{kriegler2010small}. 

Suppose $T_{\rm total}$ is the total number of homeless in Los Angeles, $T_{\rm pre}$ is the number of homeless in 244 pre-selected tracts, $\theta_{\rm ran}$ is average number of homeless per tract in all 1,810 non-pre-selected tracts. Clearly,
\begin{equation}\label{eq:T_total}
T_{\rm total} = T_{\rm pre} + 1810\cdot \theta_{\rm ran}. 
\end{equation}
The proposed semi-supervised inference framework fit into the 1,810 samples with 265 labeled and 1,545 unlabeled samples. We can apply the proposed semi-supervised least squares estimator $\hat{\theta}_{\rm SSLS}^1$ to estimate $\theta_{\rm ran}$ and use \eqref{eq:T_total} to calculate the estimate and 95\% confidence interval for $T_{\rm total}$. In contrast, the estimate via sample-mean estimator was also calculated. The results are shown in Table \ref{tb:homeless_in_LA}.
\begin{table}
	\begin{center}
		\begin{tabular}{cccc}
			\hline
			via $\hat{\theta}_{\rm SSLS}$ & 95\%-CI & via $\bar\Y$ & 95\%-CI\\
			53824 & 	[47120, 60529]	& 52527 & [45485, 59570]\\\hline
		\end{tabular}
		\caption{Estimated total number of homeless in Los Angeles County}
		\label{tb:homeless_in_LA}
	\end{center}
\end{table}
It is easy to see that the estimate via $\hat{\theta}_{\rm SSLS}^1$ is slightly larger than the one via $\bar{\Y}$. 

To further investigate and diagnose, we calculated the least squares estimator $\hat{\beta}$, the average predictor values across all 1,810 non-pre-selected tracts $\bar{\X}_{\rm full}$ and the average predictor values across 265 randomly selected tracts $\bar{\X}$. These values are listed in Table \ref{tb:diagnostic}.
\begin{table}\label{tb:diagnostic}
	\begin{center}
		\begin{tabular}{c||cccc}\hline
			& $\hat{\beta}$ &$\bar{\X}_{\rm full} - \bar{\X}$ & $\bar{\X}$ & $\bar{\X}_{\rm full}$ \\
			Intercept & 21.963 & \\
			Perc.Industrial & 0.027 & 0.143 &  61.293 & 61.149 \\
			Perc.Residential & -0.087 & -0.075 &  4.066 &  4.141\\
			Perc.Vacant & 1.404 &  -0.075 &   4.066  & 4.141\\
			Perc.Commercial & 0.338 & -0.542 & 15.130 & 15.672\\
			Perc.OwnerOcc &  -0.233 &  2.489 &  54.039 &  51.550\\
			Perc.Minority & 0.058 & 0.833 & 50.890 &  50.057\\
			MedianInc (in \$K) & 0.074 & 0.638 & 48.805 & 48.167 \\
			& \multicolumn{4}{c}{Adjustment: $\hat{\beta}_{(2)}^\top (\bar{\X}_{\rm full} - \bar{\X})$ = -0.768}
			\\\hline
		\end{tabular}
		\caption{Diagnostic Table for Los Angeles Data Example}
	\end{center}
\end{table}

We can see from Table \ref{tb:diagnostic} that due to insufficiency of sampling, there is difference between $\bar{\X}$ and $\bar{\X}_{\rm full}$, especially for the predictor \verb|Perc.OwnerOcc|. When there is association between these prectors and reponse, it is more reasonable to adjust for this discrepancy from taking the mean. Recall the proposed estimator
$$
\hat{\theta}_{\rm SSLS} = \bar \Y + \hat \beta_{(2)}^\top  \left(\bar\X_{\rm full} - \bar\X\right),\quad \text{where}\quad \bar{\X}_{\rm full} = \frac{1}{n+m} \sum_{k=1}^{n+m} X_k, \bar{\X} = \frac{1}{n}\sum_{k=1}^n X_k. $$
The difference between two estimates exactly originated from the adjustment term $\hat{\beta}_{(2)}^\top (\bar\X_{\rm full} - \bar{\X})$, which has been justified in both theoretical analysis and simulation studies in the previous sections.

\section{Proofs of The Main Results}
\label{proofs.sec}

We prove the main results in this section. The proofs of other technical results are provided in the Supplement.

\bigskip

\subsection{Proofs for Ideal Semi-supervised Inference Estimator $\hat{\theta}_{\rm LS}$}

\begin{proof}[Proof of Theorem \ref{th:theta_LS_asymptotic}]

We first show that $\hat{\theta}_{\rm LS}$ is invariant under simultaneous affine translation on both $\X$ and $\mu$. Specifically, suppose $X_k = U\cdot Z_k + \alpha$, $(k=1,\cdots, n)$ for any fixed invertible matrix $U\in \mathbb{R}^{p\times p}$ and vector $\alpha\in \mathbb{R}^p$. Then one has 
$$\vec{X}_k = \begin{bmatrix}
1 & 0\\
\alpha & U
\end{bmatrix}\vec{Z}_k, \quad \vec{\X} = \vec{\Z}\begin{bmatrix}
1 & \alpha^\top \\
0 & U^\top 
\end{bmatrix}, $$
\begin{equation*}
\begin{split}
\hat{\theta}_{\rm LS} = & \vec{\mu}^\top \left(\vec{\X}^\top \vec{\X}\right)^{-1}\vec{\X}^\top \Y\\
= & (1, \mu^\top )\left(\begin{bmatrix}
1 & 0\\
\alpha & U
\end{bmatrix}\vec{\Z}^\top \vec{\Z}\begin{bmatrix}
1 & \alpha^\top \\
0 & U^\top 
\end{bmatrix}\right)^{-1}\begin{bmatrix}
1 & 0\\
\alpha & U
\end{bmatrix}\vec{\Z}^\top \Y\\
= & (1, \mu^\top ) \begin{bmatrix}
1 & \alpha^\top \\
0 & U^\top 
\end{bmatrix}^{-1}\left(\vec\Z^\top \vec\Z\right)^{-1}\vec{\Z}^\top \Y \\
=&(1, (U^{-1}(\mu-\alpha))^{\top }) \left(\vec{\Z}^\top \vec{\Z}\right)^{-1}\vec{\Z}^\top \Y.
\end{split}
\end{equation*}
Since $\E Z_k = U^{-1}(\mu - \alpha)$, we know $\hat{\theta}_{\rm LS}$ is invariant under simultaneous affine translation on $\X$ and $\mu$. 

Based on the affine transformation invariant property, we only need to consider the situation when $\E X = \mu=0$, $\Cov(X) = I_{p}$, where $I_p$ is the $p$-by-$p$ identity matrix. Next we discuss the asymptotic behavior for $\hat\theta_{\rm LS}$.  For simplicity, we note $1_n = (\overbrace{1, \cdots, 1}^n)^\top $, $\mathbb{P}_{\vec \X} = \vec{\X}(\vec{\X}^\top \vec{\X})^{-1} \vec{\X}^\top \in \mathbb{R}^{(p+1)\times (p+1)}$ as the projection matrix onto the column space of $\vec{\X}$. $\bar{\X} = \frac{1}{n}\sum_{k=1}^n X_k$. Clearly, $1_n$ lies in the column space of $\vec{\X}$, which means $\mathbb{P}_{\vec{\X}} 1_n = 1_n$. Then,
\begin{equation}\label{eq:hat_theta_LS-theta}
\begin{split}
\hat{\theta}_{\rm LS} - \theta = & \vec \mu^\top  \hat{\beta} - \theta = \vec \mu^\top  \left(\vec{\X}^\top \vec{\X}\right)^{-1} \vec{\X}^\top \Y - \theta\\
= & \vec{\mu}^\top \left(\vec{\X}^\top \vec{\X}\right)^{-1} \vec{\X}^\top \left(\vec{\X}\beta + \Bdelta\right) - \theta = \vec{\mu}^\top \left(\vec{\X}^\top \vec{\X}\right)^{-1}\vec{\X}^\top \Bdelta\\
= & \frac{1_n^\top }{n}\vec \X\left(\vec{\X}^\top \vec{\X}\right)^{-1} \vec{\X}^\top \Bdelta - \frac{1_n^\top }{n}(\vec \X - 1_n\vec\mu^\top )\left(\vec{\X}^\top \vec{\X}\right)^{-1}\vec{\X}^\top \Bdelta\\
= & \frac{1_n^\top \mathbb{P}_{\vec{\X}}}{n} \Bdelta - \left(0, \frac{1^\top _n}{n}\X\right) \left(\vec{\X}^\top \vec{\X}\right)^{-1}\vec{\X}^\top \Bdelta\\
= & \frac{1_n^\top }{n}\Bdelta - \left(0, \frac{1_n^\top}{n}\X\right) \left(\frac{1}{n}\vec{\X}^\top \vec{\X}\right)^{-1} \left(\frac{1}{n}\vec{\X}^\top \Bdelta\right)\\
= & \bar\Bdelta - \left(0, \bar\X^\top \right) \left(\frac{1}{n}\vec{\X}^\top \vec{\X}\right)^{-1} \left(\frac{1}{n}\vec{\X}^\top \Bdelta\right),
\end{split}
\end{equation}
\begin{equation}\label{eq:(n-p-1)MSE}
\begin{split}
& \frac{n-p-1}{n}MSE\\
 = & \frac{1}{n}\|\Y - \vec \X \hat{\beta}\|_2^2 = \frac{1}{n}\left\| \Bdelta + \vec \X \beta - \vec \X \left(\vec \X^\top \vec \X\right)^{-1} \vec \X^\top (\X \beta + \Bdelta)\right\|_2^2\\
= & \frac{1}{n}\left\| \Bdelta - \vec \X(\vec{\X}^\top \vec{\X})^{-1} \vec{\X}^\top \Bdelta \right\|_2^2 = \frac{1}{n}\left(\Bdelta^\top  \Bdelta - \Bdelta^\top \vec{\X} (\vec{\X}^\top \vec{\X})^{-1}\vec{\X}^\top \Bdelta\right)\\
= & \left(\frac{1}{n}\Bdelta^\top \Bdelta - \left(\frac{1}{n}\vec{\X}^\top  \Bdelta\right)^\top \left(\frac{1}{n}\vec{\X}^\top \vec{\X}\right)^{-1}\left(\frac{1}{n}\vec{\X}^\top  \Bdelta\right)\right).
\end{split}
\end{equation}
Since $P$ is fixed and has finite second moment, by law of large number, one can show as $n\to \infty$,
$$\frac{1}{n}\Bdelta^\top \Bdelta = \frac{1}{n}\sum_{k=1}^n\delta_k^2 \overset{d}{\to} \E \delta^2 =  \tau^2, $$
\begin{equation}\label{ineq:Xdelta}
\left\| \frac{\vec \X^\top  \boldsymbol{\delta} }{n} \right\|_2^2 \overset{d}{\to} \|\E \vec{X}\delta\|_2^2 = 0,
\end{equation}
$$\frac{1}{n}\vec \X^\top \vec \X \overset{d}{\to} \E \vec X \vec X^\top  = \begin{bmatrix}
1 & 0\\
0 & \Cov(X)
\end{bmatrix}. $$
Since $\Cov(X) = I_p$ is invertible, we know
$$\left(\frac{1}{n}\vec \X^\top \vec \X\right)^{-1} \overset{d}{\to} \begin{bmatrix}
1 & 0\\
0 & \Cov(X)^{-1}
\end{bmatrix}.$$
Additionally, since $\mathbb{E}X=0$, and $X_1,\ldots, X_n$ are independent,
\begin{equation}\label{eq:thm1-1}
\begin{split}
& \mathbb{E}\left\| \frac{1_n^\top  \X }{\sqrt{n}} \right\|_2^2 = \frac{1}{n}\mathbb{E}\left(\sum_{k=1}^nX_k\right)\left(\sum_{k=1}^nX_k\right)^\top \\
= & \frac{1}{n}\mathbb{E}\sum_{k=1}^nX_k^\top X_k = \mathbb{E}\tr(XX^\top) = \tr(\Cov(X)) <\infty.\\
\end{split}
\end{equation}
Based on the asymptotic distributions above, for any $\varepsilon>0$, we have
\begin{equation*}
\begin{split}
& \mathbb{P}\left(\sqrt{n}\left(0, \bar\X^\top \right) \left(\frac{1}{n}\vec{\X}^\top \vec{\X}\right)^{-1} \left(\frac{1}{n}\vec{\X}^\top \Bdelta\right) \geq \varepsilon\right) \\
\leq & \mathbb{P}\left(\left\|\frac{1_n^\top\X}{\sqrt{n}}\right\|_2 \left\|\frac{\vec{\X}^\top\boldsymbol{\delta}}{n}\right\|_2\cdot \left\|\left(\frac{1}{n}\vec{\X}^\top\vec{\X}\right)^{-1}\right\|\geq \varepsilon\right) \\
\leq & \mathbb{P}\left(\left\|\frac{1_n^\top\X}{\sqrt{n}}\right\|_2 \geq \varepsilon/\varepsilon_n \right) + \mathbb{P}\left( \left\|\frac{\vec{\X}^\top\boldsymbol{\delta}}{n}\right\|_2 \geq \varepsilon_n/(2(\|\Cov(X)^{-1}\|+1)) \right) \\
& + \mathbb{P}\left( \left\|\left(\frac{1}{n}\vec{\X}^\top\vec{\X}\right)^{-1}\right\|\geq 2(\|\Cov(X)^{-1}\|+1)\right),
\end{split}
\end{equation*}
where $\varepsilon_n$ grows slowly with $n$ to ensure that $\mathbb{P}\left( \left\|\frac{\vec{\X}^\top\boldsymbol{\delta}}{n}\right\|_2 \geq \varepsilon_n/(2(\|\Cov(X)^{-1}\|+1)) \right) \to 0$ as $n\to \infty$. By such the argument, 
$$\forall \varepsilon>0, \quad \lim_{n\to \infty} \mathbb{P}\left(\sqrt{n}\left(0, \bar\X^\top \right) \left(\frac{1}{n}\vec{\X}^\top \vec{\X}\right)^{-1} \left(\frac{1}{n}\vec{\X}^\top \Bdelta\right) \geq \varepsilon\right)=0,$$
which means $\sqrt{n}\left(0, \bar\X^\top \right) \left(\frac{1}{n}\vec{\X}^\top \vec{\X}\right)^{-1} \left(\frac{1}{n}\vec{\X}^\top \Bdelta\right) \overset{d}{\to} 0$. Next, by central limit theorem, $$\sqrt{n}\bar{\boldsymbol{\delta}}/\tau \overset{d}{\to} N(0, 1).$$
Combining \eqref{eq:hat_theta_LS-theta}, \eqref{eq:(n-p-1)MSE} and the previous asymptotic arguments, we know
$$\sqrt{n}\left(\hat{\theta}_{\rm LS} - \theta\right)/\tau = \sqrt{n}\bar{\Bdelta}/\tau - \sqrt{n}(0, \bar{\X}^\top)\left(\frac{1}{n}\vec{\X}^\top \vec{\X}\right)^{-1} \left(\frac{1}{n}\vec{\X}^\top \Bdelta\right)/\tau \to N(0, 1),$$
$$\frac{n-p-1}{n}MSE \overset{d}{\to} \tau^2,$$
in the case that $P_X$ fixed and $n\to \infty$. 

Next, we use $C$ and $c$ to denote generic constants which does not depend on $n$ (but may depend on the distribution $P$). When $P$ further has finite fourth moment, by Berry-Esseen's CLT,
\begin{equation}\label{eq:thm1-CLT}
	\left|\mathbb{P}\left(\sqrt{n}\bar{\boldsymbol{\delta}}/\tau \geq x\right) - \Phi(x)\right| \leq \frac{C}{\sqrt{n}}.
\end{equation}
We also have a finer estimation for $\|(\tilde{\X}^\top\Bdelta)/n\|_2^2$ than the one in \eqref{ineq:Xdelta}. Note that
\begin{equation}\label{eq:thm1-2}
	\mathbb{E}\left\|\frac{\vec{\X}^\top\Bdelta}{n}\right\|_2^2 = \frac{1}{n^2}\mathbb{E}\left(\sum_{k=1}^n\vec{X}_k\delta_k\right)^\top\left(\sum_{k=1}^n\vec{X}_k\delta_k\right) = \frac{1}{n^2}\sum_{k=1}^n\mathbb{E} \vec{X}_k^\top\vec{X}_k\delta_k^2 \leq \frac{C}{n},
\end{equation}
\begin{equation}\label{eq:thm1-3}
\begin{split}
& \mathbb{E} \left\|\frac{1}{n}\vec{\X}^\top\vec{\X} - I_{p+1} \right\|_F^2 = \mathbb{E} \tr\left(\frac{1}{n}\sum_{k=1}^n\left(\vec{X}_k\vec{X}_k^\top - \mathbb{E} \vec{X}\vec{X}^\top\right)\right)^2\\
= & \frac{1}{n^2}\sum_{k=1}^n\mathbb{E}\tr\left(\vec{X}_k\vec{X}_k^\top - \mathbb{E}\vec{X}_k\vec{X}_k^\top\right)^2 = \frac{1}{n}\mathbb{E}\tr\left(\vec{X}\vec{X}^\top - \mathbb{E}\vec{X}\vec{X}^\top\right)^2 \leq \frac{C}{n}.
\end{split}
\end{equation}
By Markov's Inequality,
\begin{equation}\label{ineq:thm1-4}
\begin{split}
& \mathbb{P} \left(\sqrt{n}(0, \bar{\X}^\top)\left(\frac{1}{n}\vec{\X}^\top\vec{\X}\right)^{-1}\left(\frac{1}{n}\vec{\X}^\top\Bdelta\right) \geq \frac{C}{n^{1/4}}\right)\\
= & \mathbb{P} \left(\left\| \frac{1_n^\top\X^\top}{\sqrt{n}}\right\|_2\cdot \left\|\left(\frac{1}{n}\vec{\X}^\top\vec{\X}\right)^{-1}\right\|\cdot \left\|\frac{\vec{\X}^\top\Bdelta}{n}\right\|_2 \geq \frac{C}{n^{1/4}}\right)\\
\leq & \mathbb{P} \left(\left\|\frac{1_n^\top \X}{\sqrt{n}}\right\|_2 \geq Cn^{1/8}\right) + \mathbb{P}\left(\left\|\frac{1}{n}\vec{\X}^\top\vec{\X} - I_{p+1}\right\|_F \leq 1/2\right)\\
& + \mathbb{P}\left(\left\|\frac{\vec{\X}^\top\Bdelta}{n}\right\|_2 \geq \frac{C}{n^{3/8}}\right)\\
\leq & \frac{\mathbb{E} \|1_n^\top\X/\sqrt{n}\|_2^2}{Cn^{1/4}} + \frac{\mathbb{E}\|\frac{1}{n}\vec{\X}^\top\vec{\X}-I_{p+1}\|_F^2}{1/2} + \frac{\mathbb{E}\|\vec{\X}^\top\Bdelta/n\|_2^2}{Cn^{-3/4}}\\
\overset{\eqref{eq:thm1-1}\eqref{eq:thm1-2}\eqref{eq:thm1-3}}{\leq} & Cn^{-1/4}.
\end{split}
\end{equation}
Finally, for any $x>0$,
\begin{equation*}
\begin{split}
& \mathbb{P}\left(\frac{\sqrt{n}(\hat{\theta}_{LS} - \theta)}{\tau} \leq x\right) \\
\overset{\eqref{eq:hat_theta_LS-theta}}{\leq} & \mathbb{P}\left(\frac{\sqrt{n}\bar{\Bdelta}}{\tau} \leq x+\frac{C}{n^{1/4}}\right) \\
& + \mathbb{P}\left(-\sqrt{n}\left(0, \bar\X^\top \right) \left(-\frac{1}{n}\vec{\X}^\top \vec{\X}\right)^{-1} \left(\frac{1}{n}\vec{\X}^\top \Bdelta\right)/\tau \leq -\frac{C}{n^{1/4}}\right)\\
\overset{\eqref{ineq:thm1-4}}{\leq} & \Phi\left(x+\frac{C}{n^{1/4}}\right) + Cn^{-1/4} \leq \Phi(x) + Cn^{-1/4}.
\end{split}
\end{equation*}
Here, the last inequality is due to the fact that the cdf of the standard normal distribution $\Phi(\cdot)$ is a Lipschitz continuous function. Similarly, 
\begin{equation*}
\mathbb{P}\left(\frac{\sqrt{n}(\hat{\theta}_{LS} - \theta)}{\tau} \leq x\right) \geq \Phi(x) - Cn^{-1/4}.
\end{equation*}
These together complete the proof of this theorem.

\end{proof}

\ \par

\begin{proof}[Proof of Theorem \ref{th:theta_LS_asymptotic_p_grow}] First, based on the proof of Theorem \ref{th:theta_LS_asymptotic}, the affine transformation on $\X$ would not affect the property of $\hat\theta_{\rm LS}$. Without loss of generality, we assume that $\E X = 0$, $\Var(X) = I$. In other words, $\Z = \X$. Next, based on formulas \eqref{eq:hat_theta_LS-theta} and \eqref{eq:(n-p-1)MSE}, we have
\begin{equation*}
\begin{split}
& \sqrt{n}(\hat{\theta}_{\rm LS} - \theta)/\tau = \frac{\sqrt{n}\bar{\Bdelta}}{\tau} - \sqrt{n} (0, \bar{\X}^\top) \left(\frac{1}{n}\vec{\X}^\top \vec{\X}\right)^{-1} \left(\frac{1}{n}\vec{\X}^\top\Bdelta\right),\\
& \left|\frac{\sqrt{n}}{\tau} (0, \bar{\X}^\top) \left(\frac{1}{n}\vec{\X}^\top \vec{\X}\right)^{-1} \left(\frac{1}{n}\vec{\X}^\top\Bdelta\right)\right| \leq \left\|\frac{1_n\X^\top }{n^{3/4}}\right\|_2\cdot \lambda_{\min}^{-1}\left(\frac{1}{n}\vec{\X}^\top \vec{\X}\right) \cdot\left\|\frac{\vec{\X}^\top \Bdelta}{n^{3/4}\tau}\right\|_2,
\end{split}
\end{equation*}
then we only need to prove the following asymptotic properties in order to finish the proof of Theorem \ref{th:theta_LS_asymptotic_p_grow}:
\begin{equation}\label{eq:Bdelta_asymp}
\frac{\sqrt{n} \bar \Bdelta}{\tau} \overset{d}{\to} N(0, 1),
\end{equation}
\begin{equation}\label{eq:vector_asymp}
\left\| \frac{1_n\X}{n^{3/4}}\right\|_2 \overset{d}{\to} 0, \quad \left\|\frac{\vec{\X}^\top \Bdelta}{n^{3/4}}\right\|_2 / \tau\overset{d}{\to} 0,  
\end{equation}
\begin{equation}\label{eq:matrix_asymp}
\begin{split}
& \text{For some uniform $t_1 > t_2>0$}, \\
& P\left(t_1 \geq \lambda_{\max}\left(\frac{1}{n} \vec{\X}^\top \vec{\X}\right) \geq \lambda_{\min}\left(\frac{1}{n}\vec{\X}^\top \vec{\X}\right) \geq t_2\right) \to 1.
\end{split}
\end{equation}
Here $\lambda_{\max}, \lambda_{\min}(\cdot)$ represent the largest and least eigenvalues of the given matrix. Next we will show \eqref{eq:Bdelta_asymp}, \eqref{eq:vector_asymp} and \eqref{eq:matrix_asymp} separately.
\begin{itemize}
	\item Based on the assumption of the theorem, $\frac{\delta_1}{\tau},\cdots, \frac{\delta_n}{\tau}$ are i.i.d. samples with mean 0, variance $1$ and bounded $(2+2\varepsilon)$-th moment, \eqref{eq:Bdelta_asymp} holds by Lyapunov's central limit theorem. 
	
	\item Since $X_1, \cdots, X_k$ are i.i.d. samples with mean 0 and covariance $I_p$, we can calculate that
	$$\E \left\|\frac{1_n \X^\top }{n^{3/4}}\right\|_2^2 = \frac{1}{n^{3/2}} \cdot n\E \|X\|_2^2 = \frac{p}{n^{1/2}} \to 0,\quad \text{as } n\to 0.$$
	Since $X_1\delta_1, \cdots, X_n\delta_n$ are i.i.d. samples with mean 0 and satisfying \eqref{ineq:th3_assump2}, we have 
	$$\E \left\|\frac{\vec{\X}^\top \Bdelta}{n^{3/4}}\right\|_2^2 = \frac{1}{n^{3/2}}\cdot n \E \|\vec X \delta\|_2^2 \leq \frac{M_3}{n^{1/2}} \E \|X\|_2^2 \cdot \E \delta^2 = \frac{p}{n^{1/2}}M_3\tau^2$$
	Thus, $\E \|\frac{\vec{\X}^\top \Bdelta}{n^{3/4}}\|_2^2/\tau^2 \to 0$ as $n\to \infty$. Thus, we have \eqref{eq:vector_asymp}. 
	
	\item For \eqref{eq:matrix_asymp}, since $\E X = 0, \Cov(X) = I_p$ and Assumption \eqref{ineq:th3_assump1} holds, \eqref{eq:matrix_asymp} is directly implied by Theorem 2 in \cite{yaskov2014lower}. 
\end{itemize}
\end{proof}

\section*{Acknowledgements}
The authors thank Arun Kuchibhotla for many helpful discussions. The authors also thank the Editor, the Associate Editor, and anonymous referees for many helpful comments, which greatly help improve the presentation of this paper.

\bibliographystyle{apa}
\bibliography{reference}

\newpage


\newpage

\setcounter{page}{1}
\setcounter{section}{0}
\pagestyle{plain}

\begin{center}
	{\LARGE Supplement to ``Semi-supervised Inference: General}
	
	\medskip	
	{\LARGE Theory and Estimation of Means"	
		}
	
	\bigskip\medskip
	{Anru Zhang$^{1}$, ~~ Lawrence D. Brown$^{2}$ ~ and~ T. Tony Cai$^{2}$}

\medskip
\end{center}
\footnotetext[1]{Department of Statistics, University of Wisconsin-Madison, Madison, WI. The research of Anru Zhang was supported in part by NSF Grant DMS-1811868.}
\footnotetext[2]{Department of Statistics, The Wharton School, University of Pennsylvania, Philadelphia, PA 19104. The research of Lawrence Brown was supported in part by NSF Grant DMS-10-07657.
	The research of Tony Cai was supported in part by NSF Grants DMS-1208982 and DMS-1403708, and NIH Grant R01 CA127334.} 

\medskip

	\indent In this supplement we provide additional proofs for the main results of the paper.

\section*{Additional Proofs for Ordinary Semi-supervised Inference Estimator $\hat{\theta}_{\rm SSLS}$}

\begin{proof}[Proof of Theorems \ref{th:asymp_imperfect} and \ref{th:asymp_imperfect_growing_p}] 
	We start with the proof of \eqref{eq:asymp2_imperfect}. From the proof of Theorems \ref{th:theta_LS_asymptotic} and \ref{th:theta_LS_asymptotic_p_grow}, we have proved that
	$$\frac{MSE}{\tau^2} \to 1. $$
	By the basic property of sample covariance and Proposition \ref{th:nvar(bar_Y)}, we also have
	$$\frac{n\hat{\sigma}_Y^2}{\Var(Y)} \overset{d}{\to} 1, \quad \Var(Y) = \tau^2 + \E \beta_{(2)}^\top\Sigma \beta_{(2)}. $$
	Therefore, under either the settings of Theorems \ref{th:asymp_imperfect} or \ref{th:asymp_imperfect_growing_p},
	\begin{equation}
	\begin{split}
	& \frac{\frac{m}{m+n}MSE + \frac{n}{m+n} \hat\sigma_Y^2}{\tau^2 + \frac{n}{n+m}\Var(\beta_{(2)}^\top X)}= \frac{\frac{m}{m+n}MSE + \frac{n}{m+n} \hat\sigma_Y^2}{\frac{m}{m+n}\tau^2 + \frac{n}{m+n}\Var(Y)} \overset{d}{\to} 1, \quad \text{as } n\to \infty,
	\end{split}
	\end{equation}
	which proves \eqref{eq:asymp2_imperfect}.
	
	The proof of \eqref{eq:asymp1_imperfect} is more complicated. In the rest of proof, again we use $C$ as constants does not depends on $n$ or $m$, whose exact value may vary in different scenarios. Again, since $\hat{\theta}_{\rm SSLS}$ is affine transformation invariant, without loss of generality we can assume that $\E X = 0$, $\E XX^\top  = I_p$. Thus, $Z = X$. Similarly as \eqref{eq:hat_theta_LS-theta}, the following decomposition for $\hat{\theta}_{\rm SSLS} - \theta$ holds,
	\begin{equation}\label{eq:decomposition_theta_SPLS}
	\begin{split}
	\hat{\theta}_{\rm SSLS} - \theta = & \hat{\vec \mu}^\top \left(\vec{\X}^\top \vec{\X}\right)^{-1}\vec{\X}^\top \Y - \theta = \hat {\vec{\mu}} \left(\vec{\X}^\top \vec{\X}\right)^{-1} \vec{\X}^\top (\vec{\X}\beta + \Bdelta)-\theta\\
	= & (\hat {\vec{\mu}}^\top \beta - \theta) + \hat{\vec{\mu}}(\vec{\X}^\top \vec{\X})^{-1}\vec{\X}^\top \Bdelta\\
	= & (\hat{\vec{\mu}} - \vec{\mu})^\top \beta + \left(\frac{1_n^\top }{n}\vec{\X}\left(\vec{\X}^\top \vec{\X}\right)^{-1}\vec{\X}^\top \Bdelta\right)\\
	& + \left(\hat{\vec{\mu}} - \frac{1_n}{n}\vec{\X}\right)^\top \left(\vec{\X}^\top \vec{\X}\right)^{-1}\vec{\X}^\top \Bdelta\\
	= & (\hat{\vec{\mu}} - \vec{\mu})^\top \beta + \left(\frac{1_n^\top \mathbb{P}_{\vec{\X}}}{n}\right)\Bdelta + \left(\hat{\vec{\mu}}^\top  - \frac{1_n^\top }{n}\vec{\X}\right)\left(\vec{\X}^\top \vec{\X}\right)^{-1}\vec{\X}^\top \Bdelta\\
	= & (\hat{\vec{\mu}} - \vec{\mu})^\top \beta + \bar{\Bdelta} - \left(0, \bar\X - \hat\mu\right)^\top \left(\vec{\X}^\top \vec{\X}\right)^{-1}\vec{\X}^\top \Bdelta.
	\end{split}
	\end{equation}
	In order to prove these two theorems, we only need to show the following two asymptotic equalities:
	\begin{equation}\label{eq:th4_asymp_1}
	\frac{\left(\hat{\vec{\mu}} - \vec{\mu}\right)^\top \beta + \bar{\Bdelta}}{\sqrt{\left(\frac{\tau^2}{n} + \frac{n}{n(n+m)}\beta_{(2)}^\top \E X_cX_c^\top \beta_{(2)}\right)}} \to N(0, 1),
	\end{equation}
	\begin{equation}\label{eq:th4_asymp_2}
	\frac{\left(0, \hat{\mu} - \bar{\X}\right)^\top \left(\vec{\X}^\top \vec{\X}\right)^{-1}\vec{\X}^\top \Bdelta}{\sqrt{\tau^2/n}} \overset{d}{\to} 0.
	\end{equation}
	We show them separately below under both settings that $p$ is fixed (Theorem \ref{th:asymp_imperfect}) and $p$ grows (Theorem \ref{th:asymp_imperfect_growing_p}). For convenience, we denote $T = \beta_{(2)}^\top \E X_cX_c^\top \beta_{(2)}$, $b_j = X_{j, c}^\top  \beta_{(2)}, j=1,\cdots, m+n$. Clearly $\E b_j^2 = T$. 
	
	We first show \eqref{eq:th4_asymp_1}. 
	The left hand side of \eqref{eq:th4_asymp_1} can be further written as
	\begin{equation*}
	\begin{split}
	& \left(\hat{\vec{\mu}} - \vec{\mu}\right)^\top \beta + \bar\Bdelta = (1-1)\beta_1 + (\hat{\mu} - \mu)\beta_{(2)} + \bar\Bdelta\\
	= & \sum_{i=1}^n\left(-\frac{m}{n(n+m)}(X_i-\mu)^\top \beta_{(2)}+\frac{1}{n}\delta_i\right) + \sum_{i=n+1}^{n+m} \frac{1}{n+m}(X_i-\mu)^\top \beta_{(2)}\\
	:=  & \sum_{j=1}^n A_j^{(n)} + \sum_{j=n+1}^{n+m} B_j^{(n)} :=  S_n
	\end{split}
	\end{equation*}
	Here $A_j^{(n)} = \frac{m}{n(n+m)} b_j + \frac{1}{n}\delta_j$, $ B_j^{(n)} = \frac{1}{n+m} b_j$. It is easy to calculate that $\E A_i^{(n)} = \E B_j^{(n)} = 0$, $\E A_i^{(n)2} = \frac{\tau^2}{n^2} + \frac{m^2}{n^2(n+m)^2}T$, $\E B_j^{(n)2} = \frac{1}{(n+m)^2}T$,
	\begin{equation}
	s_n^2 :=  \E S_n^2 = \sum_{i=1}^n \E A_i^{(n)2} + \sum_{j=n+1}^{n+m}\E B_j^{(n)2} =   \frac{\tau^2}{n} + \frac{mT}{n(n+m)}.
	\end{equation} 
	Next we analyze the asymptotic distribution for $S_n$
	separately under both settings when $p$ is fixed and $p$ is growing. Specifically, we use Lindeberg-Feller central limit theorem for the fixed $p$ case under second moment condition and Lyapunov central limit theorem for growing $p$ under $(2+\kappa)$-th moment condition. 
	
	\begin{itemize}
		\item Under the setting of Theorem \ref{th:asymp_imperfect}, i.e., when $p$ and the distribution $P(Y, X_1, \cdots, X_p)$ is fixed, we check the following Lindeberg-Feller condition:
		\begin{equation}\label{ineq:Lindeberg Condition}
		\begin{split}
		\forall \varepsilon>0,  \lim_{n\to\infty} \frac{1}{s_n^2} \Big[& \sum_{j=1}^n \E \left(A_j^{(n)2} I\{A_j^{(n)2} \geq \varepsilon s_n^2\}\right)\\
		&  + \sum_{j=n+1}^{m+n}\E \left(B_j^{(n)2} I\{B_j^{(n)2} \geq \varepsilon
		s_n^2\}\right)\Big] = 0.
		\end{split}
		\end{equation}
		Here $I\{\cdot\}$ is the indicator random variable for given event. Note that, for any $x_1, x_2\in \mathbb{R}$,
		\begin{equation}
		\begin{split}
		& (x_1 + x_2)^2 I\{(x_1+x_2)^2 \geq s_n^2\} \leq 4\max(x_1^2, x_2^2) I\{4\max(x_1^2, x_2^2) \geq s_n\}\\
		\leq &  4x_1^2 I\{x_1^2 \geq s_n^2/4\} + 4x_2^2 I\{x_2^2 \geq s_n^2/4\},
		\end{split}
		\end{equation} 
		we have
		\begin{equation*}
		\begin{split}
		& \E \left(A_j^{(n)2}I\{|A_j^{(n)}|^2 \geq \varepsilon s_n^2\}\right) \leq \E \left(A_j^{(n)2} I\{A_j^{(n)2} \geq \varepsilon\tau^2/n\}\right)\\
		\leq & \E \left(4 \frac{m^2}{n^2(n+m)^2} b_j^2 I\left\{\frac{m^2}{n^2(n+m)^2} b_j^2 \geq \frac{\varepsilon\tau^2}{4n}\right\}\right)\\
		& + \E \left(4 \frac{1}{n^2}\delta_j^2 I\left\{\frac{1}{n^2}\delta_j^2 \geq \frac{\varepsilon\tau^2}{4n}\right\}\right)\\
		\leq & \frac{4}{n^2} \left(\E \left(b_j^2I\left\{b_j^2\geq \frac{n\varepsilon \tau^2}{4}\right\}\right) + \E \left(\delta_j^2I\left\{\delta_j^2\geq \frac{n\varepsilon \tau^2}{4}\right\}\right)\right).
		\end{split}
		\end{equation*}	
		Similarly one can calculate that
		\begin{equation*}
		\begin{split}
		\E \left(B_j^{(n)2}I\{|B_j^{(n)}|^2 \geq \varepsilon s_n^2\}\right) \leq \frac{1}{n(n+m)} \E \left(b_j^2 I\left\{b_j^2 \geq \frac{\varepsilon n \tau^2}{4}\right\}\right).
		\end{split}
		\end{equation*}
		Therefore, 
		\begin{equation*}
		\begin{split}
		& \frac{1}{s_n^2} \left[\sum_{j=1}^n \E \left(A_j^{(n)2} I\{A_j^{(n)2} \geq \varepsilon s_n^2\}\right) + \sum_{j=m+1}^{m+n}\E \left(B_j^{(n)2} I\{B_j^{(n)2} \geq \varepsilon s_n^2 \}\right)\right] \\
		\leq & \frac{n}{\tau^2} \cdot\left[\frac{5}{n}\E \left(b_j^2I\left\{b_j^2\geq \frac{n\varepsilon \tau^2}{4}\right\}\right) + \frac{4}{n}\E \left(\delta_j^2I\left\{\delta_j^2\geq \frac{n\varepsilon \tau^2}{4}\right\}\right) \right]\\
		\leq & \frac{5}{\tau^2}\E  \left(b^2 I\left\{b^2 \geq n\varepsilon\tau^2/4\right\}\right) + \frac{4}{\tau^2} \E \left(\delta^2 I\left\{\delta^2 \geq n\varepsilon\tau^2/4\right\}\right) \to 0.
		\end{split}
		\end{equation*}
		By Lindeberg-Feller CLT, we know $S_n/s_n\to N(0, 1)$, which implies \eqref{eq:th4_asymp_1}.
		\item Under the setting of Theorem \ref{th:asymp_imperfect_growing_p}, i.e., when the distribution $P$ is not fixed and $p$ is growing, the proof as we also have $(2+2\kappa)$-moment conditions. In this case, Lyapunov's condition for central limit theorem will be used as the main tool. One can check that
		\begin{equation*}
		\begin{split}
		\E |A_i|^{2+2\kappa} \leq & C\left(\E \left(\frac{m}{n(n+m)}b_i\right)^{2+2\kappa} + \E \left(\frac{\delta_i}{n}\right)^{2+2\kappa}\right)\\
		\overset{\eqref{ineq:th3_assump0}}{\leq} & C\left(\frac{m}{n(n+m)}\right)^{2+2\kappa} T^{1+\kappa} + C\left(\frac{\tau}{n}\right)^{2+2\kappa},
		\end{split}
		\end{equation*}
		\begin{equation*}
		\begin{split}
		\E |B_j|^{2+2\kappa} = \frac{1}{(n+m)^{2+2\kappa}} \E \left((X_i-\mu)^\top \beta_{(2)}\right)^{2+2\kappa} \overset{\eqref{ineq:th3_assump1}}{\leq}\frac{1}{(n+m)^{2+2\kappa}}T^{1+\kappa}.
		\end{split}
		\end{equation*}
		Thus,
		\begin{equation}\label{ineq:EA_i + EB_i^2+2varepsilon}
		\begin{split}
		& \sum_{i=1}^n\E |A_i^{(n)}|^{2+2\kappa} + \sum_{j=n+1}^{n+m}\E |B_j^{(n)}|^{2+2\kappa}\\
		\leq & C\left\{\left(\frac{m}{n(n+m)}\right)^{2+2\kappa}n + \left(\frac{1}{n+m}\right)^{2+2\kappa}m\right\} T^{1+\kappa} + C\frac{\tau^{2+2\kappa}}{n^{1+2\kappa}}\\
		\leq & C \frac{m(m^{1+2\kappa} + n^{1+2\kappa})}{n^{1+2\kappa}(n+m)^{2+2\kappa}}T^{1+\kappa} + C\frac{\tau^{2+2\kappa}}{n^{1+2\kappa}}\\
		\leq & C\left(\frac{m}{n^{1+2\kappa}(n+m)}T^{1+\kappa} + \frac{\tau^{2+2\kappa}}{n^{1+2\kappa}}\right)\\
		\end{split}
		\end{equation}
		On the other hand,
		\begin{equation}\label{ineq:s_n^2+2varepsilon}
		s_n^{2+2\kappa} = \left(\frac{\tau^2}{n} + \frac{mT}{n(n+m)}\right)^{1+\kappa} \geq \frac{\tau^{2+2\kappa}}{n^{1+\kappa}} + \frac{m^{1+\kappa}T^{1+\kappa}}{n^{1+\kappa}(n+m)^{1+\kappa}}. 
		\end{equation}
		Since as $n, m\to \infty$,
		$$\frac{\frac{m}{n^{1+2\kappa}(n+m)}T^{1+\kappa}}{\frac{m^{1+\kappa}T^{1+\kappa}}{n^{1+\kappa}(n+m)^{1+\kappa}}} = \left(\frac{n+m}{nm}\right)^{\kappa} \to 0, \quad \frac{\tau^{2+2\kappa}/(n^{1+2\kappa})}{\tau^{2+2\kappa}/((n^{1+\kappa})} = \frac{1}{n^{\kappa}} \to 0, $$
		combining \eqref{ineq:EA_i + EB_i^2+2varepsilon} and \eqref{ineq:s_n^2+2varepsilon}, we have
		$$\lim_{n\to\infty}\frac{1}{s_n^{2+2\kappa}} \left(\sum_{i=1}^n \E |A_i^{(n)}|^{2+2\kappa} + \sum_{j=n+1}^{n+m} \E |B_j^{(n)}|^{2+2\kappa}\right). $$
		By Lyapunov's central limit theorem, we know 
		$$\left(\sum_{i=1}^n A_i^{(n)} +\sum_{j=n+1}^{n+m}B_j^{(n)}\right)\Bigg /\sqrt{\frac{m}{n(n+m)}T + \frac{\tau^2}{n}} \to N(0, 1),$$ 
		which implies \eqref{eq:th4_asymp_1}.
	\end{itemize}
	Next, we show \eqref{eq:th4_asymp_2} under both settings of fixed $p$ and growing $p$. We can calculate that
	\begin{equation*}
	\begin{split}
	& \frac{\left|(0, \hat{\mu} - \bar{\X})^\top \left(\vec{\X}^\top \vec{\X}\right)^{-1}\vec{\X}^\top \Bdelta\right|}{\sqrt{\tau^2/n}} \leq \frac{\|\hat{\mu} - \bar{\X}\|_2 \cdot\lambda_{\min}^{-1}(\vec{\X}^\top\vec{\X})\cdot \|\vec{\X}^\top\Bdelta\|_2}{\sqrt{\tau^2/n}}\\
	\leq & n^{1/4}\|\hat{\mu} - \bar{\X}\|_2\cdot \lambda_{\min}^{-1}\left(\frac{1}{n}\vec{\X}^\top\vec{\X}\right) \cdot \left\|\frac{\vec{\X}^\top \Bdelta}{n^{3/4}\tau}\right\|_2.
	\end{split}
	\end{equation*} 
	\begin{itemize}
		\item We first consider the simpler case where $P$ is fixed, i.e., the setting in Theorem \ref{th:asymp_imperfect}. The proof is similar to the one of Theorem \ref{th:theta_LS_asymptotic}. Note that $\E X_i = 0$, $\E \vec{X} \vec{X}^\top = I_{p+1}$, $\E \vec{X}\delta = 0$, thus by law of large number,
		\begin{equation}\label{eq:asymp_fixed_p_theta_LSPS}
		\begin{split}
		\frac{1}{n} \vec{\X}^\top \Bdelta \overset{d}{\to} 0, & \quad \frac{1}{n} \vec{\X}^\top \vec{\X} \overset{d} {\to} I_p,\\
		\hat{\vec{\mu}} = \frac{1}{n+m} \vec{X}_k \to (1, 0, \cdots, 0)^\top , &\quad \frac{\vec{\X} 1_n}{n} = \frac{1}{n} \sum_{k=1}^n \vec{X}_k \to (1, 0,\cdots, 0)^\top . 
		\end{split}
		\end{equation}
		These facts together yields \eqref{eq:th4_asymp_2}.
		
		\item Now we move to the case that $p$ grows, i.e., the setting in Theorem \ref{th:asymp_imperfect_growing_p}. Similarly as the proof of Theorem \ref{th:theta_LS_asymptotic_p_grow}, we have
		\begin{equation*}\label{ineq:asymp1_theta_LSPS}
		\left\|\frac{\sum_{i=1}^n X_i}{n^{3/4}}\right\|_2 \overset{d}{\to} 0, \quad \left\|\frac{\sum_{i=1}^{n}\vec X_i \delta_i}{n^{3/4}}\right\|_2/\tau \overset{d}{\to} 0,\quad \left\|\frac{n^{1/4}\sum_{i=1}^{n+m}X_i}{(m+n)}\right\|_2 \overset{d}{\to} 0,
		\end{equation*}
		\begin{equation*}\label{ineq:asymp2_theta_LSPS}
		\begin{split}
		& \exists t_1\geq t_2>0, \quad \text{ such that }\\
		& P\left(t_1\geq \lambda_{\max}\left(\frac{1}{n}\sum_{i=1}^n \vec X_i\vec X_i^\top \right) \geq \lambda_{\min}\left(\frac{1}{n}\sum_{i=1}^n \vec X_i\vec X_i^\top \right) \geq t_2\right)\to 1. 
		\end{split}
		\end{equation*}
		Similarly these imply \eqref{eq:th4_asymp_2}.	
	\end{itemize}

	To sum up, we have finished the proof of Theorems \ref{th:asymp_imperfect} and \ref{th:asymp_imperfect_growing_p}.
\end{proof}

\section*{Additional Proofs for the Random Design Model}


\begin{proof}[Proof of Proposition \ref{th:nvar(bar_Y)}] First, we introduce some basic facts about the regression slope and total deviation are summarized in the following lemma.
	\begin{Lemma}\label{lm:basic_properties_delta}
		Let $(Y, X)\sim P$ have finite second moment, and let the matrix $\vec{\varXi}$ be non-singular. Then 
		\begin{equation*}
		\beta = \vec{\varXi}^{-1} \left(\E \vec{X} Y\right),\quad	\E \delta = 0, \quad \E \delta X = 0, \quad \theta = \vec{\mu}^\top \beta.
		\end{equation*}
	\end{Lemma}
	
	\begin{proof}[Proof of Lemma \ref{lm:basic_properties_delta}]
		Since
		\begin{equation*}
		\begin{split}
		& \E \left(Y - \vec{X}^\top \beta\right)^2 = \E Y^2 + \beta^\top  \left(\E \vec{X} \vec{X}^\top \right) \beta - 2\beta^\top  \E  \left(\vec{X} Y \right)\\
		= & \E Y^2 + \left(\beta - (\E \vec X\vec X^\top)^{-1} \E (\vec X Y)\right)^\top \left(\E \vec X\vec X^\top \right) \left(\beta - (\E \vec X\vec X^\top)^{-1} \E (\vec X Y)\right)\\
		& - \E (\vec X Y)^\top  \left(\E \vec X\vec X^\top \right)^{-1} \E (\vec X Y),
		\end{split}
		\end{equation*}
		we know $\beta = \argmin_{\gamma} \E (Y - \vec X^\top \gamma)^2 = (\E \vec X\vec X^\top )^{-1} \E (\vec X Y)$. Besides,
		\begin{equation*}
		\begin{split}
		\E (\vec X \delta) = \E \vec{X}Y - \E \vec X\vec X^\top  \beta = \E \vec{X} Y - \E \vec X \vec X^\top  \cdot \left(\E\vec X \vec X^\top \right)^{-1} \E (\vec X Y) = 0.
		\end{split}
		\end{equation*}
		Then $\E\delta = 0, \E X\delta = 0$ have been proved since $\vec X = (1, X^\top )^\top $. Finally,
		\begin{equation*}
		\begin{split}
		\vec{\mu}^\top \beta = & \E \vec{X}^\top \left(\E\vec{X}\vec{X}^\top\right)^{-1} \E\vec{X}Y = (1, \mu^\top)\cdot \begin{bmatrix}
		1 & \mu^\top\\
		\mu & \Cov(X) + \mu\mu^\top
		\end{bmatrix}^{-1} \cdot \begin{pmatrix}
		EY\\
		EX Y
		\end{pmatrix}\\
		= & (1, \overbrace{0, \ldots, 0}^p) \begin{pmatrix}
		EY\\
		EXY
		\end{pmatrix} = \E Y  =\theta,
		\end{split}
		\end{equation*} 
		which has finished the proof of this lemma. 
	\end{proof}
	
	Then we consider the proof for Lemma \ref{lm:basic_properties_delta}. $\bar \Y$ is the sample mean, which is clearly an unbiased estimator for the population mean $\theta$. In addition, since $\{Y_i\}_{i=1}^n$'s are i.i.d. samples, it can be calculated that
	\begin{equation}
	\begin{split}
	n\Var(\bar\Y) = & \Var(Y_i) = \Var(\delta_i) + \Var(\vec{X}_i \beta) + 2\Cov(\delta_i, \vec{X}_i\beta)\\
	\overset{\text{Lemma \ref{lm:basic_properties_delta}}}{=} & \tau^2 + \beta_{(2)}^\top \E (X - \mu)(X - \mu)^\top  \beta_{(2)}.
	\end{split}
	\end{equation}
\end{proof}

\section*{Proofs for the analysis of $\ell_2$-risk}

\begin{proof}[Proof of Theorems \ref{th:MSE_delta}]
	The idea of the proof for Theorem \ref{th:MSE_delta} is to first introduce a ``good event" $Q$ such that $P(Q^c)$ is exponentially small; then prove that $\E \left[n\left(\hat{\theta}_{\rm LS} - \theta\right)^21_Q\right]$ has upper bound as \eqref{eq:risk_delta_LS} and \eqref{eq:risk_delta_LS_s_n}. For convenience, for any subset $\Omega \subseteq\{1,\ldots, n\}$, we introduce the following notations
	\begin{equation}
	\vec{\vvarXi} = \frac{1}{n} \sum_{k=1}^n \vec Z_k \vec Z_k^\top , \quad \vec{\vvarXi}_{-\Omega} = \frac{1}{n} \sum_{k=1, k\notin \Omega}^n \vec Z_k Z_k^\top.
	\end{equation} 
	Also, we note $\poly(n, p)$ for some polynomial of $n$ and $p$. We also introduce the following lemmas. The proofs are postponed to the Supplement.
	
	\begin{Lemma}\label{lm:cov_concentration} Suppose $\vec{\Z} = (\vec{Z}_1, \cdots, \vec{Z}_n)^\top $ satisfies Assumption 2 \eqref{ineq:assump2_l2_risk} or Assumption 2' \eqref{ineq:assumption3_l2_risk}.
		\begin{itemize}
			\item {\bf(Theorem 5.39 in \cite{Vershynin_random_matrix})} We have the following concentration inequality,
			\begin{equation} \label{ineq:vershynin}
			P\left(\left\|\frac{1}{n}\sum_{k=1}^n \vec{Z}_k\vec{Z}_k^\top  - \E \vec{Z}_k\vec{Z}_k^\top \right\| > C\sqrt{\frac{p}{n}} + t\right) \leq 2\exp(-cnt^2). 
			\end{equation}
			Here $C, c$ are constants only depending on $M_5$ in Assumption \eqref{ineq:assump2_l2_risk} or $M_6$ in Assumption \eqref{ineq:assumption3_l2_risk}.
			\item For all $q\geq 2$, the following moment condition holds for some constant $C_q$ that only depends on $q$ under either Assumption 2 \eqref{ineq:assump2_l2_risk} or Assumption 2' \eqref{ineq:assumption3_l2_risk},
			\begin{equation}\label{ineq:Z_q-th moment}
			\E \left\|\sum_{k=1}^n Z_k\right\|_2^q \leq C_q\left(pn\right)^{q/2}.
			\end{equation}
			\item The following moment condition holds for $\sum_{k=1}^n\vec{Z}_k\delta_k$ and $2\leq q<4$:
			\begin{equation}\label{ineq:Zdelta_q-th moment}
			\E \left\|\sum_{k=1}^n \vec{Z}_k\delta_k\right\|_2^q \leq C_q (pn)^{q/2}
			\end{equation}
			under either Assumption 1+2 (\eqref{ineq:assump1_l2_risk}, \eqref{ineq:assump2_l2_risk}) or 1+2' (\eqref{ineq:assump1_l2_risk}, \eqref{ineq:assumption3_l2_risk}).
		\end{itemize}
	\end{Lemma}
	
	\begin{Lemma}\label{lm:inverse_expansion}
		Suppose $A, B$ are two squared matrices, $A, A+B$ are both invertible. Then for all $q\geq 0$, one has the following expansion for $(A+B)^{-1}$,
		\begin{equation}\label{eq:inverse_expasion}
		(A+B)^{-1} = \sum_{k=0}^{q-1} \left(-A^{-1}B\right)^{k}A^{-1} + \left(-A^{-1}B\right)^{q}(A+B)^{-1}. 
		\end{equation}
	\end{Lemma}
	
	For the proof of Theorem \ref{th:MSE_delta}, we first consider the probability that $\hat{\theta}_{\rm LS}\neq \hat{\theta}_{\rm LS}^1$. Note $\bar \mu = \frac{\max(\Y) + \min(\Y)}{2}$, then we have
	\begin{equation*}
	\begin{split}
	& P\left(\hat{\theta}_{\rm LS} \neq \hat{\theta}_{\rm LS}^1\right) = P\left(|\hat{\theta}_{\rm LS} - \bar{\mu}| > (n+\frac{1}{2})(\max(\Y) - \min(\Y))\right)\\
	\leq & P\bigg(\left\|\left(\frac{1}{n} \vec{\X}^\top  \vec{\X}\right)^{-1}\right\|\cdot \sqrt{\left\|\frac{1}{n} \vec{\X}^\top \vec{\X}\right\|} \cdot \frac{\max(\Y) - \min(\Y)}{2}\\
	& \quad > (n+\frac{1}{2}) (\max(\Y) - \min(\Y))\bigg)\\
	\overset{\eqref{ineq:vershynin}}{\leq} & \exp(-cn) ,\quad \text{for large } n.
	\end{split}
	\end{equation*}
	Set the event $Q$ as
	\begin{equation}\label{eq:Q_event}
	\begin{split}
	Q = \left\{\hat{\theta}_{\rm LS} = \hat{\theta}_{\rm LS}^1, \quad \max_{1\leq i,j,k\leq n}\left\{\left\|\vec{\vvarXi} - I_{p+1}\right\|, \left\|\vec{\vvarXi}_{-\{i,j,k\}} - I_{p+1}\right\|\right\} \leq \frac{C_1}{n^{1/4}}\right\}
	\end{split}
	\end{equation}
	for some large constant $C_1>0$. Based on Lemma \ref{lm:cov_concentration} and the fact that $\sqrt{p/n} = o(n^{-1/4})$, we have
	\begin{equation}\label{ineq:P(Q)}
	\begin{split}
	P\left(Q^c\right) \leq & P\left(\left\|\vec \Sigma - I_{p+1}\right\| > C_1n^{-1/4}\right)+\sum_{i, j, k} P\left(\left\|\vec \Sigma_{-\{i,j,k\}} - I_{p+1}\right\| > C_1n^{-1/4}\right)\\
	& + P\left(\hat{\theta}_{\rm LS}\neq \hat{\theta}_{\rm LS}^1
	\right)\\
	\leq & Cn^3\cdot \exp(-cn^{1/2}) \quad \text{ for large } n.
	\end{split}
	\end{equation}
	Recall the composition of $\hat\theta_{\rm LS} - \theta$ in \eqref{eq:hat_theta_LS-theta}, thus,
	\begin{equation}\label{eq:E1_Q hat theta-theta}
	\begin{split}
	& \E \left(1_Q(\hat{\theta}_{\rm LS} - \theta)^2\right)\\
	= & \E 1_Q \Bdelta^2 + \E \left[1_Q\left((0, \frac{1_n}{n}\Z^\top )\left(\frac{1}{n}\vec{\Z}^\top \vec{\Z}\right)^{-1}(\frac{1}{n}\vec{\Z}^\top \Bdelta)\right)^2\right]\\
	& - 2\E \left[1_Q\bar{\Bdelta}(0, \frac{1_n}{n}\Z^\top )\left(\frac{1}{n}\vec{\Z}^\top \vec{\Z}\right)^{-1}(\frac{1}{n}\vec{\Z}^\top \Bdelta)\right]\\
	= & \E 1_Q\bar \Bdelta^2 + \E \left[1_Q\left((0, \frac{1_n}{n}\Z^\top )\left(\frac{1}{n}\vec{\Z}^\top \vec{\Z}\right)^{-1}(\frac{1}{n}\vec{\Z}^\top \Bdelta)\right)^2\right]\\
	& - 2\sum_{k, l, m=1}^n \frac{1}{n^3} \E \left[1_Q \delta_k(0, Z_l^\top )\vec{\vvarXi}^{-1} \vec{Z}_m\delta_m \right]\\
	= & \E 1_Q\bar \Bdelta^2 + \E \left[1_Q\left((0, \frac{1_n}{n}\Z^\top )\left(\frac{1}{n}\vec{\Z}^\top \vec{\Z}\right)^{-1}(\frac{1}{n}\vec{\Z}^\top \Bdelta)\right)^2\right]\\
	& - \frac{2(n-1)}{n^2}\E \left[1_Q \delta_1(0, Z_1^\top )\vec{\vvarXi}^{-1} \vec{Z}_2\delta_2 \right]\\
	& - \frac{2(n-1)}{n^2}\E \left[1_Q \delta_1^2(0, Z_2^\top )\vec{\vvarXi}^{-1} \vec{Z}_1\right] - \frac{2(n-1)}{n^2}\E \left[1_Q \delta_1(0, Z_2^\top )\vec{\vvarXi}^{-1} \vec{Z}_2\delta_2 \right]\\
	& - \frac{2}{n^2} \E \left[1_Q \delta_1^2(0, Z_1^\top )\vec{\vvarXi}^{-1} \vec{Z}_1 \right] - \frac{2(n-1)(n-2)}{n^2}\E \left[1_Q \delta_1(0, Z_2^\top )\vec{\vvarXi}^{-1} \vec{Z}_2\delta_3\right].
	\end{split}
	\end{equation}
	The analyses for each of the seven terms in \eqref{eq:E1_Q hat theta-theta} are relatively complicated, which we postpone to Lemma \ref{lm:five terms} in the Supplement. Based on \eqref{eq:E1_Q hat theta-theta} and Lemma \ref{lm:five terms}, one has
	\begin{equation*}
	\begin{split}
	& \E 1_Q\left(\hat \theta_{\rm LS} - \theta\right)^2 = \frac{1}{n}\tau^2 + O(\poly(p, n)\exp(-cn^{1/2}))\\
	& + \frac{1}{n^2}\bigg(2(\E \delta^2 Z)^\top \E (ZZ^\top Z) + \left(\tr(\E Z\delta Z^\top )\right)^2\\
	&\quad\quad + 3\|\E Z\delta Z^\top \|_F^2 - \E \tr(Z\delta^2Z^\top ) + 2\tau^2\bigg).
	\end{split}
	\end{equation*}
	Besides,
	\begin{equation}\label{ineq:E1_Q^c_theta}
	\begin{split}
	& \E 1_{Q^c}\left(\hat{\theta}_{\rm LS}^1 - \theta\right)^2 \leq \E 1_{Q^c}(2n\|Y\|_\infty + \E Y)^2\\
	\leq & \poly(n) (\E Y^{2+2\varepsilon})^{\frac{1}{1+\varepsilon}} \cdot (\E  1_{Q^c})^{\frac{\varepsilon}{1+\varepsilon}} \leq \poly(n)\exp(-n^{1/2}) \leq \poly(n) \exp(-n^{1/2}).
	\end{split}
	\end{equation}
	Our final step gets back to the $\ell_2$-risk of $\hat\theta_{\rm LS}^1$:
	\begin{equation*}
	\begin{split}
	& \E \left(\hat{\theta}^1_{\rm LS} - \theta\right)^2 = \E 1_Q \left(\hat{\theta}_{\rm LS}^1 - \theta\right)^2 + \E 1_{Q^c} \left(\hat{\theta}_{\rm LS}^1 - \theta\right)^2\\
	= & \frac{1}{n}\tau^2 + O(\poly(p, n)\exp(-cn^{1/2}))\\
	& + \frac{1}{n^2}\bigg(2(\E \delta^2 Z)^\top \E (ZZ^\top Z) + \left(\tr(\E Z\delta Z^\top )\right)^2\\
	&\quad\quad\quad + 3\|\E Z\delta Z^\top \|_F^2 - \E \tr(Z\delta^2Z^\top ) + 2\tau^2\bigg).
	\end{split}
	\end{equation*}
	In fact, given $Z = \Sigma^{-1/2}X$, we have
	\begin{equation*}
	\begin{split}
	& (\E\delta^2 Z)^\top \E(ZZ^\top Z) = (\E \Sigma^{-1/2}X_c\delta^2 )^\top \E(\Sigma^{-1/2}X_cZ^\top\Sigma^{-1} X_c)\\
	= & \E\left(\delta^2 X_c\right)^\top \cdot \E\left(\Sigma^{-1} X_cX_c^\top \Sigma^{-1} X_c\right),
	\end{split}
	\end{equation*}
	\begin{equation*}
	\tr\left(\E Z \delta Z^\top\right) = \tr\left(\E \Sigma^{-1/2} X\delta X^\top \Sigma^{-1/2}\right) = \tr\left(\Sigma^{-1}\Sigma_{\delta 1}\right),
	\end{equation*}
	\begin{equation*}
	\E\tr \left( Z \delta^2 Z^\top\right) = \E \tr\left(\Sigma^{-1/2} X\delta^2 X^\top \Sigma^{-1/2}\right) = \tr\left(\Sigma^{-1}\Sigma_{\delta 2}\right).
	\end{equation*}
	Therefore, we have finished the proof of Theorem \ref{th:MSE_delta}. 
\end{proof}


\begin{proof}[Proof of Theorem \ref{th:MSE_SPLS}] Similarly to the previous proofs, we can transform $X$, $Y$ and assume $\mu = 0, \Cov(X)=I_p, X = Z$ without loss of generality. We start by introducing the following notations and decomposition in \eqref{eq:decomposition_theta_SPLS}:
\begin{equation*}
\X = [X_1 ~ \cdots ~ X_{n}]^\top,\quad 
\X_{\rm full} = [X_1 ~ \cdots ~ X_{n+m}]^\top, \quad \X_{\rm add} = [X_{n+1} ~ \cdots ~ X_{n+m}]^\top,
\end{equation*}
\begin{equation*}
\bar\X = \frac{1}{n}\sum_{k=1}^n X_k,\quad 
\bar\X_{\rm full} = \frac{1}{n+m}\sum_{k=1}^{n+m} X_k, \quad \bar{\X}_{\rm add} = \frac{1}{m}\sum_{k=n+1}^{n+m} X_k.
\end{equation*}
\begin{equation*}
\begin{split}
& \hat{\theta}_{\rm SSLS} - \theta = (\hat{\vec{\mu}} - \vec{\mu})^\top \beta + \bar{\Bdelta} + (\hat{\vec{\mu}} - \bar{\X}^\top )(\vec{\X}^\top \vec{\X})^{-1} \vec{\X}\Bdelta\\
= & \bar{\X}_{\rm full}^\top \beta_{(2)} + \bar\Bdelta + (0 ~-\bar\X + \bar{\X}_{\rm full})\left(\vec{\X}^\top \vec{\X}\right)^{-1} \vec{\X}\Bdelta.
\end{split}
\end{equation*}
Again we note 
$$\vec{\vvarXi} = \frac{1}{n}\sum_{k=1}^n \vec{X}_k\vec{X}_k^\top , \quad \vec{\vvarXi}_{-\Omega} = \frac{1}{n}\sum_{k=1, k\notin \Omega}^n \vec{X}_k\vec{X}_k^\top , \text{if  } \Omega \subseteq \{1,\cdots, n\}, $$
and define the ``good" event that
\begin{equation*}
\begin{split}
Q = \bigg\{& \hat{\theta}_{\rm SSLS} = \hat{\theta}_{\rm SSLS}^1,\\ 
& \max\left\{\left\|\vec{\vvarXi} - I_{p+1}\right\|, \left\|\vec{\vvarXi}_{-\{i,j,k\}} - I_{p+1}\right\| \forall 1\leq i, j, k \leq n\right\} \leq C_1n^{-1/4}\bigg\}.
\end{split}
\end{equation*}
Then,
\begin{equation}\label{eq:decomposition_theta_SPLS_l2}
\begin{split}
& \E (\hat{\theta}_{\rm SSLS}^1 - \theta)^2 = \E 1_Q (\hat{\theta}_{\rm SSLS}^1 - \theta)^2 + \E 1_{Q^c} (\hat{\theta}_{\rm SSLS}^1 - \theta)^2\\
= & \E \left(\bar{\X}^\top_{\rm full}\beta_{(2)} + \bar{\Bdelta}\right)^21_Q + \E \left((0 ~ -\bar{\X} + \bar{\X}_{\rm full}) \left(\vec{\X}^\top \vec{\X}\right)^{-1} \vec{\X}\Bdelta\right)^21_Q\\
& + 2 \E \left(\bar{\X}^\top _{\rm full} \beta_{(2)} + \bar{\Bdelta}\right)\left((0, -\bar{\X} + \bar{\X}_{\rm full}) \left(\vec{\X}^\top \vec{\X}\right)^{-1}\vec{\X}\Bdelta\right)1_Q + \E 1_{Q^c} (\hat{\theta}_{\rm SSLS}^1 - \theta)^2.
\end{split}
\end{equation}
In the analysis below, we analyze the four terms in \eqref{eq:decomposition_theta_SPLS_l2} separately. 

\begin{itemize}
	\item First of all, since $\delta$ and $\vec{\X}$ are with mean zero and uncorrelated,
	\begin{equation*}
	\begin{split}
	& \E \left(\bar{\X}^\top _{\rm full}\beta_{(2)} + \bar\Bdelta\right)^2 = \Var(\bar{\X}_{\rm full}^\top \beta_{(2)}) + \Var(\bar\Bdelta)\\
	= & \frac{\tau^2}{n} + \frac{\beta_{(2)}^\top \E (X-\mu)(X-\mu)^\top \beta_{(2)}}{m+n}. 
	\end{split}
	\end{equation*}
	Besides,
	\begin{equation*}
	\begin{split}
	& \E \left(\bar{\X}_{\rm full}^\top \beta_{(2)} + \bar\Bdelta\right)^21_{Q^c} \leq  \left(\E \left(\bar{\X}_{\rm full}^\top \beta_{(2)} + \bar{\Bdelta}\right)^4\right)^{1/2} \left(P(Q^c)\right)^{1/2}\\
	= & \poly(n)\left(\E (\bar{\Y} - \theta)^4\right)^{1/2} \exp(-cn^{1/2}) \leq \poly(n)\exp(-cn^{1/2}).
	\end{split}
	\end{equation*}
	Thus,
	\begin{equation}\label{eq:theta_SPLS_l2_term1}
	\begin{split}
	& \E \left(\bar{\X}^\top  \beta_{(2)} + \bar\Bdelta\right)^21_Q = \E \left(\bar{\X}^\top \beta_{(2)} + \bar\Bdelta\right)^2 - \E \left(\bar{\X}^\top \beta_{(2)} + \bar\Bdelta\right)^21_{Q^c}\\
	= & \frac{\tau^2}{n} + \frac{\beta_{(2)}^\top \E (X-\mu)(X-\mu)^\top \beta_{(2)}}{m+n} + O\left(\poly(n)\exp(-cn^{1/2})\right).
	\end{split}
	\end{equation}

	\item Secondly, 
	\begin{equation}\label{eq:theta_SPLS_l2_term2}
	\begin{split}
	& \E \left((0 ~ -\bar{\X} + \bar{\X}_{\rm full}) \left(\vec{\X}^\top \vec{\X}\right)^{-1} \vec{\X}\Bdelta\right)^21_Q\\
	\leq & 2\E \left(\|\bar{\X}\|_2^2 + \|\bar{\X}_{\rm full}\|_2^2\right)\cdot \left(\left\|\left(\vec{\X}^\top \vec{\X}\right)^{-1}\right\|\right)^21_Q \cdot \left\|\vec{\X}\Bdelta\right\|^2_2\\
	\leq & C\left(\E \|\bar{\X}\|_2^4 + \E \|\bar{\X}_{\rm full}\|_2^4 \right) \cdot \left(\frac{C}{n}\right)^2 \cdot \left(\E \|\vec{\X}\Bdelta\|_2^4\right)^{1/2}\\
	\overset{\text{Lemma \ref{lm:five terms}}}{\leq} & C\left(\frac{p^2}{n^2}\right).
	\end{split}
	\end{equation}
	
	\item The analysis of the third term in \eqref{eq:decomposition_theta_SPLS_l2} is more complicated. We first decompose it as
	\begin{equation}\label{ineq:separate_2_theta_SPLS}
	\begin{split}
	& \E \left(\bar{\X}^\top _{\rm full} \beta_{(2)} + \bar{\Bdelta}\right)\left((0, -\bar{\X} + \bar{\X}_{\rm full})^\top  \left(\vec{\X}^\top \vec{\X}\right)^{-1}\vec{\X}\Bdelta\right)1_Q\\
	= & - \E \left(\bar{\X}^\top _{\rm full}\beta_{(2)} + \bar\Bdelta\right)\left(0, \frac{m}{n(n+m)}1^\top _n\X\right)\left(\vec{\X}^\top \vec{\X}\right)^{-1}\vec{\X}\Bdelta 1_Q \\
	& + \E \left(\bar{\X}^\top _{\rm full}\beta_{(2)} + \bar\Bdelta\right)\left(0, \frac{1}{n+m}1^\top _m\X_{\rm add}\right)\left(\vec{\X}^\top \vec{\X}\right)^{-1}\vec{\X}\Bdelta 1_Q\\
	= & - \frac{m}{n(n+m)}\sum_{i,j,k=1}^n (\frac{1}{n+m}X_i^\top \beta_{(2)} + \frac{1}{n}\delta_i) \left(0, X_j\right)^\top  \left(\vec{\X}^\top  \vec{\X}\right)\vec{X}_k\delta_k 1_Q\\
	& - \frac{m}{n(n+m)^2}\sum_{i=n+1}^{n+m}\sum_{j,k=1}^n (X_i^\top \beta_{(2)}) \left(0, X_j\right)^\top  \left(\vec{\X}^\top  \vec{\X}\right)\vec{X}_k\delta_k 1_Q\\
	& + \frac{1}{n+m}\sum_{i,k=1}^n\sum_{j=n+1}^{n+m} (\frac{1}{m+n}X_i^\top \beta_{(2)} + \frac{1}{n}\delta_i) \left(0, X_j\right)^\top  \left(\vec{\X}^\top  \vec{\X}\right)\vec{X}_k\delta_k 1_Q\\
	& + \frac{1}{(n+m)^2}\sum_{i, j=n+1}^{n+m}\sum_{k=1}^n X_i^\top \beta_{(2)} \left(0, X_j\right)^\top  \left(\vec{\X}^\top  \vec{\X}\right)\vec{X}_k\delta_k 1_Q.
	\end{split}
	\end{equation}
	The evaluation of the four terms above are provided separately in Lemma \ref{lm:five terms}. Therefore,
	\begin{equation}
	\E \left(\bar{\X}^\top  \beta_{(2)} + \bar{\Bdelta}\right)\left((0, -\bar{\X} + \bar{\X}_{\rm full}) \left(\vec{\X}^\top \vec{\X}\right)^{-1}\vec{\X}\Bdelta\right)1_Q \leq C \frac{p^2}{n^2}.
	\end{equation}
	\item Similarly to \eqref{ineq:E1_Q^c_theta} in Theorem \ref{th:theta_LS_asymptotic}, one can show
	$$\E 1_{Q^c} (\hat{\theta}_{\rm SSLS} - \theta)^2 = \poly (n) \exp(-n^{1/2}). $$
\end{itemize}
Combining \eqref{eq:decomposition_theta_SPLS_l2} and the separate analyses above, we have finished the proof for this theorem.
\end{proof}

\ \par

\begin{proof}[Proof of Propositions \ref{pr:Gaussian}] 
Similarly to the proofs for the previous theorems, we can linearly transform $X$ and without loss of generality assume $\E X = 0, \Var(X) = I_p$. We then consider $\hat{\theta}_{\rm LS} - \theta= \vec{\mu}^\top \left(\vec{\X}^\top \vec{\X}\right)^{-1} \vec{\X}^\top  \Y - \theta$. Note that
$$\frac{1}{n}\vec{\X}^\top  \vec{\X} = \begin{bmatrix}
1 & \bar{\X}^\top \\
\bar{\X} & \hat\SSigma_X + \bar{\X}\bar{\X}^\top 
\end{bmatrix}, \quad \frac{1}{n}\X^\top \X = \hat\SSigma_X + \bar{\X}\bar{\X}^\top, $$
where $\bar{\X} = \frac{1}{n}\sum_{k=1}^n X_k$, $\hat\SSigma_X = \frac{1}{n} \sum_{k=1}^n(X_k - \bar{\X})(X_k - \bar{\X})^\top $. The block-wise matrix inverse formula yields
\begin{equation}
\left(\frac{1}{n}\vec{\X}^\top \vec{\X}\right)^{-1} = \begin{bmatrix}
1 + \bar\X^\top \hat\SSigma_X^{-1} \bar\X & -\bar{\X}^\top \hat\SSigma_X^{-1}\\
-\hat\SSigma_X^{-\top} \bar{\X} & \hat\SSigma_X^{-1}
\end{bmatrix}.
\end{equation} 
By the expansion in \eqref{eq:hat_theta_LS-theta}, we have
\begin{equation*}
\begin{split}
\hat{\theta}_{\rm LS} - \theta & = \bar\Bdelta - \left(0, \frac{1_n}{n}\X^\top\right) \left(\frac{1}{n}\vec{\X}^\top\vec{\X}\right)^{-1} \left(\frac{1}{n}\vec{\X}^\top\Bdelta\right)\\
& = \left(\frac{1_n^\top}{n} + \bar{\X}^\top\hat\SSigma_X^{-1}\bar{\X} \frac{1_n^\top}{n} - \frac{1}{n}\bar{\X}^\top\hat\SSigma_X^{-1} \X\right)\Bdelta.
\end{split}
\end{equation*}
When $\X\in \mathbb{R}^{n\times p}$ are i.i.d. standard normal, it is commonly known that $\bar{\X}$, $\hat\SSigma_X$ and $\Bdelta$ are all independent, and $\bar{\X}\overset{iid}{\sim} N(0, 1/n)$, $\hat\SSigma_X^{-1}$ satisfies inverse-Wishart distribution $n\cdot \mathcal{W}_p^{-1}(I_p,  n-1)$, and its expectation is $\E \hat\SSigma_X^{-1} = \frac{n}{n-p-2}I_p$. Therefore,
\begin{equation*}
\begin{split}
& \E \left(\hat\theta_{\rm LS}  - \theta\right)^2 = \E \delta^2 \cdot \E \left\|\frac{1_n^\top}{n} + \bar{\X}^\top\hat\SSigma_X^{-1}\bar{\X} \frac{1_n^\top}{n} - \frac{1}{n}\bar{\X}^\top\hat\SSigma_X^{-1} \X\right\|_2^2\\
= & \tau^2 \cdot \E \bigg(\frac{1}{n} (1 + \bar\X^\top\hat\SSigma_X^{-1}\bar{\X})^2 + \frac{1}{n}(\bar{\X}^\top\hat\SSigma_X^{-1}(\bar{\X}\bar{\X}^\top + \hat\SSigma_X)\hat\SSigma_X^{-1}\bar{\X})\\
& \quad\quad\quad - \frac{2}{n}(\bar{\X}\hat\SSigma_X^{-1}\bar{\X} + (\bar{\X}\hat\SSigma_X^{-1} \bar{\X})^2)\bigg) \\
= & \tau^2 \cdot \E \left(\frac{1}{n} + \frac{1}{n} \bar{\X}^\top\hat\SSigma_X^{-1}\bar{\X} \right) = \frac{\tau^2}{n} \left(1 +  \tr\left(\E \hat\SSigma_X^{-1} \cdot \E \bar{\X}\bar{\X}^\top \right)\right)\\
 = & \frac{\tau^2}{n}\left(1 + \tr\left(\frac{nI_p}{n-p-2}\cdot\frac{I_p}{n}\right)\right) = \frac{\tau^2}{n} + \frac{p}{n(n-p-2)} \tau^2.
\end{split}
\end{equation*}
The calculation for $\E(\hat{\theta}_{\rm SSLS} - \theta)^2$ is similar.  Since $\hat{\theta}_{\rm SSLS} - \theta= \hat{\vec{\mu}}^\top \left(\vec{\X}^\top \vec{\X}\right)^{-1} \vec{\X}^\top  \Y - \theta$, by the calculation in Theorem \ref{th:asymp_imperfect}, we have
\begin{equation*}
\begin{split}
& \hat{\theta}_{\rm SSLS} - \theta  \overset{\eqref{eq:decomposition_theta_SPLS}}{=} (\hat{\vec{\mu}} - \vec{\mu})^\top \beta + \bar{\Bdelta} + \left(\hat{\vec{\mu}}^\top  - \frac{1_n^\top }{n}\vec{\X}\right)\left(\vec{\X}^\top \vec{\X}\right)^{-1}\vec{\X}^\top \Bdelta\\
= & \bar{\X}_{\rm full}^\top \beta_{(2)} + \left(\frac{1_n^\top }{n} + \frac{m}{m+n} \bar{\X}^\top \hat\SSigma_X^{-1} \bar{\X}\frac{1_n}{n} - \frac{m}{n(m+n)}\bar{\X}^\top \hat\SSigma_X^{-1} \X \right)\Bdelta\\
& + \frac{1}{m+n} 1_m^\top  \X_{\rm add} (-\hat\SSigma_X^{-1}\bar{\X} \frac{1_n^\top }{n} + \frac{1}{n}\hat\SSigma_X^{-1} \X)\Bdelta.
\end{split}
\end{equation*}
Since $\X_{\rm add}$, $\Bdelta$ and $\X$ are all independent with mean 0, it is easy to check that any two of the three terms above are uncorrelated. Thus, 
\begin{equation*}
\begin{split}
& \E \left(\hat{\theta}_{\rm SSLS} - \theta\right)^2\\
= & \E \left(\bar\X^\top _{\rm full}\beta_{(2)}\right)^2 + \E \left[\left(\frac{1_n^\top }{n} + \frac{m}{m+n} \bar{\X}^\top \hat\SSigma_X^{-1} \bar{\X}\frac{1_n^\top }{n} - \frac{m}{n(m+n)}\bar{\X}^\top \hat\SSigma_X^{-1} \X \right)\Bdelta\right]^2\\
& + \left[\frac{1}{m+n} 1_m^\top  \X_{\rm add} (-\hat\SSigma_X^{-1}\bar{\X} \frac{1_n^\top }{n} + \frac{1}{n}\hat\SSigma_X^{-1} \X)\Bdelta\right]^2,
\end{split}
\end{equation*}
\begin{equation*}
\begin{split}
\E \left(\bar\X_{\rm full}^\top \beta_{(2)}\right)^2 = \frac{1}{m+n}\beta_{(2)}^\top  \E (X - \mu)(X-\mu)^\top  \beta_{(2)},
\end{split}
\end{equation*}
\begin{equation}
\begin{split}
& \E \left[\left(\frac{1_n^\top }{n} + \frac{m}{m+n} \bar{\X}^\top \hat\SSigma_X^{-1} \bar{\X}\frac{1_n^\top }{n} - \frac{m}{n(m+n)}\bar{\X}^\top \hat\SSigma_X^{-1} \X \right)\Bdelta\right]^2\\
= & \tau^2\cdot \E \left\|\frac{1}{n}\left(1 + \frac{m}{m+n} \bar{\X}^\top \hat\SSigma_X^{-1} \bar{\X}\right)1_n^\top  - \frac{m}{n(m+n)}\bar{\X}^\top \hat\SSigma_X^{-1} \X\right\|_2^2\\
= & \tau^2\cdot \E \Bigg\{ \frac{1}{n}(1 + \frac{m}{n+m}\bar{\X}^\top \hat\SSigma_X^{-1}\bar{\X})^2 + \frac{m^2}{n^2(m+n)^2}\bar{\X}^\top \hat\SSigma_X^{-1} (n\hat\SSigma_X + n\bar{\X}\bar{\X})\hat\SSigma_X^{-1} \bar{\X}\\
& - \frac{2}{n}\left(1 + \frac{m}{m+n} \bar{\X}^\top \hat\SSigma_X^{-1}\bar{\X}\right)\cdot \bar{\X}^\top \hat\SSigma_X^{-1}\bar{\X}\cdot\frac{m}{m+n}
\Bigg\} \\
= & \tau^2 \left(\frac{1}{n} + \frac{m^2}{(n+m)^2 n}\bar{\X}^\top \hat\SSigma_X^{-1} \bar{\X}\right),
\end{split}
\end{equation}
\begin{equation*}
\begin{split}
& \E \left[\frac{1}{m+n} 1_m^\top  \X_{\rm add} (-\hat\SSigma_X^{-1}\bar{\X} \frac{1_n^\top }{n} + \frac{1}{n}\hat\SSigma_X^{-1} \X)\Bdelta\right]^2\\
= & \frac{\tau^2}{(m+n)^2} \E \left\|1_m^\top  \X_{\rm add} (-\hat\SSigma_X^{-1}\bar{\X} \frac{1_n^\top }{n} + \frac{1}{n}\hat\SSigma_X^{-1} \X)\right\|_2^2\\
= & \frac{\tau^2}{(m+n)^2} \E \Bigg[1_m^\top  \X_{\rm add} \Big(\frac{1}{n}\hat\SSigma_X^{-1} \bar{\X}\bar{\X}^\top \hat \SSigma_X^{-1} - \frac{2}{n}\hat\SSigma_X^{-1} \bar{\X}\bar{\X}^\top \hat\SSigma_X^{-1}\\
& +  \frac{1}{n}\hat\SSigma_X^{-1}(\hat\SSigma_X + \bar{\X}\bar{\X}^\top )\hat\SSigma_X^{-1} \Big)\X_{\rm add}^\top 1_m\Bigg]\\
= & \frac{\tau^2}{(m+n)^2n} \E  1_m^\top  \X_{\rm add}\hat\SSigma_X^{-1} \X_{\rm add}^\top  1_m = \frac{\tau^2m}{(m+n)^2n} \E  X_{n+1}^\top \hat\SSigma_X^{-1} X_{n+1}.
\end{split}
\end{equation*}
To sum up,
\begin{equation*}
\begin{split}
& n\E (\hat{\theta}_{\rm SSLS} - \theta)^2 = \tau^2 + \frac{n}{m+n}\beta_{(2)}^\top \E (X- \mu)(X-\mu)^\top \beta_{(2)}\\
& + \frac{m\E \sigma^2(X)}{m+n}\left(\frac{m}{m+n}\left(n \E \bar \X^\top \hat\SSigma_X^{-1}\bar \X\right) + \frac{n}{m+n}\E X_{n+1}^\top \hat\SSigma_X^{-1} X_{n+1}\right).
\end{split}
\end{equation*}
Especially when $X \sim N(0, I)$, $\frac{1}{n}\hat\SSigma_X^{-1}$ is independent of $\bar{\X}$ and satisfies the inverse Wishart distribution. At this point, $\E \hat\SSigma_X^{-1} = \frac{nI_p}{n-p-2}$, 
$$n\E \left(\hat{\theta}_{\rm SSLS} - \theta\right)^2 = \tau^2 + \frac{n}{m+n} \beta_{(2)}^\top \E (X-\mu)(X-\mu)^\top \beta_{(2)} + \frac{m}{m+n} \cdot \frac{n}{n-p-2}\cdot \tau^2, $$
which has finished the proof of Proposition \ref{pr:Gaussian}.
\end{proof}

\section*{Proofs for Oracle Optimality Setting}

\begin{proof}[Detailed Calculation for \eqref{eq:oracle_semi_supervised_risk} (Oracle risk for $\hat{\theta}_{\rm ss}^\ast$)]
It is easy to see that $\hat{\theta}_{\rm ss}^\ast$ is an unbiased estimator for $\theta$, thus
\begin{equation*}
\begin{split}
& \E\left(\hat{\theta}_{\rm ss}^\ast - \theta\right)^2 = \Var\left(\hat{\theta}_{\rm ss}^\ast\right)\\
= & \sum_{k=1}^n \Var\left(\frac{Y_k}{n} - \frac{\xi_0(X_k)}{n} + \frac{\xi_0(X_k)}{n+m}\right) + \sum_{k=n+1}^{n+m}\Var\left(\frac{1}{n+m}\xi_0(X_k)\right)\\
= & n\E\left(\Var\left(\frac{Y_k}{n} - \frac{m}{n(n+m)}\xi_0(X_k)\Bigg| X_k\right)\right)\\
& + n \Var\left(\frac{\xi(X_k)}{n} - \frac{m}{n(n+m)}\xi_0(X_k)\right) + \frac{m}{(n+m)^2}\Var\left(\xi(X_k)\right)\\
= & n\frac{\sigma^2}{n^2} + n \frac{\sigma_\xi^2}{(n+m)^2} + \frac{m\sigma_\xi^2}{(n+m)^2}\\
= & \frac{\sigma^2}{n} + \frac{1}{n+m}\sigma_\xi^2,
\end{split}
\end{equation*}
which has proved \eqref{eq:oracle_semi_supervised_risk}.
\end{proof}

\ \par

\begin{proof}[Proof of Proposition \ref{pr:oracle_lower_bound}]
We first consider \eqref{eq:variance_oracle}. For any given $\sigma^2>0$, $P_X$, $\xi_0(\cdot)$ and $P_X$, we consider the following subset of $\mathcal{P}_{P_X, \xi_0, \sigma^2}$,
\begin{equation}
\begin{split}
\mathcal{P}'_{\xi_0, P_X, \sigma^2} = \Big\{P: \int_Y P(Y, X) = P_X, Y = \xi_0(X) + c + \varepsilon,\\ 
\varepsilon \text{ is independent from } X, \varepsilon\sim N(0, \sigma^2)\Big\}.
\end{split}
\end{equation}
Based on sample $\{X_i, Y_i\}_{i=1}^n$, known $P_X$ and $\xi_0(X)$, we can rewrite the model to
$$Y_i  - \xi_0(X_i) = c+\varepsilon_i,\quad i=1,\ldots, n, $$
where $Y_i$ and $\xi(X_i)$ are observable. By classical theory on normal mean estimation with Gaussian noise,
$$\inf_{\tilde{c}}\sup_{c\in \mathbb{R}} n\left(\tilde{c} - c\right)^2 = \sigma^2. $$
Note for the estimating problem in the original proposition, we target on estimating
\begin{equation*}
\theta = \E(Y) = c + \E\xi_0(X),
\end{equation*}
where $\E\xi_0(X)$ is known. Thus, estimating $\theta$ is equivalent to estimating $c$, which implies
\begin{equation*}
\begin{split}
& \inf_{\tilde{\theta}_n} \sup_{P \in \mathcal{P}_{\xi_0, \sigma^2}} \left[\E_P \left(n\left(\tilde{\theta}_n - \theta\right)^2\right)\right] \geq \inf_{\tilde{\theta}_n} \sup_{P \in \mathcal{P}'_{\xi_0, \sigma^2}} \left[\E_P \left(n\left(\tilde{\theta}_n - \theta\right)^2\right)\right]\\
\geq & \inf_{\tilde{c}}\sup_{c\in \mathbb{R}} n\left(\tilde{c} - c\right)^2 = \sigma^2.
\end{split}
\end{equation*}

Next we aim at the proof for \eqref{eq:variance_oracle_semi}. Suppose we are given fixed $\sigma_\xi^2, \sigma^2>0$ and linear function $\xi_0$. If $\xi_0(X)$ is a constant, $\varepsilon_\xi$ always equals 0, the the problem transform to the first situation. 

If $\xi_0(X) = aX+b$ with $a\neq 0$, since we can always normalize $Y$, without loss of generality let us assume $\xi_0(X) = X$.  We also focus on the situation for $p=1$ as the proof for $p>1$ essentially follows. Now we consider the following subset of $\mathcal{P}^{\rm ss}_{\xi_0, \sigma_Y^2, \sigma^2}$:
\begin{equation*}
\begin{split}
\mathcal{P}^{\rm ss'}_{\sigma_\xi^2, \sigma^2} = \Big\{P: & X\sim N(\mu, \sigma_\xi^2) , Y = X + c+ \varepsilon \text{ for some constants } \mu, c,\\
& \varepsilon \text{ is independent from } X, \varepsilon \sim N(0, \sigma^2)\Big\}.
\end{split}
\end{equation*}
In this case, $\E Y = \theta = c+\mu$. In order to calculate the minimax rate for estimating $\theta$, we first consider the Bayes estimator for $c$ and $\mu$ under the prior distribution $c, \mu \sim N(0, V^2)$, where $V^2\to \infty$. It is easy to see that
$$p_0(\mu, c) \varpropto \exp\left(-\frac{\mu^2+c^2}{2V^2}\right),$$
\begin{equation*}
\begin{split}
p(Y, X|\mu, c) \varpropto \exp\Bigg(-\frac{1}{2}\Bigg(&\frac{\sigma_\xi^2+\sigma^2}{\sigma_\xi^2\sigma^2}(X-\mu)^2 + \frac{1}{\sigma^2}(Y-\mu-c)^2\\
& - \frac{2}{\sigma^2}(X-\mu)(Y - \mu-c)\Bigg)\Bigg), 
\end{split}
\end{equation*}
$$p(X|\mu, c) \varpropto \exp\left(-\frac{1}{2\sigma_\xi^2}(X-\mu)^2\right). $$
Given observations $\{Y_k, X_k\}_{k=1}^{n}$ and $\{X_k\}_{k=n+1}^{n+m}$, the posterior distribution for $\mu$ and $c$ is
\begin{equation*}
\begin{split}
& \pi(\mu, c| \{Y_k, X_k\}_{k=1}^{n}, \{X_k\}_{k=n+1}^{n+m} ) \varpropto \frac{p\left(\{Y_k, X_k\}_{k=1}^{n}, \{X_k\}_{k=n+1}^{n+m}\big| \mu, c\right) p_0(\mu, c)}{p\left(\{Y_k, X_k\}_{k=1}^{n}, \{X_k\}_{k=n+1}^{n+m}\right)}\\
\varpropto & \prod_{k=1}^n \exp\left(-\frac{1}{2}\left(\frac{\sigma_\xi^2+\sigma^2}{\sigma_\xi^2\sigma^2}(X_k-\mu)^2 + \frac{(Y_k-\mu-c)^2}{\sigma^2} - \frac{2(X_k-\mu)(Y_k - \mu-c)}{\sigma^2}\right)\right)\\
& \cdot \exp\left(-\frac{\mu^2+c^2}{V^2}\right)  \cdot \prod_{k=n+1}^{n+m} \exp\left(-\frac{1}{2\sigma_\xi^2}(X-\mu)^2\right) / p\left(\{Y_k, X_k\}_{k=1}^{n}, \{X_k\}_{k=n+1}^{n+m}\right).
\end{split}
\end{equation*}
After simplification for the previous equation, when $V^2\to\infty$, the joint posterior distribution of $\mu, c$ is
$$\mu, c| \{Y_k, X_k\}_{k=1}^{n}, \{X_k\}_{k=n+1}^{n+m} \sim N\left(\begin{pmatrix}
\frac{1}{n+m}X_k, \frac{1}{n}(Y_k - X_k)
\end{pmatrix}, \begin{bmatrix}
\frac{1}{(n+m)\sigma_\xi^2} & 0\\
0 & \frac{1}{n\sigma^2}
\end{bmatrix}\right) $$
Therefore, the Bayes estimator for $\theta = \mu+c$ is
\begin{equation*}
\hat{\theta}_{bayes} = \E\left(\mu+c\Big| \{Y_k, X_k\}_{k=1}^{n}, \{X_k\}_{k=n+1}^{n+m}\right) = \bar\Y - \frac{1}{n} \sum_{k=1}^n X_k + \frac{1}{n+m}\sum_{k=1}^{n+m}X_k.
\end{equation*}
Similarly to the calculation for \eqref{eq:oracle_semi_supervised_risk}, it is easy to check that $\hat{\theta}_{bayes}$ has constant risk for all different values of $c$ and $\mu$:
\begin{equation*}
n\E\left(\hat{\theta}_{bayes} - \theta\right)^2 = \sigma^2 + \frac{n}{n+m}\sigma_\xi^2.
\end{equation*}
This implies that $\theta_{bayes}$ is the minimax estimator for $\theta$ in distribution class $\mathcal{P}^{\rm ss '}_{\sigma_\xi^2, \sigma^2}$. To sum up, we have finished the proof for this proposition. 
\end{proof}

\ \par

\begin{proof}[Proof of Proposition \ref{pr:oracle_lower_bound-efficiency}] 
	The proof of this proposition follows the ideas from \cite{Vaart2002semiparametric} and \cite{bickel1991efficient}. 
	Let $k(v) = 2(1+e^{-v})^{-1}$. For notational convenience, define $Z = Y - \xi^0(X)$ so that $E_{P^0}(Z|X=x) = 0$. Let $\gamma\in \mathbb{R}$ denote a generic parameter, and next we introduce a class of distribution $P_{\gamma}$ on $(X, Y)$. We fix the marginal distribution $P_{\gamma, X}=P_{X}^0$, then define the conditional distribution at $\gamma$ of $Z$ given $X = x$ via
	$$dP_{\gamma, Z}(Z|X=x) = c(\gamma, x) k(\gamma Z/\sigma^2) dP_Z^0(Z|x)$$
	where $c(\gamma, x)$ denotes the normalizing constant for the conditional distribution at $\gamma$ of $Z$ given $X=x$. Note that
	$$\frac{d}{d\gamma}\mathbb{E}_{P_{\gamma}}(Y|X=x)\Bigg|_{\gamma=0} = (1/\sigma^2(x))\mathbb{E}_{P_{\gamma=0}}(Z^2|x),$$
	it can be checked that for $\gamma = O(1/\sqrt{n})$,
	$$\mathbb{E}_{P_{\gamma}}(Y|X=x) = \xi^0(x)+\gamma\sigma^2(x)/\sigma^2 + O(1/\sqrt{n}).$$
	Then, the Fisher information at $\gamma = 0$ for a sample of size $n$ from the family $\{P_\gamma\}$ is $I = n\mathbb{E}(\sigma^2(X)/\sigma^4) = n/\sigma^2$. As in \cite{Vaart2002semiparametric}, we have obtained \eqref{eq:variance_oracle_semi}. 
\end{proof}

\ \par

\begin{proof}[Proof of Theorem \ref{th:oracle_upper_bound}] 
	For any $q\geq0$, we denote
$$X^{(q)\bullet} = (X_1, \ldots, X_p, g_1(X), \dots, g_q(X)),$$  
$$\vec{X}^{(q)\bullet} = (1, X_1, \ldots, X_p, g_1(X), \dots, g_q(X)). $$
Suppose 
\begin{equation*}
\tau_{(q)}^2 = \argmin_{\beta^{(q)}\in \mathbb{R}^{1+p+q}} \E_{(Y, X)\sim P}\left(Y - (\beta^{(q)})^\top\vec{X}^{(q)\bullet} \right)^2.
\end{equation*}
Clearly, $\tau_{(q)}^2$ is an non-increasing sequence of $q$. Based on either Assumption (i) or (ii) of Proposition \ref{th:oracle_upper_bound},
\begin{equation}\label{eq:limit_tau_q}
\lim_{q\to \infty} \tau^2_{(q)} = \E\left(Y - \E(Y|X)\right)^2 = \sigma^2.
\end{equation}
By Proposition \ref{th:nvar(bar_Y)}, $\tau_{(q)}^2 + \Var((\beta^{(q)})^\top \vec{X}^{(q)\bullet}) = \Var(Y)$. By the law of total variance, $\sigma^2 + \Var(\xi(X)) = \Var(Y)$. Suppose $\hat{\theta}_{\rm LS}^{(q)}$, $\hat{\theta}_{\rm SSLS}^{(q)}$ are the least squares estimator and semi-supervised least squares estimator with the basis $(X_1, \ldots, X_p, g_1(X), \ldots, g_q(X))$. Corresponding, suppose $(\hat{\theta}_{\rm LS}^{(q)})^1$ and $(\hat{\theta}_{\rm SSLS}^{(q)})^1$ as the refined estimators based on \eqref{eq:truncation}. Based on Theorems \ref{th:MSE_delta} and \ref{th:MSE_SPLS}, for fixed $q>0$, 
\begin{equation*}
\limsup_{n\to \infty} n\E\left((\hat{\theta}_{\rm LS}^{(q)})^1 - \theta \right)^2 = \tau_{(q)}^2, 
\end{equation*}
\begin{equation*}
\limsup_{n\to \infty} n\E \left((\hat{\theta}_{\rm SSLS}^{(q)})^1 - \theta\right)^2 = \tau_{(q)}^2 + \rho\Var((\beta^{(q)})^\top \vec{X}^{(q)\bullet}) = (1-\rho)\tau_{(q)}^2 + \rho\Var(Y).
\end{equation*}
By \eqref{eq:limit_tau_q},
\begin{equation*}
\lim_{q\to \infty} \limsup_{n\to \infty} n\E\left((\hat{\theta}_{\rm LS}^{(q)})^1 - \theta \right)^2 = \sigma^2, 
\end{equation*}
\begin{equation*}
\lim_{q\to \infty} \limsup_{n\to \infty} n\E \left((\hat{\theta}_{\rm SSLS}^{(q)})^1 - \theta\right)^2 = (1-\rho)\sigma^2 + \rho\Var(Y) = \sigma^2 + \rho \Var(\xi(X)).
\end{equation*}
Therefore, there exists sequence $\{q_n\}$ growing slowly enough that guarantees \eqref{eq:oracle_upper_bound} and \eqref{eq:oracle_upper_bound_semi}. Finally, the asymptotic distribution results hold similarly which we do no repeat here. 
\end{proof}

\section*{Proofs for Application in Average Treatment Effect}

\begin{proof}[Proof of Theorem \ref{th:d_SPLS_asymptotic}] 
	We shall note that $\hat{d}_{\rm SSLS} = \hat{\vec{\mu}}^\top  \hat \beta_t - \hat{\vec{\mu}}^\top  \hat{\beta}_c$. Based on \eqref{eq:decomposition_theta_SPLS}, we have the following extensions for these two terms separately
$$\hat{\vec{\mu}}^\top \hat \beta_t - \theta_t  = \left(\hat{\vec{\mu}}^\top  - \vec{\mu}\right)^\top \beta_t + \bar{\Bdelta}_t - \left(0, ~ \bar{\X}_t - \hat \mu\right)^\top \left(\vec{\X}_t^\top \vec{\X}_t\right)^{-1} \vec{\X}_t^\top \Bdelta_t, $$
$$\hat{\vec{\mu}}^\top \hat \beta_c - \theta_c  = \left(\hat{\vec{\mu}}^\top  - \vec{\mu}\right)^\top \beta_c + \bar{\Bdelta}_c - \left(0, ~ \bar{\X}_c - \hat\mu\right)^\top \left(\vec{\X}_c^\top \vec{\X}_c\right)^{-1} \vec{\X}_c^\top \Bdelta_c. $$
Thus $\hat{d}_{\rm SSLS}-d$ has the following decomposition
\begin{equation}\label{eq:d_decomposition}
\begin{split}
& \hat{d}_{\rm SSLS} - d = (\hat{\vec{\mu}}^\top \hat \beta_t - \theta_t) - (\hat{\vec{\mu}}^\top \hat \beta_c - \theta_c )\\
= & \bar{\Bdelta}_t - \bar{\Bdelta}_c + \left(\hat{\vec{\mu}}^\top  - \vec{\mu}\right)^\top (\beta_t - \beta_c)\\
& - \left(0, ~ \bar{\X}_t - \hat\mu\right)^\top \left(\vec{\X}_t^\top \vec{\X}_t\right)^{-1} \vec{\X}_t^\top \Bdelta_t + \left(0, ~ \bar{\X}_c - \hat\mu\right)^\top \left(\vec{\X}_c^\top \vec{\X}_c\right)^{-1} \vec{\X}_c^\top \Bdelta_c.
\end{split}
\end{equation}
Essentially the same as Theorem \ref{th:theta_LS_asymptotic}, one can show
\begin{equation}\label{eq:d_asymptotic_1}
\frac{\bar\Bdelta_t - \bar\Bdelta_c + (\hat{\vec{\mu}} - \vec{\mu})^\top (\beta_t - \beta_c)}{V} \to N(0, 1),
\end{equation}
\begin{equation}\label{eq:d_asymptotic_2}
\frac{\left(0, ~ \bar{\X}_t - \hat\mu\right)^\top \left(\vec{\X}_t^\top \vec{\X}_t\right)^{-1} \vec{\X}_t^\top \Bdelta_t}{\sqrt{\tau_t^2/n_t}} \overset{d}{\to} 0,\quad \frac{\left(0, ~ \bar{\X}_c - \hat\mu\right)^\top \left(\vec{\X}_c^\top \vec{\X}_t\right)^{-1} \vec{\X}_c^\top \Bdelta_t}{\sqrt{\tau_c^2/n_c}} \overset{d}{\to} 0,
\end{equation}
Combining \eqref{eq:d_asymptotic_1}, \eqref{eq:d_asymptotic_2} and \eqref{eq:d_decomposition}, we have
$$\frac{\hat{d}_{\rm SSLS} - d}{V} \to N(0, 1).$$
Next we show the asymptotic property for $\hat{V}$. Based on the proof of Theorem \ref{th:asymp_imperfect}, we have already shown
\begin{equation*}
\lim_{n_t \to \infty} \frac{MSE_t}{\tau_t^2} \overset{d}{\to} 1,\quad \lim_{n_c \to \infty} \frac{MSE_c}{\tau_c^2} \overset{d}{\to} 1.
\end{equation*}
Besides, $\hat\beta_{t, (2)} \overset{d}{\to} \beta_{t, (2)}$, $\hat\beta_{c, (2)} \overset{d}{\to} \beta_{c, (2)}$, $\hat{\SSigma}_X \overset{d}{\to} \E(X-\mu)(X-\mu)^\top$ as $n_t, n_c \to \infty$. Thus, whenever $V^2>0$,
\begin{equation*}
\hat{V}^2/V^2 \overset{d}{\to} 1, \quad \text{as } n_t, n_c \to \infty.
\end{equation*}
\end{proof}

\section*{Proof of Technical Lemmas}

We collect all technical proofs in this section.

\begin{proof}[Proof of Lemma \ref{lm:cov_concentration}]
\begin{itemize}
	\item Part 1 directly follows from Theorem 5.39 in \cite{vershynin2012close}. \item For Part 2, it can be calculated that
	\begin{equation*}
	\begin{split}
	\E \left\|\sum_{k=1}^n Z_k\right\|_2^q = & \E \left(\sum_{i=1}^p \left(\sum_{k=1}^nZ_{ki}\right)^2\right)^{q/2} \overset{\text{H\"older's ineq}}{\leq} \E \sum_{i=1}^p \left|\sum_{k=1}^n Z_{ki}\right|^{q}\cdot p^{q/2-1}.\\
	\end{split}
	\end{equation*}
	By Marcinkiewicz-Zygmund inequality \citep{chow2012probability}, under either Assumption 2 or 2', we have
	\begin{equation*}
	\begin{split}
	\E \left|\sum_{k=1}^n Z_{ki}\right|^q \overset{\text{M-Z ineq}}{\leq} &  C_q \E \left(\sum_{k=1}^n |Z_{ki}|^2\right)^{q/2}\\
	\overset{\text{H\"older's ineq}}{\leq} & C_q n^{q/2-1} \sum_{k=1}^n \E |Z_{ki}|^{q} \leq C_q n^{q/2-1}, \quad i=1,\cdots, p.
	\end{split}
	\end{equation*}
	Thus, we conclude that \eqref{ineq:Z_q-th moment} holds.
	
	\item Finally we consider Part 3. Recall the fact that $\E \delta = 0$, $\E Z_k \delta =0$. The proof is similar to Part 2. When $2\leq q< 4$, under either Assumption 2 or 2',
	$$\E \left|Z_{ki} \delta_k \right|^q \overset{\text{H\"older's ineq}}{\leq} \left(\E |Z_{ki}|^{\frac{4q}{4-q}}\right)^{\frac{4-q}{4}} \left(\E \delta_k^4\right)^{\frac{q}{4}} \leq C_q < \infty,  $$
	$$\E \left| \delta_k \right|^q \leq \left(\E \delta_k^4\right)^{\frac{q}{4}} \leq C_q < \infty. $$
	Thus, by Marcinkiewicz-Zygmund inequality \citep{chow2012probability},
	\begin{equation*}
	\begin{split}
	& \E \left|\sum_{k=1}^n Z_{ki}\delta_k\right|^q \overset{\text{M-Z ineq}}{\leq} C_q \E \left(\sum_{k=1}^n |Z_{ki}\delta_k|^2\right)^{q/2}\\
	\overset{\text{H\"older's ineq}}{\leq} & C_q n^{q/2-1} \sum_{k=1}^n \E |Z_{ki}\delta_k|^{q} \leq C_q n^{q/2-1}, \quad i=1,\cdots, p.
	\end{split}
	\end{equation*}
	\begin{equation*}
	\begin{split}
	& \E \left|\sum_{k=1}^n \delta_k\right|^q \overset{\text{M-Z ineq}}{\leq} C_q \E \left(\sum_{k=1}^n |\delta_k|^2\right)^{q/2}\\
	\overset{\text{H\"older's ineq}}{\leq} & C_q n^{q/2-1} \sum_{k=1}^n \E |\delta_k|^{q} \leq C_q n^{q/2-1}, \quad i=1,\cdots, p.
	\end{split}
	\end{equation*}
	Therefore,
	\begin{equation*}
	\begin{split}
	\E \left\|\sum_{k=1}^n \vec{Z}_k\delta_k\right\|_2^q = & \E \left(\left(\sum_{k=1}^n \delta_k\right)^2 +\sum_{i=1}^p \left( \sum_{k=1}^n\vec{Z}_{ki}\delta_k\right)^2\right)^{q/2}\\ \overset{\text{H\"older's ineq}}{\leq} & \E \left(\left|\sum_{k=1}^n\delta_k\right|^q + \sum_{i=1}^p \left|\sum_{k=1}^n Z_{ki}\right|^{q}\right)\cdot (p+1)^{q/2-1}\\
	\leq & C_q (p+1)^{q/2}n^{q/2} \leq C_q(pn)^{q/2},
	\end{split}
	\end{equation*}
	which has shown \eqref{ineq:Zdelta_q-th moment}.
\end{itemize}
\end{proof}

\begin{proof}[Proof of Lemma \ref{lm:inverse_expansion}] 
	Since
\begin{equation*}
\begin{split}
I = & \sum_{k=0}^{q-1} \left((-A^{-1}B)^k - (-A^{-1}B)^{k+1}\right) + (-A^{-1}B)^q\\
= & \sum_{k=0}^{q-1}\left(-A^{-1}B\right)^{k}(I + A^{-1}B) + (-A^{-1}B)^q\\
= & \sum_{k=0}^{q-1}\left(-A^{-1}B\right)^kA^{-1} (A+B) + \left(-A^{-1}B\right)^q
\end{split}
\end{equation*}
Right multiply $(A+B)^{-1}$ to the equation above, we obtain \eqref{eq:inverse_expasion}.
\end{proof}

\ \par

\begin{Lemma}[Separate Analysis of \eqref{eq:E1_Q hat theta-theta}]\label{lm:five terms} Under the setting of the proof for Theorem \ref{th:MSE_delta}, one has
	\begin{equation}\label{ineq:lm_1}
	\E \left[1_Q\delta_1^2(0, Z_2^\top )\vec{\vvarXi}^{-1}\vec Z_1\right] = -\frac{1}{n}\tr\left(\left(\E \delta_1^2Z_1\right)^\top \cdot \E \left(Z_2Z_2Z_2^\top \right)\right) - \frac{\tau^2}{n} + O\left(\frac{p^2}{n^{5/4}}\right)
	,
	\end{equation}
	\begin{equation}\label{ineq:lm_2}
	\E \left[1_Q \delta_1(0, Z_1^\top )\vec\vvarXi^{-1} \vec Z_2\delta_2\right] = O\left(\exp(-cn)\cdot \poly(n, p)\right),
	\end{equation}
	\begin{equation}\label{ineq:lm_3}
	\E \left[1_Q\delta_1(0, Z_2^\top )\vec{\vvarXi}^{-1}\vec{Z}_2\delta_2\right] = -\frac{1}{n}\|\E Z\delta Z^\top \|_F^2 + O\left(\frac{p^2}{n^{5/4}}\right),
	\end{equation}
	\begin{equation}\label{ineq:lm_4}
	\E \left[1_Q\delta_1^2(0, Z_1^\top )\vec{\vvarXi}^{-1}\vec{Z}_1\delta_1\right] = \E \delta^2Z^\top Z+O\left(\frac{p}{n^{1/4}}\right),
	\end{equation}
	\begin{equation}\label{ineq:lm_5}
	\E \left[1_Q\delta_1(0, Z_2^\top )\vec{\vvarXi}^{-1}\vec Z_3\delta_3\right] = O\left(\frac{p^4}{n^3}\right),
	\end{equation}
	\begin{equation}\label{ineq:lm_6}
	\E \left[1_Q\bar \Bdelta^2\right] = \frac{\tau^2}{n} + O\left(\exp(-cn^{1/2})\poly(n)\right),
	\end{equation}
	\begin{equation}\label{ineq:lm_7}
	\begin{split}
	& \E \left[1_Q\left(\left(0, \frac{1_n}{n}\Z^\top \right)\vec{\vvarXi}^{-1}\left(\frac{1}{n}\vec{\Z}^\top \Bdelta\right)\right)^2\right]\\
	= & \frac{1}{n^2} \left(\tr(\E  Z\delta^2Z^\top ) + \left(\tr(\E Z\delta Z^\top )\right)^2 + \|\E Z\delta Z^\top \|_F^2\right) + O\left(\frac{p^2}{n^{2+1/4}}\right).
	\end{split}
	\end{equation}
\end{Lemma}

\begin{proof}[Proof of Lemma \ref{lm:five terms}] We analyze \eqref{ineq:lm_1} - \eqref{ineq:lm_7} separately in the next seven parts.
\begin{enumerate}
	\item Recall $\vec{\vvarXi}_{-\{1, 2\}} = \frac{1}{n}\sum_{k=3}^n \vec{X}_k\vec{X}_k^\top$, we also denote $\vec{\vvarXi}_{1,2} = \frac{1}{n}(\vec{X}_1\vec{X}_1^\top + \vec{X}_2 \vec{X}_2^\top)$, $\vec{\vvarXi}_{1,2,3} = \frac{1}{n}(\vec{X}_1\vec{X}_1^\top + \vec{X}_2 \vec{X}_2^\top + \vec{X}_3\vec{X}_3^\top)$.
	Under the event $Q$, $\vec{\vvarXi}^{-1}$ and $ \vec{\vvarXi}^{-1}_{-\{1, 2\}}$ are invertible. By Lemma \ref{lm:inverse_expansion}, we can further calculate that
	\begin{equation}\label{eq:E delta_1 Z_2 Sigma Z_1}
	\begin{split}
	& \E \left[1_Q \delta_1^2 (0, Z_2^\top ) \vec{\vvarXi}^{-1}\vec{Z}_1\right]\\
	= & \E \Big[1_Q \delta_1^2 (0, Z_2^\top ) \Big(\vec{\vvarXi}^{-1}_{-\{1,2\}} - \vec{\vvarXi}^{-1}_{-\{1,2\}}\frac{1}{n}(\vec Z_1\vec Z_1^\top +\vec Z_2\vec Z_2^\top )\vec{\vvarXi}^{-1}_{-\{1,2\}}\\
	& + \vec{\vvarXi}^{-1}_{-\{1,2\}}\vec\vvarXi_{1,2}\vec{\vvarXi}^{-1}_{-\{1,2\}}\vec\vvarXi_{1,2}\vec{\vvarXi}^{-1} \Big)\vec{Z}_1\Big].
	\end{split}
	\end{equation}
	We will calculate each term in \eqref{eq:E delta_1 Z_2 Sigma Z_1} separately below. To get around the difficulty that $Q$ is dependent of $Z_1, Z_2$, we introduce another event
	$$Q' = \left\{\|\vec{\vvarXi}_{-\{1,2\}} - I\| \leq Cn^{-1/4}\right\}. $$
	Based on Lemma \ref{lm:cov_concentration} and $p = o(n^{1/2})$, we have $P(Q') \geq 1 - \exp(-cn^{1/2})$ for some constant $c>0$, $Q\subseteq Q'$ and $Q'$ is independent of $Z_1$ and $Z_2$. Then
	\begin{equation}\label{eq:E delta_1^2 Z_2 Sigma-12 Z_1}
	\begin{split}
	& \left|\E \left[1_Q\delta_1^2(0, Z_2^\top )\vec{\vvarXi}_{-\{1,2\}}^{-1}\vec{Z}_1\right]\right|\\
	\leq & \left|\E \left[1_{Q'}\delta_1^2(0, Z_2^\top )\vec{\vvarXi}_{-\{1,2\}}^{-1}\vec{Z}_1\right]\right| + \left|\E \left[1_{Q'\backslash Q}\delta_1^2(0, Z_2^\top )\vec{\vvarXi}_{-\{1,2\}}^{-1}\vec{Z}_1\right]\right|\\
	\leq & \left|\E \left\{\E _{Z_2}\left[1_{Q'}\delta_1^2(0, Z_2^\top )\vec{\vvarXi}_{-\{1,2\}}^{-1}\vec{Z}_1\right]\Big| Z_1, Z_3,\ldots, Z_n\right\}\right|\\
	& + \E 1^2_{Q'\backslash Q} \cdot \E \left[1_{Q'\backslash Q}\delta_1\|Z_2\|_2\|\vec{\vvarXi}_{-\{1,2\}}^{-1}\|\|\vec{Z}_1\|_2\right] \\
	\overset{\text{C-Z}}{\leq} & 0 + \cdot \left(\E  \delta^4\right)^{\frac{1}{2}} \left(\E 1_{Q'}\right)^{\frac{1}{8}} \left(\E \|Z_2\|^8\right)^{\frac{1}{8}} \E \left(\|\vec Z_1\|^8\right)^{\frac{1}{8}} \cdot \left(\E \left[1_{Q'\backslash Q} \vec{\vvarXi}_{-\{1,2\}}^{-1}\right]^8\right)^{\frac{1}{8}}\\
	\leq & O\left(\exp(-cn)\cdot \poly(n, p)\right).
	\end{split}
	\end{equation}
	Here ``C-Z" represents Cauchy-Schwarz inequality. Note that
	\begin{equation*}
	\begin{split}
	& (\E \delta_1^2Z_1)^\top \cdot \E (Z_2Z_2^\top Z_2) + \tau^2 = (\tau^2, \E \delta_1^2Z_1) \cdot \begin{pmatrix}
	1\\
	\E Z_2Z_2^\top Z_2
	\end{pmatrix}\\
	= & \E \delta_1^2\vec Z_1^\top  \cdot \E \vec Z_2 \vec Z_2^\top \begin{pmatrix}
	0\\
	Z_2
	\end{pmatrix}
	= \E \delta_1^2(0, Z_2^\top )^\top \vec Z_2\vec Z_2^\top \vec Z_1,
	\end{split}
	\end{equation*}
	we also have
	\begin{equation*}
	\begin{split}
	& \Bigg|\E \left[1_Q\delta_1^2(0, Z_2^\top )\vec{\vvarXi}_{-\{1,2\}}^{-1}\frac{1}{n}\left(\vec Z_2\vec Z_2^\top \right)\vec{\vvarXi}_{-\{1,2\}}^{-1}\vec{Z}_1\right]\\
	& - \frac{1}{n}\left(\E \delta_1^2Z_1\right)^\top \cdot \E \left(Z_2 Z_2Z_2^\top \right) - \frac{\tau^2}{n} \Bigg|\\
	\leq & \frac{1}{n}\left|\E \left[1_{Q'}\delta_1^2(0, Z_2^\top )\vec{\vvarXi}_{-\{1,2\}}^{-1}\vec Z_2 \vec Z_2^\top \vec{\vvarXi}_{-\{1, 2\}}^{-1}\vec Z_1\right] - \E \delta_1^2(0, Z_2^\top )\vec Z_2\vec Z_2^\top  \vec Z_1\right|\\
	& + \frac{1}{n}\left|\E \left[1_{Q'\backslash Q}\delta_1^2(0, Z_2^\top )\vec{\vvarXi}_{-\{1,2\}}^{-1}\vec Z_2 \vec Z_2^\top \vec{\vvarXi}_{-\{1, 2\}}^{-1}\vec Z_1\right]\right|\\
	\leq & \frac{1}{n} \left|\E \left[1_{Q'}\delta_1^2(0, Z_2^\top )(\vec{\vvarXi}_{-\{1,2\}}-I)\vec Z_2\vec Z_2^\top \vec{\vvarXi}_{-\{1,2\}}^{-1}\vec Z_1\right]\right|\\
	& + \frac{1}{n}\left|\E \left[1_{Q'}\delta_1(0, Z_2^\top )I\vec Z_2\vec Z_2^\top (I - \vec{\vvarXi}_{-\{1,2\}}^{-1})\vec Z_1\right]\right|\\
	& + \frac{1}{n}\left|\E 1_{(Q')^c}\delta_1^2(0, Z_2^\top )\vec Z_2\vec Z_2^\top \vec Z_1\right|\\
	& + \frac{1}{n}\left|\E \left[1_{Q'\backslash Q}\delta_1^2(0, Z_2^\top )\vec{\vvarXi}_{-\{1,2\}}^{-1}\vec Z_1 \vec Z_1^\top \vec{\vvarXi}_{-\{1, 2\}}^{-1}\vec Z_1\right]\right|.\\
	\end{split}
	\end{equation*}
	Similarly as the procedure before, one can show that the formula above is no more than $O\left(\frac{p^2}{n^{5/4}}\right) + O\left(\exp(-cn)\cdot \poly(n, p)\right).$
	Thus,
	\begin{equation}\label{ineq:lemma_term2}
	\begin{split}
	& \E \left[1_Q\delta_1^2(0, Z_2^\top )\vec{\vvarXi}_{-\{1,2\}}^{-1}\frac{1}{n}\left(\vec Z_1\vec Z_1^\top \right)\vec{\vvarXi}_{-\{1,2\}}^{-1}\vec{Z}_1\right]\\
	= & \frac{1}{n}\left(\E \delta_1^2Z_1\right)^\top \cdot \E \left(Z_2 Z_2Z_2^\top \right) + \frac{\tau^2}{n} + O\left(\frac{p^2}{n^{5/4}}\right). 
	\end{split}
	\end{equation}
	Similarly to the calculation of \eqref{eq:E delta_1^2 Z_2 Sigma-12 Z_1} we can calculate that
	\begin{equation}\label{ineq:lemma_term3}
	\begin{split}
	& \E \left[1_Q \delta_1^2 (0, Z_2^\top ) \left( \vec{\vvarXi}^{-1}_{-\{1,2\}}\frac{1}{n}(\vec Z_1\vec Z_1^\top )\vec{\vvarXi}^{-1}_{-\{1,2\}} \right)\vec{Z}_1\right]\\
	= & O\left(\exp(-cn)\poly(n, p)\right).
	\end{split}
	\end{equation}
	\begin{equation}\label{ineq:lemma_term4}
	\begin{split}
	& \left|\E \left[1_Q \delta_1^2 (0, Z_2^\top ) \left( \vec{\vvarXi}^{-1}_{-\{1,2\}}\vec\vvarXi_{1,2}\vec{\vvarXi}^{-1}_{-\{1,2\}}\vec\vvarXi_{1,2}\vec{\vvarXi}^{-1} \right)\vec{Z}_1\right]\right|\\
	\leq & \left|\E \left[1_Q\delta_1^2\|Z_2\|_2(1+cn^{-1/4})^3\|\vec{\vvarXi}_{1,2}\|^2\|\vec Z_1\|_2\right]\right| \leq O\left( \frac{p^3}{n^2}\right).
	\end{split}
	\end{equation}
	Summarizing \eqref{eq:E delta_1 Z_2 Sigma Z_1}, \eqref{eq:E delta_1^2 Z_2 Sigma-12 Z_1}, \eqref{ineq:lemma_term2}, \eqref{ineq:lemma_term3} and \eqref{ineq:lemma_term4}, we obtain \eqref{ineq:lm_1}. 
	
	\item Similarly to the calculation of \eqref{ineq:lm_1}, we have
	\begin{equation}\label{eq:lm_2_decompose}
	\begin{split}
	& \E \left[1_Q \delta_1(0, Z_1^\top ) \vec{\vvarXi}^{-1}\vec{Z}_2\delta_2\right]\\
	= & \E \Big[1_Q \delta_1(0, Z_1^\top ) \Big(\vec{\vvarXi}^{-1}_{-\{1,2\}} - \vec{\vvarXi}^{-1}_{-\{1,2\}}\frac{1}{n}(\vec Z_1\vec Z_1^\top +\vec Z_2\vec Z_2^\top )\vec{\vvarXi}^{-1}_{-\{1,2\}}\\
	& + \vec{\vvarXi}^{-1}_{-\{1,2\}}\vec\vvarXi_{1,2}\vec{\vvarXi}^{-1}_{-\{1,2\}}\vec\vvarXi_{1,2}\vec{\vvarXi}^{-1} \Big)\vec{Z}_2\delta_2\Big].
	\end{split}
	\end{equation}
	We can calculate each term of \eqref{eq:lm_2_decompose} separately and similarly as the calculation for \eqref{ineq:lm_1}, then finish the proof of \eqref{ineq:lm_2}.
	\item Similarly to the calculation of \eqref{ineq:lm_1} and \eqref{ineq:lm_2}, we have
	\begin{equation}\label{eq:lm_3_decompose}
	\begin{split}
	& \E \left[1_Q \delta_1(0, Z_2^\top ) \vec{\vvarXi}^{-1}\vec{Z}_2\delta_2\right]\\
	= & \E \Big[1_Q \delta_1(0, Z_2^\top ) \Big(\vec{\vvarXi}^{-1}_{-\{1,2\}} - \vec{\vvarXi}^{-1}_{-\{1,2\}}\frac{1}{n}(\vec Z_1\vec Z_1^\top +\vec Z_2\vec Z_2^\top )\vec{\vvarXi}^{-1}_{-\{1,2\}}\\
	& + \vec{\vvarXi}^{-1}_{-\{1,2\}}\vec\vvarXi_{1,2}\vec{\vvarXi}^{-1}_{-\{1,2\}}\vec\vvarXi_{1,2}\vec{\vvarXi}^{-1} \Big)\vec{Z}_2\delta_2\Big]
	\end{split}
	\end{equation}
	Again based on the decomposition \eqref{eq:lm_3_decompose}, we can similarly prove \eqref{ineq:lm_3}.	
	
	\item \eqref{ineq:lm_4} can be calculated similarly based on the following idea,
	\begin{equation*}
	\begin{split}
	& \left|\E \left[1_Q\delta_1(0, Z_1^\top )\vec{\vvarXi}^{-1}\vec Z_1\delta_1\right] - \E \delta^2Z^\top Z \right|\\
	\leq & \left|\E \left[1_Q\delta_1^2(0, Z_1^\top )\left(\vec{\vvarXi}^{-1} - I\right)\vec Z_1\right]\right| + \left|\E 1_{Q^c}\delta^2Z^\top Z\right|\\
	\leq & O\left(\frac{p}{n^{1/4}}\right) + O\left(\exp(-cn^{1/2})\poly(p, n)\right)\\
	= & O\left(\frac{p}{n^{1/4}}\right).
	\end{split}
	\end{equation*}
	\item Note that we have the following decomposition,
	\begin{equation*}
	\begin{split}
	& \E \left[1_Q\delta_1(0, Z_2^\top )\vec{\vvarXi}^{-1}\vec Z_3 \delta_3\right]\\
	= & \E \Bigg[1_Q \delta_1(0, Z_2^\top )\Big(\vec{\vvarXi}_{-\{123\}}^{-1} - \vec{\vvarXi}_{-\{123\}}^{-1}\vec{\vvarXi}_{123}\vec{\vvarXi}_{-\{123\}}^{-1} \\
	& \quad + \vec{\vvarXi}_{-\{123\}}^{-1}\vec{\vvarXi}_{123}\vec{\vvarXi}_{-\{123\}}^{-1}\vec\vvarXi_{123}\vec{\vvarXi}_{-\{123\}}^{-1}\\
	& \quad +\vec{\vvarXi}_{-\{123\}}^{-1}\vec{\vvarXi}_{123}\vec{\vvarXi}_{-\{123\}}^{-1}\vec\vvarXi_{123}\vec{\vvarXi}_{-\{123\}}^{-1}\vec{\vvarXi}_{123}\vec{\vvarXi}^{-1}\Big)\vec Z_3\delta_3\Bigg].
	\end{split}
	\end{equation*}
	Since $\E \delta_1 = 0$, $\E Z_2=0$, $\E \vec Z_3 \delta_3=0$, similarly as the calculation before, we have
	$$\E \left[1_Q\delta_1(0, Z_2^\top )\vec{\vvarXi}^{-1}_{-\{123\}}\vec Z_3\delta_3\right] = O\left(\exp(-cn^{1/2})\poly(p, n)\right)$$
	$$\E \left[1_Q\delta_1(0, Z_2^\top )\vec{\vvarXi}^{-1}_{-\{123\}}\vec{\vvarXi}_{123}\vec{\vvarXi}^{-1}_{-\{123\}}\vec Z_3\delta_3\right] = O\left(\exp(-cn^{1/2})\poly(p, n)\right)$$
	\begin{equation*}
	\begin{split}
	& \E \left[1_Q\delta_1(0, Z_2^\top )\vec{\vvarXi}^{-1}_{-\{123\}}\vec{\vvarXi}_{123}\vec{\vvarXi}^{-1}_{-\{123\}}\vec{\vvarXi}_{123}\vec{\vvarXi}^{-1}_{-\{123\}}\vec Z_3\delta_3\right]\\
	= & O\left(\exp(-cn^{1/2})\poly(p, n)\right)
	\end{split}
	\end{equation*}
	\begin{equation*}
	\begin{split}
	& \E \left[1_Q\delta_1(0, Z_2^\top )\vec{\vvarXi}^{-1}_{-\{123\}}\vec{\vvarXi}_{123}\vec{\vvarXi}^{-1}_{-\{123\}}\vec{\vvarXi}_{123}\vec{\vvarXi}^{-1}_{-\{123\}}\vec{\vvarXi}_{123}\vec{\vvarXi}^{-1}\vec Z_3\delta_3\right]\\
	\leq & O\left(\frac{p^4}{n^3}\right).
	\end{split}
	\end{equation*}	
	\item Since $\E [\bar \Bdelta^2] = \frac{\E \delta^2}{n} = \frac{\tau^2}{n}$, we have
	\begin{equation}
	\begin{split}
	\left|\E \left[1_Q\bar \Bdelta^2\right] - \frac{\tau^2}{n}\right| = & \left|\E 1_{Q^c}\bar\Bdelta^2\right|\leq \sqrt{\E 1_{Q^c}^2 \cdot \E \bar\Bdelta^4}\\
	\leq & C\exp(-cn^{1/2})\poly(n),
	\end{split}
	\end{equation}
	which implies \eqref{ineq:lm_6}.
	
	\item We can calculate that
	\begin{equation}
	\begin{split}
	& \left|\E \left[1_Q\left(\left(0, 1_n\Z^\top \right)\vec{\vvarXi}^{-1}\vec{\Z}^\top \Bdelta\right)^2\right] - \E \left((0, 1_n\Z^\top ) \vec \Z^\top \Bdelta\right)^2\right|\\
	\leq & \left|\E \left[1_Q\left((0, 1_n\Z^\top )\vec{\vvarXi}^{-1}\vec{\Z}^\top \Bdelta\right)^2 - 1_Q\left((0, 1_n\Z^\top )I\vec{\Z}^\top \Bdelta\right)^2\right]\right|\\
	& + \left|\E 1_{Q^c}\left((0, 1_n\Z^\top )\vec{\Z}^\top \Bdelta\right)^2\right|\\
	\leq & \left|\E \left[1_Q (0, 1_n\Z^\top )(\vec{\vvarXi}^{-1}-I)\vec{\Z}^\top \Bdelta \cdot 1_Q (0, 1_n\Z^\top )(\vec{\vvarXi}^{-1}+I)\vec{\Z}^\top \Bdelta\right]\right|\\
	& + \E 1_{Q^c} \|1_n \Z^\top\|_2^2 \cdot \|\vec{\Z}^\top \Bdelta\|_2^2\\  
	\leq & \E 1_Q \|1_n \Z^\top \|_2^2 \cdot \|\vec{\Z}^\top \Bdelta\|_2^2 \|\vec{\vvarXi}^{-1} - I\| \cdot \|\vec{\vvarXi}^{-1} + I\| + \E 1_{Q^c} \|1_n \Z^\top\|_2^2 \cdot \|\vec{\Z}^\top \Bdelta\|_2^2\\
	\leq & \left(\E \|\vec{Z}^\top \Bdelta\|_2^3 \right)^{\frac{2}{3}}\cdot \left(\E 1_Q \|\vec{\vvarXi}^{-1} - I\|^6 \|\vec{\vvarXi}^{-1} + I \|^6 \right)^{\frac{1}{6}}\cdot \left(\E\| 1_n \Z^\top\|_2^{12}\right)^{\frac{1}{6}}\\
	& + \left(\E 1_{Q^c}\right)^{\frac{1}{6}}\left(\E\|\vec{\Z}^\top \Bdelta\|_2^3\right)^{\frac{2}{3}}\left(\E\|1_n \Z^\top \|_2^{12}\right)^{\frac{1}{6}}\\
	\leq & C(pn)^2n^{-1/4},
	\end{split}
	\end{equation}
	\begin{equation*}
	\begin{split}
	& \E \left((0, 1_n\Z^\top )\vec{\Z}^\top \Bdelta\right)^2 = \sum_{i,j,k,l=1}^n (0, Z_i^\top )\vec{Z}_j\delta_j (0, Z_k^\top )\vec{Z}_l\delta_l\\
	= & \sum_{i=1}^n \E \left(Z_i^\top  Z_i\delta_i\right)^2\\
	& +\sum_{1\leq i\neq j\leq n}\left(Z_i^\top  Z_j\delta_jZ_i^\top  Z_j\delta_j + Z_i^\top  Z_i\delta_iZ_j^\top  Z_j\delta_j + Z_i^\top  Z_j\delta_jZ_j^\top  Z_i\delta_i\right)\\
	= & O\left(np^2\right) + n(n-1)\bigg\{\tr\left(\E Z_iZ_i^\top  \cdot \E Z_j\delta_j^2Z_j^\top \right)\\
	& \quad  + \left(\tr\left(Z\delta Z^\top \right)\right)^2 + \tr\left(\E Z\delta Z^\top \right)^2\bigg\}\\
	= & n^2 \left(\tr(Z\delta^2Z^\top) + \left(\tr(\E Z\delta Z^\top )\right)^2 + \|\tr(\E Z\delta Z^\top )\|_F^2\right) + O\left(np^2\right).
	\end{split}
	\end{equation*}
	Combine the two equalities above, we obtain \eqref{ineq:lm_7}.
\end{enumerate}
\end{proof}

\ \par

\begin{Lemma}[Separate Analysis in proof of Theorem \ref{th:MSE_SPLS}]\label{lm:theta_SPLS_four_terms} Under the setting in Theorem \ref{th:MSE_SPLS}, we have
	\begin{equation}\label{eq:lm_SPLS_term1}
	\E \sum_{i,j,k=1}^n X_i^\top \beta_{(2)} \left(0, X_j\right)^\top  \left(\vec{\X}^\top  \vec{\X}\right)^{-1}\vec{X}_k\delta_k 1_Q = O\left(p^2\right),
	\end{equation}
	\begin{equation}\label{eq:lm_SPLS_term2}
	\E \sum_{i,j,k=1}^n \delta_i \left(0, X_j\right)^\top  \left(\vec{\X}^\top  \vec{\X}\right)^{-1}\vec{X}_k\delta_k 1_Q = O\left(p^2\right),
	\end{equation}
	\begin{equation}\label{eq:lm_SPLS_term3}
	\E \sum_{i=n+1}^{n+m}\sum_{j,k=1}^n (X_i^\top \beta_{(2)}) \left(0, X_j\right)^\top  \left(\vec{\X}^\top  \vec{\X}\right)^{-1}\vec{X}_k\delta_k 1_Q = 0,
	\end{equation}
	\begin{equation}\label{eq:lm_SPLS_term4}
	\E \sum_{i,k=1}^n\sum_{j=n+1}^{n+m} (\frac{1}{m+n}X_i^\top \beta_{(2)} + \frac{1}{n}\delta_i) \left(0, X_j\right)^\top  \left(\vec{\X}^\top  \vec{\X}\right)^{-1}\vec{X}_k\delta_k 1_Q = 0,
	\end{equation}
	\begin{equation}\label{eq:lm_SPLS_term5}
	\E \sum_{i,j=n+1}^{n+m}\sum_{k=1}^n X_i^\top \beta_{(2)} \left(0, X_j\right)^\top  \left(\vec{\X}^\top  \vec{\X}\right)^{-1}\vec{X}_k\delta_k 1_Q = O\left(p^2\right).
	\end{equation}
\end{Lemma}	
\begin{proof}[Proof of Lemma \ref{lm:theta_SPLS_four_terms}]
We first consider \eqref{eq:lm_SPLS_term1}. By the fact that $X_1,\cdots, X_n$ are i.i.d. distributed, we have
\begin{equation}\label{eq:term1_decompose}
\begin{split}
& \E \sum_{i,j,k=1}^n \E  X_i^\top \beta_{(2)} \left(0, X_j\right)^\top  \left(\vec{\X}^\top  \vec{\X}\right)^{-1}\vec{X}_k\delta_k 1_Q \\
= & n(n-1)(n-2)\E X_1^\top \beta_{(2)} (0, X_2)^\top \left(\vec{\X}^\top \vec{\X}\right)^{-1} \vec{X}_3\delta_31_Q\\
& + n\E X_1^\top \beta_{(2)} (0, X_1)^\top \left(\vec{\X}^\top \vec{\X}\right)^{-1} \vec{X}_1\delta_11_Q\\
& + n(n-1)\E X_1^\top \beta_{(2)} (0, X_2)^\top \left(\vec{\X}^\top \vec{\X}\right)^{-1} \vec{X}_1\delta_11_Q\\
& + n(n-1)\E X_1^\top \beta_{(2)} (0, X_2)^\top \left(\vec{\X}^\top \vec{\X}\right)^{-1} \vec{X}_2\delta_21_Q\\
& + n(n-1)\E X_1^\top \beta_{(2)} (0, X_1)^\top \left(\vec{\X}^\top \vec{\X}\right)^{-1} \vec{X}_2\delta_21_Q.\\
\end{split}
\end{equation}
Note the expansion of 
\begin{equation*}
\begin{split}
\vec{\vvarXi} \overset{\text{Lemma \ref{lm:inverse_expansion}}}{=} & \Big(\vec{\vvarXi}_{-\{123\}}^{-1} - \vec{\vvarXi}_{-\{123\}}^{-1}\vec{\vvarXi}_{123}\vec{\vvarXi}_{-\{123\}}^{-1}\\
& \quad + \vec{\vvarXi}_{-\{123\}}^{-1}\vec{\vvarXi}_{123}\vec{\vvarXi}_{-\{123\}}^{-1}\vec\vvarXi_{123}\vec{\vvarXi}_{-\{123\}}^{-1}\\
& \quad - \vec{\vvarXi}_{-\{123\}}^{-1}\vec{\vvarXi}_{123}\vec{\vvarXi}_{-\{123\}}^{-1}\vec\vvarXi_{123}\vec{\vvarXi}_{-\{123\}}^{-1}\vec{\vvarXi}_{123}\vec{\vvarXi}^{-1}\Big),
\end{split}
\end{equation*}
we have
\begin{equation*}
\begin{split}
& \E X_1^\top \beta_{(2)}(0, X_2)^\top \left(\vec{\X}^\top \vec{\X}\right)^{-1} \vec{X}_3\delta_31_Q\\
= &	\frac{1}{n}\E X_1^\top \beta_{(2)}(0, X_2)^\top \vec{\vvarXi}_{-\{123\}}^{-1} \vec{X}_3\delta_31_Q\\
& - \frac{1}{n}\E X_1^\top \beta_{(2)}(0, X_2)^\top \vec{\vvarXi}_{-\{123\}}^{-1}\vec{\vvarXi}_{\{123\}}\vec{\vvarXi}_{-\{123\}}^{-1} \vec{X}_3\delta_31_Q \\
& + \frac{1}{n}\E X_1^\top \beta_{(2)}(0, X_2)^\top \vec{\vvarXi}_{-\{123\}}^{-1}\vec{\vvarXi}_{\{123\}}\vec{\vvarXi}_{-\{123\}}^{-1}\vec{\vvarXi}_{\{123\}}\vec{\vvarXi}_{-\{123\}}^{-1} \vec{X}_3\delta_31_Q \\
& + \frac{1}{n}\E X_1^\top \beta_{(2)}(0, X_2)^\top \vec{\vvarXi}_{-\{123\}}^{-1}\vec{\vvarXi}_{\{123\}}\vec{\vvarXi}_{-\{123\}}^{-1}\vec{\vvarXi}_{\{123\}}\vec{\vvarXi}_{-\{123\}}^{-1} \vec{\vvarXi}_{\{123\}}\vec{\vvarXi}^{-1}\vec{X}_3\delta_31_Q.
\end{split}
\end{equation*}
Similarly to the proof of Lemma \ref{lm:five terms}, we can compute that
\begin{equation*}
\begin{split}
& \frac{1}{n}\E X_1^\top \beta_{(2)}(0, X_2)^\top \vec{\vvarXi}_{-\{123\}}^{-1} \vec{\X}_3\delta_31_Q\\
& - \frac{1}{n}\E X_1^\top \beta_{(2)}(0, X_2)^\top \vec{\vvarXi}_{-\{123\}}^{-1}\vec{\vvarXi}_{\{123\}}\vec{\vvarXi}_{-\{123\}}^{-1} \vec{X}_3\delta_31_Q\\
& + \frac{1}{n}\E X_1^\top \beta_{(2)}(0, X_2)^\top \vec{\vvarXi}_{-\{123\}}^{-1}\vec{\vvarXi}_{\{123\}}\vec{\vvarXi}_{-\{123\}}^{-1}\vec{\vvarXi}_{\{123\}}\vec{\vvarXi}_{-\{123\}}^{-1} \vec{X}_3\delta_31_Q \\
= & \poly(p, n)\exp(-cn^{1/2}),
\end{split}
\end{equation*}
\begin{equation*}
\begin{split}
& \frac{1}{n}\E X_1^\top \beta_{(2)}(0, X_2)^\top \vec{\vvarXi}_{-\{123\}}^{-1}\vec{\vvarXi}_{\{123\}}\vec{\vvarXi}_{-\{123\}}^{-1}\vec{\vvarXi}_{\{123\}}\vec{\vvarXi}_{-\{123\}}^{-1} \vec{\vvarXi}_{\{123\}}\vec{\vvarXi}^{-1}\vec{X}_3\delta_31_Q\\
 = &  O\left(\frac{p^4}{n^4}\right).
\end{split}
\end{equation*}

Similarly for the other terms in \eqref{eq:term1_decompose}, we can compute that
\begin{equation*}
\begin{split}
& \E X_1^\top \beta_{(2)} (0, X_1)^\top \left(\vec{\X}^\top \vec{\X}\right)^{-1} \vec{X}_1\delta_11_Q = \frac{1}{n}\E X_1^\top \beta_{(2)} (0, X_1)^\top \vec{\vvarXi}^{-1} \vec{X}_1\delta_11_Q\\
= & \frac{1}{n}\E X_1^\top \beta_{(2)} (0, X_1)^\top \left(\vec{\vvarXi}^{-1}_{-\{1\}} - \vec{\vvarXi}^{-1}_{-\{1\}}\vec{\vvarXi}_{\{1\}}\vec{\vvarXi}^{-1}\right) \vec{X}_1\delta_11_Q\\
= & O\left(\frac{p^2}{n^2}\right),
\end{split}
\end{equation*}
\begin{equation*}
\begin{split}
& \E X_1^\top \beta_{(2)} (0, X_2)^\top \left(\vec{\X}^\top \vec{\X}\right)^{-1} \vec{\X}_1\delta_11_Q + \E X_1^\top \beta_{(2)} (0, X_2)^\top \left(\vec{\X}^\top \vec{\X}\right)^{-1} \vec{\X}_2\delta_21_Q\\
& + \E X_1^\top \beta_{(2)} (0, X_1)^\top \left(\vec{\X}^\top \vec{\X}\right)^{-1} \vec{\X}_2\delta_21_Q = O\left(\frac{p^3}{n^3}\right).
\end{split}
\end{equation*}
Combining the inequalities above, decomposition \eqref{eq:term1_decompose} along with the fact that $p = o(n^{1/2})$, we can get \eqref{eq:lm_SPLS_term1}. 

Next, the proofs of \eqref{eq:lm_SPLS_term2} and \eqref{eq:lm_SPLS_term5} are essentially the same as \eqref{eq:lm_SPLS_term1}, which we do not repeat here. The proofs to \eqref{eq:lm_SPLS_term3} and \eqref{eq:lm_SPLS_term4} follows from the setting that $\{X_i\}_{i=n+1}^{n+m}$ are with mean zero and independent of $\{(\delta_i, X_i)\}_{i=1}^{n}$.
\end{proof}

 \end{document}